\documentclass[aps,prb,11pt,superscriptaddress,onecolumn,amsmath,amssymb,floatfix,eqsecnum,nofootinbib]{revtex4} 
\usepackage{graphicx} 
\usepackage{psfrag} 
\usepackage{amsmath,amssymb} 
\usepackage{colordvi} 

\newcommand{\beq}{\begin{equation}} 
\newcommand{\eeq}[1]{\label{#1} \end{equation}} 
\newcommand{\beqn}{\begin{eqnarray}} 
\newcommand{\eeqn}{\end{eqnarray}} 
\newcommand{\been}{\begin{enumerate}} 
\newcommand{\een}{\end{enumerate}}

\def\ket#1{\vert#1\rangle}      
\def\ipr#1#2{\langle#1\vert#2\rangle} 
\def\me#1#2#3{{\langle#1\vert#2\vert#3\rangle}} 
 
\def\s12{spin-${\frac{1}{2}}$}

\def \overleftright#1{#1 {\kern-8pt\hbox{{\raise10pt\hbox 
{$\scriptstyle{\leftrightarrow}$}}}}} 
 
\def\ie{{\it i.e.,\ }}

\def\sqr#1#2{{\vcenter{\vbox{\hrule height.#2pt \hbox{\vrule 
width.#2pt height#1pt \kern#1pt \vrule width.#2pt} \hrule height.#2pt}}}} 

\def \overleftright#1{#1 {\kern-8pt\hbox{{\raise10pt\hbox 
{$\scriptstyle{\leftrightarrow}$}}}}} 

\def\pmb#1{\setbox0=\hbox{$#1$}%
\kern-.025em\copy0\kern-\wd0 
\kern.05em\copy0\kern-\wd0 
\kern-.025em\raise.0433em\box0 }

\def\q2{{Q^2}} 
\def\gtwid{\raise.3ex\hbox{$>$\kern-.75em\lower1ex\hbox{$\sim$}}} 
\def\ltwid{\raise.3ex\hbox{$<$\kern-.75em\lower1ex\hbox{$\sim$}}} 
\def\12{\frac{1}{2}} 
\def\part{\partial}

\def\low#1{\lower.5ex\hbox{${}_#1$}} 
 
\def\psl{\raise.15ex\hbox{$/$}\kern-.57em\hbox{$\partial$}} 
\def\partt{\raise.15ex\hbox{$\widetilde$}{\kern-.37em\hbox{$\partial$}}}

\newcommand{\vsquare}[1][1.5]{ 
\setlength{\unitlength}{#1 mm} 
\begin{picture}(1,1)(0,0) 
\put(0,0){\line(1,0){1}} 
\put(0,0){\line(0,1){1}} 
\put(1,1){\line(-1,0){1}} 
\put(1,1){\line(0,-1){1}} 
\put(1,0){\vector(0,1){1}} 
\end{picture} 
} 
\newcommand{\mysquare}[1][1.5]{ 
\setlength{\unitlength}{#1 mm} 
\begin{picture}(1,1)(0,0) 
\put(0,0){\line(1,0){1}} 
\put(0,0){\line(0,1){1}} 
\put(1,1){\line(-1,0){1}} 
\put(1,1){\line(0,-1){1}} 
\end{picture} 
} 
 
\newcommand{\dmin}{\Delta^{\!\! -}} 
\newcommand{\dplus}{\Delta^{\!\! +}} 
 
\begin{document} 
 
\title{Topological Order and Conformal Quantum Critical Points} 
\author{Eddy Ardonne} 
\email{ardonne@uiuc.edu} 
\affiliation{Department of Physics, University of Illinois at Urbana-Champaign, 1110 
W.\ Green St.\ , Urbana, IL  61801-3080, USA} 
\author{Paul Fendley} 
\email{fendley@virginia.edu} 
\affiliation{Department of Physics, University of Virginia, Charlottesville, VA 22901-4714, USA} 
\author{Eduardo Fradkin} 
\email{efradkin@uiuc.edu} 
\affiliation{Department of Physics, University of Illinois at Urbana-Champaign, 1110 
W.\ Green St.\ , Urbana, IL  61801-3080, USA} 
 
\date{\today} 
 
\begin{abstract} 
 
We discuss a certain class of two-dimensional quantum systems which
exhibit conventional order and topological order, as well as
two-dimensional quantum critical points separating these phases.  All
of the ground-state equal-time correlators of these theories are equal
to correlation functions of a {\em local} two-dimensional classical
model. The critical points therefore exhibit a time-independent form of
conformal invariance.  These theories characterize the universality
classes of two-dimensional quantum dimer models and of quantum
generalizations of the eight-vertex model, as well as ${\mathbb Z}_2$
and non-abelian gauge theories.  The conformal quantum critical points
are relatives of the Lifshitz points of three-dimensional anisotropic
classical systems such as smectic liquid crystals.  In particular, the
ground-state wave functional of these {\em quantum Lifshitz points} is
just the statistical (Gibbs) weight of the ordinary 2D free boson, the
2D Gaussian model. The full phase diagram for the quantum eight-vertex
model exhibits quantum critical lines with continuously-varying
critical exponents separating phases with long-range order from a
${\mathbb Z}_2$ deconfined topologically-ordered liquid phase. We show
how similar ideas also apply to a well-known field theory with
non-Abelian symmetry, the strong-coupling limit of $2+1$-dimensional
Yang-Mills gauge theory with a Chern-Simons term. The ground state of
this theory is relevant for recent theories of topological quantum computation.
\end{abstract} 
 
\maketitle

\section{Introduction} 
 
During the past decade and a half there has been an intense search for
new kinds of theories describing quantum condensed-matter
systems. Many experimental results have implied that
strongly-correlated fermionic systems exhibit qualitatively new types
of physical behavior. The now-classic example of this is the
fractional quantum Hall effect, where one of the striking consequences
of strong correlations is that the Laughlin quasiparticles have
fractional charge and fractional statistics, even though though the
microscopic degrees of freedom are electrons with integer charge and
fermionic statistics [\onlinecite{kivelson01}].
 
Traditionally one classifies different phases in terms of order
parameters which give a global characterization of the physical
state. In turn, the local fluctuations of this order parameter field
drive the phase transitions between ordered and disordered states of
these systems. This viewpoint, pioneered by Landau and his school,
has been extremely successful in condensed matter physics and in other
areas of physics, such as particle physics, through the powerful
underlying concept of spontaneous symmetry breaking. Much of the
structure of modern theory of critical phenomena is based on this
point of view [\onlinecite{lubensky95,cardy96}].
 
However, there are many different experimentally-realizable
phases (and even more realizable theoretically!) in the fractional
quantum Hall effect, but no {\em local} order parameter distinguishes
between them. These phases are incompressible liquid states which
have a fully gapped spectrum and do not break any symmetries of the
Hamiltonian.  The lack of a local order parameter led to many
interesting discussions of the off-diagonal long-range order in the
Hall effect [\onlinecite{prange:QHE}].  One particularly elegant way
of characterizing the order in the fractional quantum Hall effect is
as {\em topological order} [\onlinecite{wen90}]. The topological order
parameters are non-local; they are expectation values of operators
which are lines or loops. Because of this, they can (and do) depend on
topology: their value depends on the genus of the two-dimensional
surface on which the electrons live. One interesting characteristic of
a {\em topological phase} is that the correlation functions in the
ground state do not depend on the locations of the operators, but only
on how the loops braid through each other. In addition, the
degeneracies of these non-symmetry breaking ground states on
topologically non-trivial manifolds are determined by the topology of
these manifolds [\onlinecite{wen-niu90}].

Although so far the only unambiguous experimental realizations of
topological phases are in the fractional quantum Hall effect, there
has been considerable effort to find, both theoretically and
experimentally, condensed matter systems whose phase diagrams may
exhibit topological ground states.  Much of the current work involves
studying fractionalized phases in time-reversal invariant systems (see
e.g.\ [\onlinecite{senthil00,moessner01a}]).  One reason is that the
``normal state" of high-temperature superconductors lacks an
electron-like quasiparticle state in its spectrum.  There are
reasons to believe that frustrated magnets may also exhibit
fractionalized behavior as well.

A particularly well-known and simple model with a topological phase is
the {\em quantum dimer model}, which was invented as a way of modeling
the short-range resonating-valence-bond theory of superconductivity
[\onlinecite{kivelson87}]. The degrees of freedom of this
two-dimensional model are classical dimers living on a two-dimensional
lattice. With a special choice of Hamiltonian (called the RK point),
the exact ground-state wave function can be found
[\onlinecite{rokhsar88}].  When the dimers are on the square lattice,
the result is a critical point. If one deforms this special
Hamiltonian, one generically obtains ordered phases. However, a
topological phase occurs in the quantum dimer model on the triangular
lattice [\onlinecite{moessner01a}].  When the quantum dimer model is
in a topological phase, an effect analogous to fractionalization
occurs [\onlinecite{kivelson87}]. This is called spin-charge
separation. One can view the dimers as being created by
nearest-neighbor pairs of lattice electrons in a spin-singlet
state. Even though the fundamental degrees of freedom (the electrons)
have both spin and charge, one finds that the basic excitations have
either charge (holons) or spin (spinons), but not both. To prove this
occurs, one must show that if one breaks apart an electron pair
(dimer) into two holons or two spinons, they are deconfined.  For the
triangular-lattice quantum dimer model, this was shown in Ref.\
[\onlinecite{moessner01a}]; the analogous statement in terms of
holon-holon correlators was proven in Ref.\
[\onlinecite{fendley02}]. At a quantum critical phase transition
between an ordered/confining phase and a disordered/deconfining phase
(or between different confining states), confinement is lost: the RK
point is deconfining [\onlinecite{fradkin90b,fradkin91,moessner02a}].

The notion of spin-charge separation is one of the basic assumptions
behind the RVB theories of high-temperature superconductivity
[\onlinecite{anderson87,kivelson87,baskaran88,affleck88,kotliar88,ioffe89,palee98,kalmeyer88,wen89}],
which effectively can be regarded as strongly-coupled lattice gauge
theories. In $2+1$-dimensional systems spin-charge separation can only
take place if these gauge theories are in a deconfined phase
[\onlinecite{read-sachdev90,fradkin91,mudry94}]. In $2+1$-dimensions
this is only possible for discrete gauge symmetries. For a continuous
gauge group, say $U(1)$ or $SU(2)$, $2+1$-dimensional gauge theories
are always in a confining phase, unless the matter fields carry a
charge higher than the fundamental charge so that the gauge symmetry
is broken to a discrete subgroup
[\onlinecite{fradkin-shenker79}]. Thus, the only consistent scenarios
for spin-charge separation necessarily involve an effective discrete
gauge symmetry, which in practice reduces to the simplest case
${\mathbb Z}_2$. Of particular interest is the fact that the
low-energy sector of the deconfined phases of discrete gauge theories
are the simplest topological field theories
[\onlinecite{witten88,preskill90,dewild95}].

Many of these ideas have their origin in the conceptual description of
confined phases of gauge theories as monopole condensates, and of
their deconfined states as ``string
condensates" [\onlinecite{polyakov-book}].  In gauge theories it has
long been known that their phases cannot characterized by a local
order parameter, since local symmetries cannot be spontaneously
broken. The phases of gauge theories are understood instead in terms
of the behavior of generally non-local operators such as Wilson loops
and disorder operators [\onlinecite{thooft78,fradkin78}], a concept
borrowed from the theory of the two-dimensional Ising
magnet [\onlinecite{kadanoff-ceva71}].

Interesting as they are, the applicability of these ideas to the
problem of high-temperature superconductivity and other strongly
correlated systems is still very much an open problem. Topological
fractionalized ground states are not the only possible explanation of
the unusual physics of the cuprates. In fact, when constructing local
microscopic models of strongly-correlated systems which are suspected
to have fractionalized phases, many theorists have found that instead
these models have a strong tendency to exhibit spatially-ordered
states, {\it a.k.a.\/} ``valence bond crystals", which appear to
compete with possible deconfined states. It is now clear that
the regimes of strongly-correlated systems which may favor
fractionalized phases also favor, and perhaps more strongly,
non-magnetic spatially ordered states of different types, including
staggered flux states [\onlinecite{affleck88}], or $d$-density wave
states [\onlinecite{ddw}], and electronic analogs of liquid
crystalline phases [\onlinecite{nature,subir-rmp}]. By now there are a
number of examples of models with short-range interactions whose phase
diagrams contain both fractionalized and spatially-ordered phases
[\onlinecite{moessner01a,moessner01b,balents02,senthil02a}].  It has
recently been proposed that deconfined critical points may describe
the quantum phase transitions between ordered N\'eel states and
valence bond crystals [\onlinecite{vishwanath03b}].

For several reasons, most of the studies of topological order and 
quantum critical points have focused on examples with two spatial 
dimensions. The experimental reason is that the Hall effect is 
two-dimensional, and typical strongly-correlated systems, such as  
the cuprate high-temperature superconductors, are often effectively 
two-dimensional. Theoretically, it is because in two dimensions 
particles can have exotic statistics interpolating between bosonic and 
fermionic [\onlinecite{wilczekbook}].  A common characteristic of 
topological phases in two dimensions is the presence of exotic 
statistics, which occur in the fractional quantum Hall effect 
[\onlinecite{prange:QHE}].  The statistics can even be non-abelian: in 
some cases, the change in the wave function depends on the order in 
which particles are exchanged [\onlinecite{moore-read91}]. Systems with 
non-abelian statistics are particularly interesting because they are 
useful for error correction in quantum computers 
[\onlinecite{kitaev97,freedman01a,freedman01b,preskill02b,ioffe03}]. 

In this paper we will discuss models with topological phases and
ordered phases, as well as quantum phase transitions separating
them. There has also been a great deal of interest in quantum critical
points in and of themselves [\onlinecite{sachdev-book}]. At a quantum
critical point, the physics is of course scale invariant, but it need
not be Lorentz invariant. The quantum critical points discussed in
this paper have dynamical critical exponent $z=2$, instead of the
usual $z=1$ of a Lorentz-invariant theory.  This allows for some
striking new physics. The action of these $z=2$ quantum critical
points is invariant under time-independent conformal transformations
of the two-dimensional space.  A remarkable consequence is that the
{\em ground-state wave functionals} of the field theories discussed
here are conformally invariant in space. This means that the ground
state wave functional is invariant under any angle-preserving
coordinate transformations of space.
For two-dimensional space, there is an infinite set of such
transformations, as is familiar from studies of two-dimensional
conformal field theory[\onlinecite{yellow}]. This sort of behavior is
not common at all: the action of a field theory at a critical point is
often scale invariant (and also conformally invariant), 
but the ground-state wave functional itself in general is
not. We dub critical points with this behavior {\em conformal quantum
critical points}.

One of the consequences of the conformal {\em invariance} of the
ground state wave function is that all the equal-time correlators of
the quantum theory are equal to suitable correlation functions of
observables of a two-dimensional Euclidean conformal field theory. We
will exploit this connection in this paper quite extensively. However,
just as important, conformal invariance of the wave function implies
that the ground state of this $2+1$-dimensional theory at a conformal
quantum critical point must have zero resistance to shear stress in
the two-dimensional plane. 

This can be seen as follows. Consider an
infinitesimal local distortion of the geometry of the two-dimensional
plane represented by an infinitesimal change $\delta g_{ij}(x)$of the
two-dimensional metric, as is conventional in the theory of
elasticity[\onlinecite{lubensky95}]\footnote{Recall that the change in the metric, given by the strain tensor, is quadratic in the local deformation of the system.}.  Let $\ket{\Psi}$ be the ground
state wave function for the undistorted plane and $\ket{\Psi(g)}$ be
the ground state wave function in the distorted plane with
two-dimensional metric $g_{ij}(x)=\delta_{ij}+\delta g_{ij}(x)$. Under
this distortion the Hamiltonian of the system changes by an amount
\begin{equation}
\delta H(g)=\int d^2 x \; \frac{\delta H}{\delta g_{ij}(x)} \; 
\delta g_{ij}(x)+\ldots
\label{deltaH}
\end{equation}
To first order in perturbation theory in $\delta H$, the change of the
ground state wave energy is
\begin{equation}
\delta E_0=\frac{\me{\Psi}{\; \delta H(g)\; }{\Psi}}{\ipr{\Psi}{\Psi}}
\equiv \langle \delta H(g) \rangle= \int d^2x \;
\displaystyle{\Big\langle {\frac{\delta H }{\delta g_{ij}(x)}\Big\rangle}}
 \delta g_{ij}(x)+\ldots
\label{change}
\end{equation}
where $E_0$ is the exact ground state energy of the distorted system. 
On the other hand, the change of the norm of the ground state wave function $\|\Psi\|$ is, to all orders in perturbation theory, given by[\onlinecite{Baym90}]
\begin{equation}
\|\Psi\|^2=\frac{\partial E_0}{\partial\varepsilon_0}
\end{equation}
where $\varepsilon_0$ is the ground state energy of the undistorted system.
Thus, the change of the norm $\|\Psi\|$ is determined by the ($ 2 \times 2$) {\em stress tensor} $T_{ij}$ of the $2+1$-dimensional theory
\begin{equation}
T_{ij}(x)=\Big\langle \frac{\delta H\; }{\delta g_{ij}(x)}\Big\rangle\ .
\label{stress}
\end{equation}
On the other hand, we can regard $\| \Psi\|^2$ as the partition
function $Z$ of a two-dimensional Euclidean conformal field
theory. This theory has an Euclidean {\em stress-energy} tensor,
$T^{\rm cft}_{ij}$, defined by[\onlinecite{yellow}]
\begin{equation}
T^{\rm cft}_{ij}=-\Big\langle \frac{\delta S_{\rm cft}}{\delta g_{ij}(x)}\Big\rangle=\frac{\delta \ln Z}{\delta g_{ij}(x)}\ ,
\label{2Dcft}
\end{equation}
which essentially coincides with the {\em stress} tensor of the
$2+1$-dimensional quantum field theory defined above. 
Scale
invariance, rotational invariance and conservation require that
$T_{ij}$ be a conserved (divergence free) symmetric traceless tensor.
Consequently, the effective Hamiltonian (as well as the action) at
this quantum critical point can depend on the spatial gradients of the
field only through the ``spatial curvature", {\it e.g.\/}
$(\nabla^2\varphi)^2$ in a scalar field theory. In other words, at a
conformal quantum critical point for a scalar theory, the stiffness vanishes:
the usual $(\nabla\varphi)^2$ term is not possible. This means that the dynamical
critical exponent of this quantum critical theory
must be $z=2$. We call such theories quantum Lifshitz theories;
we will discuss such critical points in detail.

In this paper
we discuss lattice models which exhibit both ordered/confined
phases and disordered/deconfined phases. We will also discuss the
field-theory description of these phases and of the phase
transitions. To simplify matters, and to be able to obtain exact
results, we will introduce models whose ground-state wave function
will be known exactly and whose properties we will be able to
determine quite explicitly. In this sense, these models are a
generalization of the quantum dimer model at the RK point. The basis
of the Hilbert space of these models is the configuration space of
a two-dimensional classical statistical-mechanical system or
Euclidean field theory.  Each of these basis states is defined to be
orthogonal with respect to the others. An arbitrary state in this
Hilbert space can therefore be described as some linear combination of
these basis elements.  Describing the Hilbert space in such a fashion
is not particularly novel.  The unusual feature of the models we will
discuss is that the ground-state wave function can be expressed as in
terms of the action or Boltzmann weights of a {\em local}
two-dimensional classical theory.  The normalization of the wave
function will then be the partition function or functional integral of
the classical two-dimensional model. This special property is why the
wave functionals at the critical points are have a time-independent
conformal invariance at their critical points. The field theory of these
conformal quantum critical points can be extended to describe nearby
ordered and disordered phases, including their confinement
properties. We will study this quite explicitly in a
quantum generalization of the eight-vertex model.
However, much of the
physics we discuss should apply to topological phases and ($z=2$)
quantum critical points in general.

We will also study theories with a continuous non-abelian symmetry. We
show that, interestingly enough, it is very difficult to construct a
non-trivial conformal quantum critical point with such a symmetry. We
do find a Hamiltonian whose ground state is the doubled Chern-Simons
theory of Ref.\ [\onlinecite{freedman03,shivajinew}]. This is a
time-reversal invariant theory of interest in topological quantum
computation and in (ordinary) supercondutivity; it is in a gapped
topological phase.
 
In section \ref{sec:scale-wf} we discuss the simplest model with a
scale-invariant critical wave function, the quantum dimer model at the
RK point. Here we also introduce the quantum Lifshitz model, the
effective field theory of these new quantum critical points. In
section \ref{sec:2dwavefn}, we generalize the relation between the
quantum dimer model and the scalar field theory discussed in section
\ref{sec:scale-wf} to include perturbations which drive the system in
to a quantum disordered/deconfined phase or to a ordered/confined phase.  In
section \ref{sec:q8v}, we define the quantum eight-vertex model by
finding a Hamiltonian whose ground-state wave function is related to
the classical eight-vertex model. This will allow us to find
quantum critical lines with variable critical exponents separating a
${\mathbb Z}_2$-ordered phase from a topologically-ordered phase. It
will also allow us to place a number of previously-known models, in
particular that of Ref.\ [\onlinecite{kitaev97}], in a more general
setting.  We show in detail how to use the known results from the
Baxter solution of the classical model to map out the critical
behavior of the quantum theory. In particular we analyze in detail the
confinement and deconfinement properties of the different phases and
at criticality. In section \ref{sec:YMCS}, we study the non-abelian case,
and see that the strongly-coupled limit of Yang-Mills theory with a
Chern-Simons term has a wave functional local in two-dimensional
classical fields [\onlinecite{witten92}]. This theory is in a phase
with topological order. In three appendices we give details of the
correlators of the quantum Lifshitz field theory (Appendix
\ref{app:gaussian}), and of the gauge-theory construction of the
quantum six-vertex (Appendix \ref{app:gauge}) and eight-vertex 
(Appendix \ref{app:z2gauge}) models.

\section{Scale-Invariant Wave Functions and Quantum Criticality}
\label{sec:scale-wf}
 
The simplest lattice model we discuss is the quantum dimer model; the
simplest field theory we dub the quantum Lifshitz theory. They both 
provide very nice illustrations of the properties discussed in the
introduction.  In the quantum dimer model, 
the space of states consists of close-packed hard-core 
dimers on a two-dimensional lattice. A quantum Hamiltonian 
therefore is an operator acting on this space of dimers, taking any 
dimer configuration to some linear combination of configurations. In 
every configuration exactly one dimer must touch every site, so any 
off-diagonal term in the Hamiltonian must necessarily move more than 
one dimer. The simplest such operator is called a ``plaquette flip'': 
if one has two dimers on opposite sites of one plaquette, one can 
rotate the dimers around the plaquette without effecting any other 
dimers. For example, for the $i$th plaquette on the square lattice one 
has 
\begin{equation} 
\begin{picture}(350,30) 
\thicklines 
\put(-40,6){$\hat{F}_i:$} 
\put(0,0){\line(0,1){20}} 
\put(0.3,0){\circle*{4}} 
\put(0.3,20){\circle*{4}} 
\put(20,0){\line(0,1){20}} 
\put(20.3,20){\circle*{4}} 
\put(20.3,0){\circle*{4}} 
\put(40,8){$\longrightarrow$} 
\put(70,0){\line(1,0){20}} 
\put(70,20){\line(1,0){20}} 
\put(70.3,0){\circle*{4}} 
\put(70.3,20){\circle*{4}} 
\put(90.3,0){\circle*{4}} 
\put(90.3,20){\circle*{4}} 
\put(170,6){and} 
\put(240,0){\line(1,0){20}} 
\put(240,20){\line(1,0){20}} 
\put(240.3,0){\circle*{4}} 
\put(240.3,20){\circle*{4}} 
\put(260.3,0){\circle*{4}} 
\put(260.3,20){\circle*{4}} 
\put(280,8){$\longrightarrow$} 
\put(310.3,0){\circle*{4}} 
\put(310.3,20){\circle*{4}} 
\put(310,0){\line(0,1){20}} 
\put(330,0){\line(0,1){20}} 
\put(330.3,0){\circle*{4}} 
\put(330.3,20){\circle*{4}} 
\end{picture} 
\end{equation} 
The operator $\hat{F}_i$ is defined as zero on any other dimer 
configuration around a plaquette (i.e.\ if the $i$th plaquette 
is not flippable). We define the operator 
$\hat{V}_i$ as the identity if the plaquette is flippable, and zero otherwise.

The Rokhsar-Kivelson Hamiltonian for the quantum dimer model 
[\onlinecite{rokhsar88}] 
\begin{equation} 
H_{RK} = \sum_i (\hat{V}_i - \hat{F}_i) 
\label{HRK} 
\end{equation} 
has the remarkable property that one can find its ground states 
exactly.  They have energy zero, and every state (in a given sub-sector 
labeled by global conserved quantities) appears with equal amplitude 
in its ground-state wave function.  These properties follow from the 
facts that $H_{RK}$ is self-adjoint, and $(\hat{V}_i - \hat{F}_i)^2 = 
2(\hat{V}_i - \hat{F}_i)$.  Hamiltonians of the form $H = \sum_i 
Q^\dagger_i Q_i$ necessarily have eigenvalues $E$ obeying $E\ge 
0$. Moreover, if one can find a state annihilated by all the $Q_i$, 
then it is necessarily a ground state. The equal-amplitude sum over 
all states is indeed such a state. In the Schr\"odinger picture, the 
wave function for this state is easy to write down. Define $Z$ as the 
number of all dimer configurations in some finite volume. This is 
precisely the classical partition function of two-dimensional dimers 
with all configurations weighted equally. Then the properly-normalized 
ground-state wave function 
for any basis state $|C\rangle$ in the Hilbert space (\ie any classical dimer configuration $C$) is 
\begin{equation} 
|\Psi_0\rangle=\frac{1}{\sqrt{Z}} \sum_C |C\rangle \Rightarrow 
\Psi_0(C) = \frac{1}{\sqrt{Z}} 
\end{equation} 
The wave function of the quantum system is indeed related to the 
classical system.

One can extend this sort of analysis to compute equal-time correlators 
in the ground state. One finds simply that these correlators are given 
by the correlation functions of the two-dimensional classical 
theory. Thus for $H_{RK}$ for dimers on the square lattice, one finds 
algebraic decay of the correlation functions [\onlinecite{rokhsar88}]. 
This model is then interpreted as a critical point between two ordered 
phases of the dimers [\onlinecite{read-sachdev90,fradkin91}]. However, 
for the analogous Hamiltonian on the triangular lattice, the classical 
two-dimensional correlators are exponentially decaying 
[\onlinecite{moessner01a}]. One can also show that spinon-type 
excitations (sites without a dimer) are deconfined on the triangular 
lattice [\onlinecite{fendley02}]. This means the quantum dimer model 
with $H_{RK}$ on the triangular lattice is interpreted as being in a 
``liquid'' phase, which has a mass gap and exponential decay of 
interactions, but which has no non-zero local order parameter. 
 
Such a relation between two-dimensional quantum theories is not limited 
to lattice models, nor are the ground-state wave functions required 
to be equal-amplitude sums over all configurations. 
We will construct now a simple (non-Lorentz 
invariant) two-dimensional quantum critical field theory \ie a theory 
whose ground state wave function represents a two-dimensional 
conformal theory. 
 
Consider a free boson $\varphi(x,t)$ in two spatial dimensions and one 
time dimension. Instead of the usual Hamiltonian quadratic in derivatives, 
we use one which has been conjectured by Henley 
[\onlinecite{henley97a}] to belong to the same universality class as 
the square-lattice quantum-dimer model.  It is 
\begin{equation} 
H =  \int d^2x \; \left[\frac{\Pi^2}{2} + \frac{\kappa^2}{2}(\nabla^2\varphi)^2\right] 
\label{eq:Hboson} 
\end{equation} 
where $\Pi=\dot\varphi$ as usual. 
The associated Euclidean action for the field $\varphi$ is 
\begin{equation} 
S=\int d^3x \left[\frac{1}{2} \left(\partial_\tau \varphi\right)^2+\frac{\kappa^2}{2}\left(\nabla^2 \varphi\right)^2\right] 
\label{3D-Euclidean} 
\end{equation} 
This system, Eq.\ (\ref{3D-Euclidean}), also arises in
three-dimensional classical statistical mechanics in the field-theory
description of Lifshitz points [\onlinecite{lubensky95}], for example
in (smectic) liquid crystals. For this reason we will call the system
with Hamiltonian (\ref{eq:Hboson}) the {\em quantum Lifshitz model}.
Of particular relevance to our discussion is the long-ago observation
by Grinstein [\onlinecite{grinstein81a}] that this system is analogous
to the two-dimensional Euclidean free boson in that it represents a
line of fixed points parametrized by $\kappa$.

Let us rederive this result by quantizing the Hamiltonian (\ref{eq:Hboson}). 
We impose the canonical commutation relations 
\begin{equation} 
[\varphi(\vec x),\Pi(\vec x^{\prime})]=i\delta(\vec x -\vec x^{\prime}) 
\label{eq:ccr} 
\end{equation} 
so in the Schr{\"o}dinger picture the canonical momentum 
is the functional derivative 
$\Pi(\vec x)= -i{\delta}/{\delta\varphi(\vec x)} $. 
The Schr{\"o}dinger equation for the wave functional $\Psi[\varphi]$ is 
then 
\begin{equation} 
\int d^2 x\, 
 \left[-\12 \left(\frac{\delta}{\delta\varphi}\right)^2 + \frac{\kappa^2}{2}(\nabla^2\varphi)^2 
\right]\Psi[\varphi] = E\Psi[\varphi] . 
\label{eq:SE} 
\end{equation} 
We can find the ground-state wave function in the same fashion as we did for 
$H_{RK}$. Indeed, if we define 
\begin{equation} 
Q(x)\equiv \frac{1}{\sqrt{2}} \left(\frac{\delta}{\delta\varphi} + 
 \kappa \; \nabla^2\varphi \right)\ , \qquad\quad 
Q^\dagger(x)\equiv \frac{1}{\sqrt{2}} \left(-\frac{\delta}{\delta\varphi} 
+ \kappa\; \nabla^2\varphi \right) , 
\label{eq:QAbelian} 
\end{equation} 
the (normal-ordered) quantum Hamiltonian is then 
\begin{equation} 
H= \12  \int d^2x \; \left\{Q^\dagger(\vec x), Q(\vec x)\right\} 
-\varepsilon_{\rm vac} V 
\equiv \int d^2x \; Q^\dagger(\vec x) Q(\vec x) 
\label{eq:HQQ} 
\end{equation} 
which is Hermitian and positive.  Here $V$ is the spatial volume
(area) of the system, and we have normal-ordered the Hamiltonian by
subtracting off the (UV divergent) zero-point energy density
\begin{equation}
\varepsilon_{\rm vac}=-\frac{\kappa}{2} \lim_{\vec y \to \vec 
x} \nabla_x^2 \delta(\vec x-\vec y)>0 \ .
\end{equation}
Any state annihilated by $Q(x)$ for all $x$ must be a 
zero-energy ground state. 
The corresponding ground-state wave functional 
$\ipr{[\varphi]}{{\rm vac}}=\Psi_0[\varphi]$ satisfies 
$Q\Psi_0 [\varphi] =0 $, where $Q$ is defined in Eq.\ (\ref{eq:QAbelian}). 
This is simply a first-order functional differential equation, and is 
easily solved, giving 
\begin{equation} 
\Psi_0[\varphi] = \frac{1}{\sqrt{{\mathcal Z}}}e^{\displaystyle{-\frac{\kappa}{2} \int d^2x \left(\nabla \varphi(x)\right)^2}} 
\label{eq:Psi0} 
\end{equation} 
where ${\mathcal Z}$ is the normalization 
\begin{equation} 
{\mathcal Z}=\int [{\mathcal D} \varphi] \; 
e^{\displaystyle{-\kappa \int d^2x \; \left(\nabla \varphi\right)^2}}\ . 
\label{eq:Zboson} 
\end{equation} 
The probability of finding the ground state in the configuration 
$\ket{[\varphi]}$ is therefore 
\begin{equation} 
\big\vert \Psi_0[\varphi]\big\vert^2=\frac{1}{{\mathcal Z}} e^{\displaystyle{-\kappa \int d^2x \; \left(\nabla \varphi\right)^2}}\ . 
\label{eq:probab} 
\end{equation} 
Consequently, the ground state expectation value of products of 
Hermitian local operators ${\mathcal O}[\varphi(\vec x)]$ reduces to 
expressions of the form 
\begin{equation} 
\me{{\rm vac}}{ {\mathcal O} [\varphi( {\vec x}_1)] \ldots {\mathcal O}[\varphi( {\vec x}_n)]}{{\rm vac}} 
=\frac{1}{\mathcal Z} \int [{\mathcal D} \varphi] \;  {\mathcal O}[\varphi(\vec x_1)] \ldots {\mathcal O}[\varphi(\vec x_n)] \; 
e^{\displaystyle{-\kappa \int d^2x \; \left(\nabla \varphi\right)^2}} \ . 
\label{eq:mapping} 
\end{equation} 
 
This two-dimensional quantum theory has a deep relation with a
two-dimensional classical theory: the ground-state expectation value
of all local observables are mapped one-to-one to correlators of a
two-dimensional massless Euclidean free boson. The latter is a
well-known conformal field theory, and its correlation functions are
easily determined (for convenience we give them explicitly in Appendix
\ref{app:gaussian}).  This two-dimensional critical field theory is
conformally-invariant, so the equal-time correlators of the quantum
theory must reflect this. This scalar field theory is therefore not
only quantum critical but it also has a time-independent conformal
invariance.
 
The equal-time expectation values of the ``charge operators"
${\mathcal O}[\varphi]$, as well as the correlation functions of the
dual vortex (or ``magnetic") operators discussed in Appendix
\ref{app:gaussian}, exhibit a power-law behavior as a function of
distance, as expected at a quantum critical point.  As shown also in
Appendix \ref{app:gaussian}, their autocorrelation functions also
exhibit scale invariance albeit with a dynamic critical exponent
$z=2$.  This behavior of the equal-time correlator was shown earlier
to occur in the quantum dimer model on the square lattice at the RK
point: there is a massless ``resonon'' excitation, and the equal-time
correlation functions for two static holons has a power law behavior
equal to that of the monomer correlation function in the classical 2D
dimer model on the square lattice.

However, not all theories whose ground state can be found in this
fashion need be critical with power-law behavior. As noted above, the
quantum dimer model on the triangular lattice is not. As we will
discuss in section \ref{sec:2dwavefn}, adding say a mass-like term to
$Q(x)$ in the scalar field theory gives a theory with
exponentially-decaying correlations in the ground state. Thus one can
understand phase transitions in such theories as well, and we will
explore several of these in this paper. Nevertheless, we expect that
when the quantum phase transition is continuous, the quantum critical
points of generic theories of this type will have the basic structure
of the quantum Lifshitz model. As discussed in the introduction, only
quantum Lifshitz points can be conformal quantum critical
points.\footnote{The role of conformal invariance in 2D classical
dynamics with $z=2$ and anisotropic 3D classical Lifshitz points was
considered recently in Refs.\ [\onlinecite{henkel02,henkel03}].}
 
In the next 
sections we will show that generalizations of this $z=2$ quantum 
Lifshitz Hamiltonian also describe the quantum phase transitions 
between generalizations of the valence bond crystal states and quantum 
disordered states which describe deconfined topological fluid phases 
(provided these quantum phase transitions are continuous). Notice that 
phase transitions from deconfined to confined, uniform and 
translationally-invariant states are described by the standard Lorentz 
invariant $z=1$ critical point of gauge theories 
[\onlinecite{fradkin78,fradkin-shenker79,kogut79}]. 
 
These examples show that one can obtain precise information about some 
2d quantum systems in terms of known properties of 2d classical 
systems. The trick of doing so is in finding a set of $Q_i$ (or $Q(x)$ 
in the continuum) which annihilates the equal-amplitude state (or some 
other desired state), and then defining $H = \sum_i Q^\dagger_i Q_i$ 
[\onlinecite{arovas91}].  This seems like it should be possible to do for any 
classical 2d theory, and indeed, there are many known examples of this 
sort. However, it is not clear for a given 2d classical theory one can 
always find a Hamiltonian which is both local and ergodic (in this 
context, ergodic means that the Hamiltonian will eventually take the 
system through all of phase space with a given set of conserved 
quantum numbers). It is also not clear that even if such a Hamiltonian 
exists, whether it will have any physical relevance. 
 
Moreover, this simple relation of the ground-state wave function of a 
2d quantum system to a 2d classical system is not at all generic: the 
quantum dimer model with $H_{RK}$ and this quantum Lifshitz field 
theory are quite special. To illustrate this, let us discuss briefly 
the ground-state wave functions of standard quantum field theories at 
a (quantum) critical point. Consider first the most common case, the 
Lorentz invariant $\varphi^4$ field theory at criticality. Below $D=4$ 
space-time dimensions this critical theory is controlled by its 
non-trivial Wilson-Fisher fixed point. The resulting theory is 
massless and in general it has an anomalous dimension $\eta \neq 
0$. Scale and Lorentz invariance fully dictate the behavior of {\it 
all} the correlation functions at this fixed point.  General fixed 
point theories are scale invariant and, in addition, they exhibit an 
enhanced, generally finite-dimensional, conformal symmetry. It is a 
very special feature of $D=1+1$-dimensional Lorentz-invariant fixed 
point theories that they exhibit a much larger, infinite-dimensional, 
conformal invariance. This enhanced symmetry leads to a plethora of 
critical behaviors in $1+1$ dimensions.  In contrast, there are 
relatively few known distinct critical points in higher dimensions for 
Lorentz-invariant field theories. 
 
It is well known that the knowledge of all the equal-time 
correlation functions determines completely the form of the ground 
state wave function, {\it i.e.\/} in the Schr\"odinger representation 
of the field theory [\onlinecite{symanzik81,symanzik83,fradkin93a}]. 
For a general theory, the ground-state wave function is a 
non-local and non-analytic functional of the field configuration. 
Thus, at the Wilson-Fisher fixed point, which describes theories with only a global conformal (scale) invariance, 
the structure of the ground state wave function is quite complicated.  For instance, 
the probability of a {\rm constant} field configuration $\varphi(\vec 
x)=\varphi$ at a critical point has the universal form [\onlinecite{fradkin93a}] 
\begin{equation} 
\left\vert \Psi_{\rm vac}(\varphi)\right\vert^2=A \; 
e^{\displaystyle{-B \; \vert \varphi \vert^{1+\delta}}} 
\label{eq:constant} 
\end{equation} 
For a Lorentz-invariant $\varphi^4$ theory the universal critical 
exponent is given by $\delta=(d+2-\eta)/(d-2+\eta)$, and $A$ and $B$ 
are two non-universal constants. Therefore, at criticality the wave 
function in general is a non-analytic non-local functional of the 
field configuration. 
In contrast, the ground-state wave function of a $1+1$-dimensional 
relativistic interacting fermions (a Luttinger liquid), which is a 
conformal field theory, has a universal non-local non-analytic 
Jastrow-like power law factorized form [\onlinecite{fradkin93b}] consistent 
with the form found by the Bethe-ansatz solution of the 
Calogero-Sutherland 
model [\onlinecite{sutherland71,haldane88,shastri88}]. This structure is a 
consequence of the (local) conformal invariance of the $1+1$-dimensional 
theory. Even in $1+1$ dimensions, the wave function is generically non-local. 
 
\section{Dimers, fermions and the quantum Lifshitz field theory} 
\label{sec:2dwavefn} 
 
In Section \ref{sec:scale-wf}, we showed how to find the exact ground-state wave 
functions of the quantum dimer model with Rokhsar-Kivelson Hamiltonian 
$H_{RK}$, and the quantum Lifshitz scalar field theory.  In this 
section, we will describe their properties in more detail, and some 
simple generalizations. 
 
\subsection{From the square to the triangular lattice}
\label{sec:from} 
 
The ground states of $H_{RK}$ are the sum over all classical dimer 
configurations in a sector with equal amplitudes [\onlinecite{rokhsar88}]. 
There are a number of useful generalizations to models where the 
ground state is still a sum over all states in a sector, but not 
necessarily with equal amplitudes. One interesting case is 
a quantum dimer model which interpolates between 
the square and triangular lattices. 
 
\begin{figure}[h] 
\begin{center} 
\psfrag{a}{$\longleftrightarrow$} 
\psfrag{h}{\hspace{-.5cm} horiz} 
\psfrag{v}{\hspace{-.4cm} vert} 
\psfrag{s}{\hspace{-.6cm} square} 
\includegraphics[width= .7\textwidth]{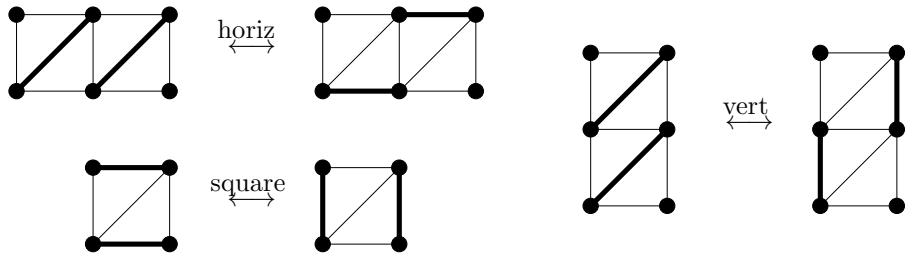} 
\caption{Dimer flips on the triangular lattice} 
\label{triflips} 
\end{center} 
\end{figure}

A triangular lattice can be made 
from a square lattice by adding bonds across all the diagonals in one 
direction. The classical dimer model on the square lattice can be 
deformed continuously into the triangular-lattice model by assigning a 
variable weight $w$ for allowing dimers along these diagonals: for 
$w=0$ we have the original square-lattice model, while $w=1$ gives the 
triangular-lattice one [\onlinecite{fendley02}]. 
There is a two-dimensional quantum Hamiltonian which has the 
$w$-dependent classical dimer model as its ground state. For dimers on 
opposite sides of a plaquette of the original square lattice, the 
Hamiltonian remains $H_{RK}$. In addition, however, parallel dimers on 
adjacent diagonals can also be flipped [\onlinecite{moessner01a}], see 
figure \ref{triflips}. 
Like the flip on the square lattice, this flip can be done without violating 
the close-packing and hard-core constraints. 
 
When the off-diagonal terms in the Hamiltonian $H_w$ consist solely of these flips, 
$H_w$ breaks up into two-by-two blocks like
$H_{RK}$ does.  In the ground state dimers along diagonals should get a 
weight $w$, so the Hamiltonian $H_{w}$ must explicitly depend on $w$. 
To construct a Hamiltonian with the desired ground state, we find a set 
of operators $Q_i$, each of which annihilates this state. 
For example, denote a configuration with dimers on adjacent diagonals as
$|1\rangle$, and $|2\rangle$ as the configuration to which it is flipped,
as shown in fig.\ \ref{triflips}.
The $Q_i$ acting on these two configurations is then 
\begin{equation} 
Q_i = \frac{1}{w^2+w^{-2}} 
\begin{pmatrix} 
w^{-2}&-1\\ 
-1&w^2 
\end{pmatrix} 
\label{Qtriang} 
\end{equation} 
where the first row and column correspond to state $|1\rangle$, while
the second corresponds to state $|2\rangle$. There are
three types of $Q_i$: $Q^{(\hbox{square})}_i$ acts on dimers on
opposite sides of a plaquette of the original square lattice, while
$Q^{(\hbox{horiz})}_i$ and $Q^{(\hbox{vert})}_i$ are associated to
flips involving dimers on the `diagonal' links; see figure
\ref{triflips}. The operator
$Q^{(\hbox{square})}_i$ is given by (\ref{Qtriang}) with $w=1$.  We
then take
\begin{equation} 
H_w = \sum_i \left[Q^{(\hbox{square})}_i + 
w Q^{(\hbox{vert})}_i + wQ^{(\hbox{horiz})}_i \right] 
\end{equation} 
where the sum is over all plaquettes $i$.  Each $Q_i$ is a projection
operator, so $Q_i=Q^\dagger_i =(Q_i)^2$.
 
We have defined the operator $Q_i$ so that it
annihilates the state $w^2|1\rangle\ +\ |2\rangle$.  
The ground state of $H_w$ is then the sum over classical dimer model 
states with each state weighted by $w^{\mathcal D}$, where ${\mathcal 
D}$ is the number of dimers along diagonals in that state.  More 
precisely, one must find the conserved quantities for a given value of 
$w$ and boundary conditions; the sum over all states with weight 
$w^{\mathcal D}$ in that sector is an eigenstate of $H_{RK}$ with zero 
energy.  The ground-state wave function 
for a configuration with ${\mathcal D}$ diagonal dimers is then 
\begin{equation} 
\Psi_0[{\mathcal D}] = \frac{w^{\mathcal D}}{\sqrt{Z(w^2)}} 
\label{psiD} 
\end{equation} 
where $Z(w)$ is the partition function for the classical dimer model 
with diagonal dimers receiving weight $w$. Note the $w^2$ in the 
argument of $Z$ in the denominator: this is because in quantum 
mechanics probabilities are given by $|\Psi_0|^2$. For the square or 
triangular lattice, this is unimportant, because $w=0$ or $w=1$, both 
of which have $w^2=w$.  For $w=1$, we have the equal-amplitude sum 
over all dimer states of the triangular lattice, the model discussed 
in Ref.\ [\onlinecite{moessner01a}].  For $w = 0$, we recover a slight 
generalization of the original square-lattice quantum-dimer model of 
Rokhsar and Kivelson [\onlinecite{rokhsar88}].  In this limit, this 
Hamiltonian reduces to $H_{RK}$ plus a potential term forbidding 
dimers on adjacent diagonal links.  Isolated diagonal dimers are still 
allowed for $w\to 0$, but since none of them can be flipped, the 
Hamiltonian does not affect them at all. Thus they can be viewed as fixed 
zero-energy defects in the square-lattice quantum-dimer model. The 
ground-state wave function for a given set of defects is the equal-amplitude 
over all configurations of dimers on the sites without defects.

Since we know the exact ground-state wave function for any $w$, one 
would like to compute the correlation functions in this model. In most 
two-dimensional lattice models, even those solvable by the Bethe 
ansatz, this is extraordinarily difficult or impossible. However, the 
classical dimer model is special in that one can do such computations, 
because like the two-dimensional Ising model, it is essentially 
free-fermionic.  Precisely, its partition function and correlators 
can be written in terms of the Pfaffian (the 
square root of the determinant) of known matrices 
[\onlinecite{fishersteph}].  One can rewrite the Pfaffians in terms of 
a functional integral over Grassmann variables at every site on the 
lattice [\onlinecite{samuel}]. The action in the case of equal 
Boltzmann weights is quadratic in the Grassmann variables, so one can 
compute easily any ground-state correlation function using the dimers, 
because the dimers can be written in terms of the fermions. This was 
discussed for the triangular lattice in 
[\onlinecite{moessner01a}]. The correlators of spinon-like or 
holon-like excitations are much more complicated, but the computation 
was done for $w=0$ in [\onlinecite{fishersteph}], and for arbitrary 
$w$ in [\onlinecite{fendley02}]. On the lattice, the holon is a defect 
or monomer, a site without a dimer.  The holon-creation operators are 
not local in terms of the fermionic variables, in a manner reminiscent 
of how the spin and fermion operators are non-local with respect to 
each other in the two-dimensional Ising model.  The 
holon two-point function is valuable in that it gives an order 
parameter for the phase with topological order: if it is non-vanishing 
as two holons are taken far apart, the holons are deconfined and 
we are indeed in a topological phase. 
The existence of topological order was previously 
established for the triangular lattice $w=1$ 
[\onlinecite{moessner01a,moessner02a}]; in [\onlinecite{fendley02}] 
the explicit correlator was computed, and indicates the 
topologically-ordered phase exists for any non-zero $w$. 
 
The ground state of the quantum dimer model with $H_{w}$ is therefore 
well understood for any $w$. There are no exact results for the 
excited states, however. In fact, since $H_w$ does not have any action 
upon empty sites, one can give the holon any gap desired without 
changing the ground states.  It is therefore useful to find a 
continuum limit and study the field theory describing this model. 
In other words, we would like to understand a field theory with 
partition function equal to the continuum limit of partition function 
$Z(w)$ in Eq.\ (\ref{psiD}). 
Since ground-state correlators for the square lattice are 
algebraically decaying, the $w=0$ model is critical and should have a 
sensible continuum limit.  Indeed, when $w=0$, the Grassmann variables 
turn into a single free massless Dirac fermion field; its action is 
the usual rotationally-invariant kinetic term.  The dimer correlation 
length $\xi$ was computed exactly as a function of $w$; for the 
triangular lattice it is about one lattice spacing, while $\xi$ 
diverges as $1/w$ for $w$ small [\onlinecite{fendley02}].  Thus there 
is a field-theory description of the continuum limit of the classical 
dimer model valid as long as $w$ is scaled to zero with the lattice 
spacing $a$ such that $w/a$ remains finite. Since the action is still 
quadratic in the Grassmann variables, the resulting fermionic field 
theory remains free. However, the Dirac fermion receives a mass 
proportional to $w/a$ [\onlinecite{fendley02}]. 
 
There are several useful aspects of taking the continuum limit, apart 
from finding the excited-state spectrum.  Correlators
are easier to compute: for $w=0$ one can use conformal field 
theory [\onlinecite{yellow}], while in the scaling limit $w\to 0$ one 
can use form-factor techniques [\onlinecite{formfac}]. Another useful 
fact is that (ignoring boundary conditions), a Dirac fermion can be 
described in terms of two decoupled Ising field theories.  In the continuum, 
the holon can be written in terms of the product of the spin field in 
one Ising model with the disorder field in the other Ising model. 
Taking $w$ away from zero amounts to giving one Ising order field an 
expectation value, and the other disorder field an expectation value. 
[\onlinecite{fendley02}]. Thus one can see directly in continuum that 
the holon order parameter is non-vanishing for $w\ne 0$. We will see 
in the next section that one can also understand the physics of the quantum 
Lifshitz critical line for all $\kappa$ in terms of two (coupled) 
Ising models. 
 
\subsection{The critical field theory} 
\label{critft} 
 
We would therefore like to find a natural-looking quantum field-theory 
Hamiltonian which has as its ground-state wave functional 
\begin{equation} 
\Psi_0[\psi] = \frac{e^{-S_{Dirac}[\psi]}}{\sqrt{Z_{Dirac}}} 
\label{psidirac} 
\end{equation} 
where $S_{Dirac}$ is the usual action for a rotationally-invariant
action for a free Dirac fermion in two Euclidean dimensions.  Since
this wave functional involves Grassman numbers, the easiest way to
think of $|\Psi_0|^2$ as a weight in the path integral defining
all correlators.  While it is possible to find a Hamiltonian acting on
this fermionic basis, it is more convenient and more intuitive to
instead use bosonic variables. It is more convenient because
correlators in a massless Dirac fermion theory (including those
involving the product of spin fields) can be bosonized, meaning that
they can be written in terms of correlators of free scalar fields
[\onlinecite{kadanoff79,yellow}]. It is more intuitive because the
classical dimer model on the square lattice has a simple description
in terms of a ``height'' variable [\onlinecite{henley97a}].  A height
is an integer-valued variable, which typically in the continuum limit
turns into a scalar field. This description will allow us also to make
contact with the quantum Lifshitz model.
 
Recently, Moessner {\it et al.\/} [\onlinecite{moessner02a}] 
generalized Henley's argument of Ref. [\onlinecite{henley97a}], and 
used the connection between quantum dimer models and their dual 
quantum roughening (height) models 
[\onlinecite{fradkin90a,fradkin90b,fradkin91,read-sachdev89,read-sachdev90,zheng-sachdev89,levitov90,elser89}] to 
argue that the Hamiltonian of Eq.\ (\ref{eq:Hboson}) actually defines 
the universality class of quantum critical points between valence bond 
crystal phases. The nature of the phase transition between valence 
bond crystal states is the focus of much current research. Quite 
recent results by Vishwanath {\it et al.\/} 
[\onlinecite{vishwanath03}] \footnote{We became aware of Ref. [\onlinecite{vishwanath03}] as this paper 
was being completed.}, and by Fradkin {\it et al.\/} 
[\onlinecite{fradkin03}], show that the transition between valence 
bond crystals is generically first order, as expected from a simple 
Landau argument. Nevertheless, when the transition is continuous, it 
is described by the Hamiltonian of Eq.\ (\ref{eq:Hboson}) which must 
be regarded as a (rather rich) multicritical point. 
 
To see how the height description arises, let us go back to the 
square-lattice quantum dimer model, where $H_w$ reduces to $H_{RK}$. 
To map the square-lattice classical dimer model onto a height model, 
one first assigns a height variable to each plaquette.  In going 
around a vertex on the even sub-lattice clockwise, the height changes 
by $+3$ if a dimer is present on the link between the plaquettes, and 
by $-1$ if no dimer is present on that link. On the odd sub-lattice, 
the heights change by $-3$ and $+1$ respectively. The flip operator 
$\hat{F}_i$ on a plaquette $i$ changes the height on that plaquette by 
either $\pm 4$. To take the continuum limit, is convenient to turn 
this into a model with heights on the sites. We define $h$ on each 
site to be the average value of the four plaquette heights around that 
site\footnote{This construction has been employed extensively in 
statistical mechanics models of classical dimers and loops; see for 
instance [\onlinecite{kondev-henley94}].}; see figure \ref{heights}. 
\begin{figure}[ht] 
\begin{center} 
\setlength{\unitlength}{1 mm} 
\begin{picture}(80,20)(0,-8) 
\multiput(0,5)(25,0){4}{\line(1,0){10}} 
\multiput(5,0)(25,0){4}{\line(0,1){10}} 
\linethickness{1 mm} 
\put(5,5){\line(0,1){5}} 
\put(30,5){\line(1,0){5}} 
\put(55,5){\line(0,-1){5}} 
\put(80,5){\line(-1,0){5}} 
\multiput(2,7)(25,0){4}{$0$} 
\put(7,7){$3$} 
\put(32,7){$-1$} 
\put(57,7){$-1$} 
\put(82,7){$-1$} 
\put(7,1){$2$} 
\put(32,1){$2$} 
\put(57,1){$-2$} 
\put(82,1){$-2$} 
\put(2,1){$1$} 
\put(27,1){$1$} 
\put(52,1){$1$} 
\put(74,1){$-3$} 
\put(1,-8){$h=3/2$} 
\put(26,-8){$h=1/2$} 
\put(50,-8){$h=-1/2$} 
\put(76,-8){$h=-3/2$} 
\end{picture} 
\caption{4 height configurations} 
\label{heights} 
\end{center} 
\end{figure} 
To avoid overcounting configurations, we identify the 
height $h$ with $h+4$. 
The flip operator $\hat F$ 
corresponds to changing $h\to h \pm 1$ on all four sites around a 
plaquette (the $\pm$ depending on the sub-lattice). Columnar order for 
dimers corresponds to an expectation value for $h$, while staggered 
order for dimers corresponds to an expectation value for $\partial 
h$. One can obtain a Hamiltonian with ordered ground states by 
allowing the coefficients of the two terms in Eq.\ (\ref{HRK}) to be 
different. If the coefficient of $\hat F_i$ is larger, this favors 
columnar order; if the coefficient of $\hat V_i$ is larger, staggered 
order is favored. It is widely assumed that $H_{RK}$, where 
the coefficients are equal, describes a phase transition between the two 
kinds of order [\onlinecite{fradkin90a,fradkin90b,read-sachdev90,sachdev-park02}]. 
However, this has never been proven, and there 
exists the possibility of an intermediate ``plaquette" phase[\onlinecite{sachdev-jalabert90}]. 
 
We would now like to take the continuum limit of the quantum dimer 
model in its height description. In this limit, we identify the height 
$h$ with a scalar field $4\varphi(x)$. Like the height, the scalar 
field must be periodic, so we identify $\varphi$ with $\varphi +1$. 
We have already noted that the continuum correlators of the 
square-lattice quantum dimer model are those of a massless Dirac 
fermion. The correlators in the ground state of the quantum 
Hamiltonian in terms of $\varphi$ must be identical.  The quantum 
Lifshitz Hamiltonian Eq.\ (\ref{eq:Hboson}) and Eq.\ (\ref{eq:HQQ}) has 
ground-state correlators of the form given in Eq. (\ref{eq:mapping}).  When 
$\kappa^{-1}=2\pi$, these correlators are precisely those of a Dirac 
fermion; this is a result of the widely-known procedure known as 
bosonization [\onlinecite{yellow}]. In fact, correlators of many 
two-dimensional critical classical statistical mechanical systems, not 
just free fermions, can be written in terms of exponentials of a free 
boson [\onlinecite{kadanoff79}]; we collect some of these results 
in Appendix \ref{app:gaussian}. In the next section we will display 
quantum lattice models whose continuum limit corresponds to all values 
of $\kappa$.

The Hamiltonian $H_{RK}$ on the square lattice in the continuum limit 
is therefore identified with the quantum Lifshitz Hamiltonian with 
$\kappa^{-1}=2\pi$. This also allows a qualitative understanding of 
the physics away from the RK point [\onlinecite{henley97a}]. A phase 
with staggered order should have an expectation value of $\partial_x 
\varphi$ or $\partial_y \varphi$ in the continuum theory. Adding a 
term $(\nabla\varphi)^2$ to the Hamiltonian with negative coefficient 
will drive the system into such an ordered phase.  Adding this term 
with positive coefficient will favor a constant value of $\varphi$, 
driving the system into columnar order. Adding terms like 
$\cos(2\pi\varphi)$ will also drive the system into a phase with 
columnar order. Thus one expects that at a critical point like that 
described by $H_{RK}$ on the square lattice, the coefficients of 
$(\nabla\varphi)^2$ and $\cos(2n\pi\varphi)$ will vanish. This leaves 
Eq.\ (\ref{eq:Hboson}) as the simplest non-trivial Hamiltonian with the 
desired properties. The requirement that the ground state be 
equivalent to a free fermion then fixes the coefficient $\kappa$; note 
that if desired $\kappa$ can scaled out of the Hamiltonian by 
redefining the compactification relation to be $\varphi\sim \varphi + 
\sqrt{\kappa}$. 
 
It is not at all clear whether critical $2+1$-dimensional field theory 
and the continuum limit of $H_{RK}$ are identical for excited states, 
although the above heuristic argument is very suggestive.  The 
Hamiltonian for the scalar field theory Eq.\ (\ref{eq:HQQ}) is purely 
quadratic in the field $\varphi$, so one can obtain essentially any 
desired information exactly.  $H_{RK}$ is not so simple on the 
lattice, but one may hope the two models are in the same universality 
class.  In any event, we can easily extract all the excited-state 
energies for the quantum Lifshitz field theory. The operators $Q(x)$ 
and $Q^\dagger(x)$ are essentially harmonic-oscillator creation and 
annihilation operators: the equal-time commutation relation 
Eq.\ (\ref{eq:ccr}) implies that 
\begin{equation} 
\left[ Q(\vec x), Q^\dagger(\vec y) \right] = \kappa \; 
\nabla^2\delta^{(2)} (\vec{x} - \vec{y}) 
\label{QQccr} 
\end{equation} 
The ground state is indeed annihilated by all $Q(x)$, so we can create 
excited states by acting with $Q^\dagger(x)$. The commutation 
relations Eq.\ (\ref{QQccr}) mean that the dispersion relation is $E = 
\kappa p^2$. This theory is gapless but not Lorentz-invariant: 
the dynamical critical exponent is $z=2$. 
 
The exponent $z=2$ can also be seen by looking at the classical action 
associated with this Hamiltonian. This will also allow us to make 
contact with the earlier three-dimensional statistical-mechanical 
results. The action consistent with the Hamiltonian of 
Eq.\ (\ref{eq:Hboson}) and the canonical commutation relations of 
Eq.\ (\ref{eq:ccr}) is 
\begin{equation} 
{\mathcal S}=\int d^3x \left[\12 \left(\partial_t 
\varphi\right)^2-\frac{\kappa^2}{2}(\nabla^2\varphi)^2\right] 
\label{eq:Sboson} 
\end{equation} 
Clearly, this action is not Lorentz invariant and has $z=2$. 
It is rotationally invariant only in the XY plane. 
Defining the imaginary time $\tau=it$ gives the Euclidean 
action (\ref{3D-Euclidean}). 
The imaginary time axis $\tau$ can be regarded as the $z$-coordinate 
of a three-dimensional classical system in which $\varphi(\vec x,\tau)$ 
is an angle-like variable and the action represents the spin-wave 
approximation of an anisotropic classical $XY$ model. In general one 
would have expected a term proportional to the operator $\left(\nabla 
\varphi\right)^2$ with a finite positive stiffness in the plane. This is 
so in the $XY$ ferromagnetic phase. On the other hand, if the 
stiffness becomes negative, there is an instability to a modulated 
helical phase. The action of Eq.\ (\ref{3D-Euclidean}) 
represents the Lifshitz point, the critical point of this phase 
transition [\onlinecite{grinstein81a,grinstein82a}] where the 
stiffness vanishes. This effective action also plays a central role in 
the smectic A-C transition [\onlinecite{grinstein81b}] 
and in other classical liquid crystal phase transitions associated 
with the spontaneous partial breaking of translation and/or rotational 
invariance [\onlinecite{lubensky95,degennes98}].  The compactified 
version of the problem (\ie the identification $\varphi\sim \varphi + 1$) 
has also been considered in this context, for example in 
Ref.\ [\onlinecite{grinstein81a}]. The choice of period in general 
depends on the physical context of the problem. 
 
The square-lattice quantum dimer model and the scalar field theory 
with Hamiltonian Eq.\ (\ref{eq:HQQ}) are both at critical points, in that 
the correlators in the ground state are algebraically decaying. 
The quantum 
dynamics implied by this Hamiltonian must be {\em compatible} with the 
2D time-independent conformal invariance. In particular, the spectrum 
of the quantum theory must be gapless and, as we learned from this 
example, the dynamic critical exponent must be $z=2$. Notice however, 
that $z=2$ alone does not guarantee a gapless (or even critical) 
theory. Indeed, instructive counter-examples to this statement are 
well known in the theory of (the absence of) quantum 
roughening [\onlinecite{fisher83,fradkin83}] where quantum fluctuations 
destroy the critical behavior and lead to an ordered state through an 
order-from-disorder mechanism. 
 
\subsection{The off-critical field theory} 
\label{offcritft} 
 
We have thus shown that the continuum limit of the square-lattice 
quantum dimer model is described by the quantum Lifshitz model at a 
special point $\kappa^{-1}=2\pi$. We also argued that at least 
some deformations of the two models result in ordered 
phases. However, we saw at the beginning of this section that not all 
deformations of the square-lattice quantum dimer model result in an 
ordered phase. Allowing dimers across the diagonals with Hamiltonian 
$H_w$ results in a 
topologically-ordered phase, where the order parameter is not local.

In this subsection, we find a bosonic field theory describing the 
topological phase in the 
continuum limit. We showed above that in the scaling limit $w\to 0$ 
with $w/a$ finite, the ground-state wave function Eq.\ (\ref{psidirac}) 
can be written in terms of a free massive Dirac fermion of mass 
proportional to $w/a$. In two dimensions, the bosonic version of a 
massive Dirac fermion is the sine-Gordon model at a particular 
coupling. Precisely, the two-dimensional fermion action is equivalent 
to 
\begin{equation} 
S_{2d} =  \int d^2 x \left[ {\kappa}(\nabla\varphi)^2 - \lambda 
\cos(2\pi\varphi) \right] . 
\label{sgaction} 
\end{equation} 
For the free-fermion case, we have $\kappa=1/2\pi$; we will discuss the more general case in section \ref{sec:q8v}. 
In the fermion language, different 
values of $\kappa$ correspond to adding a four-fermion coupling to $S_{2d}$. To 
find a Hamiltonian with this two-dimensional action describing the 
ground state, we again find an operator $Q(x)$ and define the 
Hamiltonian via Eq.\ (\ref{eq:HQQ}).  The operator 
\begin{equation} 
Q (x) \equiv \frac{1}{\sqrt{2}} \left(\frac{\delta}{\delta\varphi} + 
\kappa \; \nabla^2\varphi  + \frac{\lambda}{2\pi} \sin(2\pi\varphi)\right) 
\label{Qsg} 
\end{equation} 
annihilates the wave functional $\Psi \propto e^{-S_{2d}}$. 
Because of the extra term in $Q$, the 
commutator $\left[ Q(\vec x), Q^\dagger(\vec y) \right]$ is not a simple 
c-number, but in fact depends on the field configuration $\varphi$, 
\begin{equation} 
\left[ Q(\vec x), Q^\dagger(\vec y) \right] = \kappa 
\nabla^2\left(\delta^{(2)} (\vec{x} - \vec{y}) \right) 
+ \lambda \sin(2\pi \varphi(\vec{x})) \delta^{(2)}(\vec{x}-\vec{y}) \ . 
\label{QQccrpot} 
\end{equation} 
Thus, normal-ordering the Hamiltonian is not just an innocent ground 
state energy shift: the two parts of (\ref{eq:HQQ}) are not the same 
here. To obtain the desired ground-state wave functional, we must 
define $H$ of the form $\int Q^{\dagger} Q$. 
The three-dimensional version of this model was discussed in 
Ref.\ [\onlinecite{grinstein81a}].

This Hamiltonian is not quadratic in the field $\varphi$ except at the
critical point $m=0$, so that even with this fine tuning this model
cannot be solved simply. Since $Q$ and $Q^\dagger$ do not have simple
commutation relations, we cannot simply find the spectrum of this
theory.  Of course, one can compute properties in the fermionic
picture, but as noted before, computations involving spin fields are
non-trivial in this basis as well. However, since there are
dimensionful parameters in the Hamiltonian and no spontaneous breaking
of a continuous symmetry, it seems likely that the Hamiltonian is
gapped. Moreover, in the limit 
with $\lambda/\kappa$ finite, the action Eq.\ (\ref{sgaction}) reduces
to that of a free massive boson. Then one can solve the model
explicitly, and the quantum Hamiltonian indeed has a gap. When
$\lambda$ is reduced to a finite value, the gap should
remain\footnote{The equal-time correlators of the vertex operators
defined and computed in Appendix \ref{app:gaussian} can also be
computed away from the critical point by either by a naive
semi-classical argument, which predicts the simple exponential decay
we just discussed, or in a more sophisticated way by means of form
factors [\onlinecite{formfac}]. Apart from a subtle bound state
structure in the spectral functions, the more sophisticated approach
confirms the essence of the naive semi-classical result. On the other
hand, the $z=2$ character of the critical theory suggests that
time-dependent correlation functions must be consistent with this
fact, and that the correlators must be functions of $x^2$ or $t$, and
that the time dependent Euclidean auto-correlation functions may
obtained from the equal-time correlator by replacing $x^2
\leftrightarrow |t|$. These arguments are consistent with the
renormalization-group results of Ref.\ [\onlinecite{grinstein81a}]}.

There are many terms in the Hamiltonian of this field theory, and 
their coefficients must be fine-tuned to enable us to compute the 
ground-state wave function explicitly. There are terms like 
$\cos(4\pi\varphi)$ and $(\nabla\varphi)^2$ which, as noted above, 
tend to order the system. However, the exact lattice results for $H_w$ 
from [\onlinecite{fendley02}] show that the ground-state correlators 
for $w$ small are those of the bosonic Hamiltonian with these special 
couplings. Thus this model is not ordered but rather topologically 
ordered. Moreover, since the model is gapped, we expect that its 
physical properties are robust and persist even when the coefficients 
are tuned away from this special point. Thus there must exist a 
topological phase, not just an isolated point. An interesting open 
problem is to understand how large (in coupling constant space) the 
topological phase is, as compared to the ordered phases.

\section{A quantum eight-vertex model} 
\label{sec:q8v} 
 
In the last section, we discussed a $2+1$-dimensional theory whose 
ground-state wave function is simply described in terms of a classical 
two-dimensional bosonic field. With vanishing potential and a 
particular value of the coupling $\kappa$, it is believed to describe 
the continuum limit of the quantum dimer model on the square lattice. 
In this section, we will study lattice models which in the continuum 
limit allow arbitrary values of $\kappa$. These models also have a 
quantum critical line separating an ordered phase from a 
topologically-ordered phase. 
 
\begin{figure}[ht] 
\begin{center} 
\includegraphics[width= .9\textwidth]{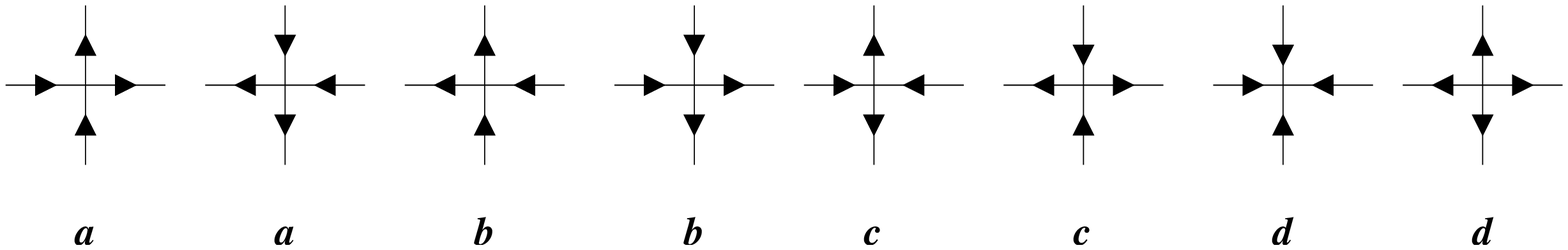} 
\caption{The eight vertices and their Boltzmann weights  } 
\label{fig8v} 
\end{center} 
\end{figure}

\begin{figure}[ht] 
\begin{center} 
\includegraphics[width= .4\textwidth]{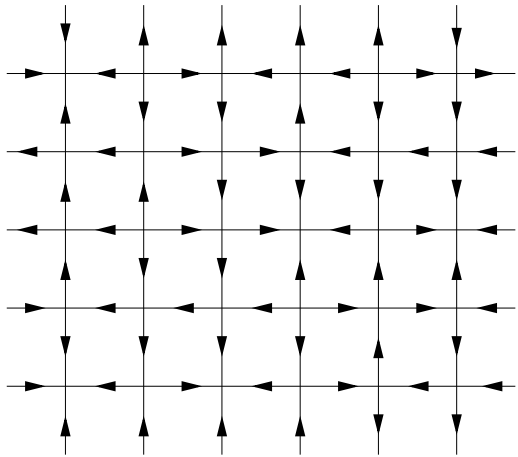} 
\caption{A typical configuration in the eight-vertex model} 
\label{fig8vtypical} 
\end{center} 
\end{figure}

The degrees of freedom in our model are those of the classical 
two-dimensional eight-vertex model. These are arrows placed on the 
links of a square lattice, with the restriction that the number of 
arrows pointing in at each vertex is even. This means that there are 
eight possible configurations at each vertex, which we display 
in figure (\ref{fig8v}). 
The classical Boltzmann weights for a given vertex in the zero-field 
eight-vertex model are usually denoted by $a$, $b$, $c$ and 
$d$, as shown in the figure. Since we are interested in rotationally-invariant 
theories, we set $a=b$ in the following; moreover, since we can rescale all the weights by a constant, we set $a=b=1$. A typical 
configuration is displayed in figure (\ref{fig8vtypical}); 
the Boltzmann weight of such a configuration is given by the product of 
Boltzmann weights of the vertices. 
 
The classical eight-vertex model is integrable, and many of its 
properties can be derived exactly [\onlinecite{baxbook}].  For 
$a=b=1$, it has ordered phases for $c>d+2$ and $d>c+2$. In these 
phases the ${\mathbb Z}_2$ symmetry of flipping all the arrows is 
spontaneously broken. 
Critical lines with continuously varying exponents at $c=d+2$ 
and $d=c+2$ separate the ordered phases from the disordered one $|c-d|<2$. 
The  correlation length diverges as [\onlinecite{baxbook}] 
\begin{equation} 
\xi \sim \big| |c-d| -2 \big|^{-\pi/(2\mu)}\ , \qquad 
\mu \equiv 2 \tan^{-1}(\sqrt{cd})\ ,
\label{critexp} 
\end{equation} 
near these critical lines. (For $\pi/\mu$ an even integer this is 
multiplied by $\log | |c-d| -2 |$ .)  When $c=0,d\leq 2$ or $d=0,c\leq 2$, 
the model 
is also critical; in fact the partition function on this line can be 
mapped onto that for the order-disorder critical line. For $d=0$, the 
exponent in Eq.\ (\ref{critexp}) diverges: there is a 
Kosterlitz-Thouless transition as one brings $c$ through $2$. Another 
useful result for the classical correlation length is that it is zero 
on the line $c=d$; this is the state of maximal disorder.

An order parameter which will be useful later comes by rewriting the 
model in terms of an Ising spin at the center of each plaquette.  This 
description is best thought of as two Ising models, with spins 
$\tau(A)$ on one sublattice, and $\tau(B)$ on the other. Then the 
Boltzmann weights can be written in terms of two Ising couplings 
between nearest sites on the $A$ lattice and on the $B$ lattice, and a 
four-spin coupling between the two $A$ and two $B$ spins around a site 
of the original lattice. The polarization operator of the eight-vertex 
model becomes $\tau(A)\tau(B)$. One finds that its expectation value 
is non-vanishing in the ordered phase, and vanishes in the disordered 
phase $|c-d|<2$ [\onlinecite{baxbook}]. One also can define 
Ne\'el-like staggered order parameters, in terms of $\tau(A)$ and 
$\tau(B)$ individually, which do not vanish in the ordered phase. 
Along the line $cd=1$, the four-spin coupling vanishes, so the 
eight-vertex model turns into two decoupled Ising models; the model 
here in this case be solved by using Pfaffian techniques 
[\onlinecite{FanWu}].  Along the line $c=d$, the two Ising couplings 
vanish, leaving only the four-spin coupling. Thus this line in the 
classical model has an extra ${\mathbb Z}_2$ gauge symmetry.

The classical eight-vertex model has a number of useful dualities 
[\onlinecite{baxbook}].  They can be described by defining the 
combinations $W_1=(a+b)/2$, $W_2=(a-b)/2$, $W_3=(c+d)/2$ and 
$W_4=(c-d)/2$. The partition function is invariant under the exchange 
of any two of the $W_j$ and under the $W_j\to -W_j$ for any $j$. These 
dualities, for example, map the critical line $c=d+2$ to the critical 
line $d=0,c\le 2$ by exchanging $W_1$ with $W_4$. In Ising language 
this amounts to performing Kramers-Wannier duality on one of the two 
types of Ising spins.  The line $c=d+2$ is invariant under the 
exchange $W_1\leftrightarrow W_3$. In Ising language, this duality 
amounts to taking the Kramers-Wannier dual of both types of Ising 
spins. Denoting the dual spins as $\mu(A)$ and $\mu(B)$, duality means 
therefore that in the disordered phases, the expectation value 
$\langle \tau(A) \tau(B)\rangle$ is non-vanishing. 
 
\subsection{Construction of the quantum eight-vertex Hamiltonian} 
\label{sec:8vqh} 
 
We now define a quantum Hamiltonian acting on a Hilbert space whose
basis elements are the states of this classical eight-vertex model.
To define such a Hamiltonian, we first need the analog of the flip
operator in the quantum dimer model. A flip operator needs to be
ergodic: by flips on various plaquettes one should be able to reach
all the states with the same global conserved quantities. The simplest
such operator for the eight-vertex model is the operator which
reverses all the arrows around a given plaquette. We write this flip
operator $\hat{\mathcal F}_i$ explicitly in gauge-theory language in
appendix \ref{app:z2gauge}.  Note that as opposed to the quantum dimer
model on the square or triangular lattice, all configurations in the
quantum eight-vertex model are flippable\footnote{However, the
quantum dimer models on the Kagome [\onlinecite{misguich}] and Fisher 
[\onlinecite{moessner02d}] lattices have Hamiltonians 
where all plaquettes are flippable.}: $\hat{\mathcal F}_i$ preserves the
restriction that an even number of arrows be pointing in or out at
each vertex.
 
The simplest Hamiltonian has no potential energy, just a flip term. It 
is convenient to write this in terms of a projection operator: for $I$ 
the identity matrix, we have $(I-\hat{\mathcal F}_i)^2 
=2(I-\hat{\mathcal F}_i)$. Then the Hamiltonian 
\begin{equation} 
H_{c=d=1}= \sum_i (I-\hat{\mathcal F}_i) 
\label{hkitaev} 
\end{equation} 
has a ground state corresponding to the equal-amplitude sum over all 
eight-vertex model states. In terms of the Boltzmann weights 
introduced above, this is the state with $a=b=c=d=1$. A Hamiltonian 
with the same ground state was introduced by Kitaev 
[\onlinecite{kitaev97}]. There the eight-vertex-model restriction of 
having an even number of arrows in and out at each vertex was not 
required a priori, but instead a term was introduced giving a positive 
energy to vertices not obeying the restriction. The zero-energy ground 
state can therefore include no such vertices, so the ground state for 
the model of [\onlinecite{kitaev97}] is indeed the sum over the 
eight-vertex-model configurations with equal weights. Because every 
plaquette is flippable and all configurations have equal weights, 
different $\hat{\mathcal F}_i$ commute.  Therefore all the terms in 
Eq.\ (\ref{hkitaev}) commute with each other, so they can be 
simultaneously diagonalized and their eigenstates can easily be found, 
as for the quantum Lifshitz field theory.  The model is gapped, and is 
in a topologically-ordered phase [\onlinecite{kitaev97}]. This follows 
as well from the results for the classical eight-vertex model 
discussed above: for $c=d=1$ all Ising couplings vanish resulting in 
two decoupled Ising models at infinite temperature. At this point the 
order parameter vanishes, but the non-local order parameter does 
not. We can thus interpret the product of dual Ising variables 
$\mu(A)\mu(B)$ as a topological order parameter. 
 
This ground state can also be mapped onto the ground state of a 
${\mathbb Z}_2$ gauge theory deep in its deconfined phase. In Appendix 
\ref{app:z2gauge}, we give a detailed derivation of the $ {\mathbb 
Z}_2$ gauge theory of the full quantum eight-vertex model we define 
below, as well as the its dual theory which we will use to 
characterize some of the phases. In Appendix \ref{app:gauge} we 
discuss the $U(1)$ gauge theory description of the quantum six-vertex 
model, which describes the limit $d=0$, and by duality, the lines 
$c=0$, $c^2=d^2+2$ and $d^2=c^2+2$. 
 
We now find a two-parameter quantum Hamiltonian whose ground state is
a sum over the states of the eight-vertex model with amplitudes given
by the classical Boltzmann weights with arbitrary $c$ and $d$.  We are
still keeping $a=b=1$ to preserve two-dimensional rotational
invariance, but this restriction can be relaxed if desired.  Our model
is neither the simplest nor the most natural extension of the
classical eight-vertex model: a simpler Hamiltonian was proposed by
Chakravarty [\onlinecite{chakravarty}] in the context of $d$-density
waves. Although we believe that our Hamiltonian and Chakravarty's
describe the same physics, we are not aware of any simple mapping
between these models.  Another lattice model related to the quantum
eight-vertex model discussed here was introduced in Ref.\
[\onlinecite{ioffe02}].  The main virtue of the construction that we
use here is the structure of the ground-state wave function.

Finding a quantum Hamiltonian with a known ground state is
straightforward to do by using the trick discussed above (and in Ref.\
[\onlinecite{arovas91}]). Namely, we find a Hamiltonian of the form
\begin{equation} \label{q8vham} 
H_{q8v} = \sum_i w_i {\mathcal Q}_i 
\end{equation} 
where ${\mathcal Q}_i={\mathcal Q}_i^\dagger \propto {\mathcal Q}_i^2$. To 
yield the desired ground state, each operator ${\mathcal Q}_i$ must 
annihilate the sum over states with each state weighted by $c^{N_c} 
d^{N_d}$, where $N_c$ and $N_d$ are the number of $c$ and $d$ type 
vertices in that state. In particular, we look for 
a ${\mathcal Q}_i$ of the form 
\begin{equation} 
{\mathcal Q}_i = \sum_i \left[\hat{\mathcal V}_i - \hat{\mathcal F}_i\right] 
\label{q8v} 
\end{equation} 
where $\hat{\mathcal V}_i$ is diagonal and depends on the Boltzmann
weights for the four vertices at the corners of the plaquette $i$.
Since $(\hat{\mathcal F}_i)^2=I$, this Hamiltonian breaks into $2$ by
$2$ blocks like $H_{RK}$ for the quantum dimer model.  If we choose
the potential ${\mathcal V}_i$ so that the blocks are of the form
\begin{equation} 
\begin{pmatrix} 
v & -1\\ 
-1 & v^{-1} 
\end{pmatrix} \ , 
\end{equation} 
${\mathcal Q}_i$ will have the desired properties.

To find $\hat{\mathcal V}_i$, let $n_c$ be the number of $c$-vertices 
at the corners of the plaquette $i$, and let $\widetilde{n}_c$ be the 
number of $c$-vertices around the plaquette after it is flipped by 
$\hat{\mathcal F}_i$.  Likewise, let $n_d$ be the number of 
$d$-vertices around the plaquette, while $\widetilde{n}_d$ is the 
number of $d$ vertices in the flipped configuration.  Note that 
$\hat{\mathcal F}_i$ always flips a $c$ or $d$ vertex to an $a$ or $b$ 
vertex, and vice versa.  Consequently, we take the operators $\hat 
{\mathcal V}_i$ to be of the form 
\begin{equation} 
\hat{\mathcal V}_i = 
c^{\widetilde{n}_c - n_c} d^{\widetilde{n}_d - n_d} 
\label{Vpot} \ . 
\end{equation} 
Explicitly, we can write the projectors as follows 
\begin{equation} 
\mathcal{Q}_i = \begin{pmatrix} 
c^{\widetilde{n}_c - n_c} d^{\widetilde{n}_d - n_d} & -1 \\ 
-1 & c^{\widetilde{n}_c - n_c} d^{\widetilde{n}_d - n_d} \\ 
\end{pmatrix} \ . 
\end{equation} 
This ${\cal Q}_i$ is indeed proportional to a projection operator: the
Hamiltonian (\ref{q8vham}) has the classical eight-vertex model as an
eigenstate.  This holds for any choice of the $w_i$, but in the
${\mathbb Z}_2$ gauge-theory language of appendix \ref{app:z2gauge},
it is natural to set all $w_i=1$.  The amplitude of the ground-state
wave function of this $H_{q8v}$ on a state with $N_c$ $c$-vertices and
$N_d$ $d$-vertices is
\begin{equation} 
\Psi_0[N_c,N_d] = \frac{c^{N_c} d^{N_d}}{\sqrt{Z(c^2,d^2)}} \ , 
\end{equation} 
where $Z(c,d)$ is the partition function of the classical two-dimensional 
eight-vertex model with weights $a=1$, $b=1$, $c$ and $d$. The arguments 
in the denominator are $c^2$ and $d^2$ because averages in the quantum 
model are calculated with respect to $|\Psi_0|^2$. 
 
In section \ref{sec:2dwavefn}, we discussed how by taking the limit
$w\to 0$ in $H_w$, one can recover a slightly-generalized
square-lattice quantum dimer model which allows defects with no
dynamics. Similarly, here one can find a quantum six-vertex model with
some defects allowed by taking the limit $d \rightarrow
0$.\footnote{This Hamiltonian in the special case $c=1$, $d=0$ was
discussed in Ref.\ [\onlinecite{balents03b}].  Moreover, the quantum
six-vertex model should be in the same universality class as
the``supersymmetric'' $XY$ model introduced some time ago
[\onlinecite{girvin93}]; this model defines a similar Hamiltonian
acting on the 2D classical XY model, and presumably can also be
understood in terms of a mapping to the quantum Lifshitz model.
Another quantum six-vertex model has been proposed as a model of a
planar pyrochlore lattice in Ref.\ [\onlinecite{moessner01c}]. This
six-vertex model is not of the Rokhsar-Kivelson type; it is the
six-vertex limit of the simpler quantum eight-vertex model we discuss
in section \ref{another}.} Configurations with
$n_d>\widetilde{n}_d$ on any plaquette will receive infinite potential
energy and so are disallowed.  Some useful facts are that $n_c+n_d +
\widetilde{n}_c +\widetilde{n}_d =4$, and $n_c-\widetilde{n}_c =0$ mod
2 and $n_d-\widetilde{n}_d =0$ mod 2.  
When $d\to 0$, configurations with
$n_d=\widetilde{n}_d=0$ are the flippable plaquettes of the six-vertex
model, and are obviously included in the ground-state wave
function. Configurations with $n_d=\widetilde{n}_d=1$ or
$n_d=\widetilde{n}_d=2$ are not suppressed, but the flip therefore
preserves the number of $d$ vertices around each plaquette. Thus in
the $d\to 0$ limit we can view these $d$ vertices as defects in the
quantum six-vertex model. The ground-state wave function is a sum over
all allowed states with a given set of plaquettes with defects. The
amplitude of each configuration in this sum is proportional to
$c^{N_c}$. We discuss the relation between the quantum six- and
eight-vertex models in more detail in the Appendices \ref{app:gauge}
and \ref{app:z2gauge}.

\subsection{The phase diagram} 
\label{sec:8vphases} 
 
\begin{figure} 
\begin{center} 
\psfrag{bz2}{Broken ${\mathbb Z}_2$} 
\psfrag{ubz2}{Unbroken ${\mathbb Z}_2$} 
\psfrag{d}{$d^2$} 
\psfrag{c}{$c^2$} 
\psfrag{confining}{Confining} 
\psfrag{deconfining}{Deconfining} 
\psfrag{kitaev}{Kitaev} 
\psfrag{kt}{KT} 
\psfrag{2}{$2$} 
\psfrag{6-vertex}{$6$-vertex} 
\psfrag{ising}{Dual $6$-vertex} 
\psfrag{I}{$I$} 
\psfrag{II}{$II$} 
\psfrag{III}{$III$} 
\psfrag{o}{Ordered} 
\psfrag{qd}{Quantum Disordered} 
\includegraphics[width=0.6 \textwidth]{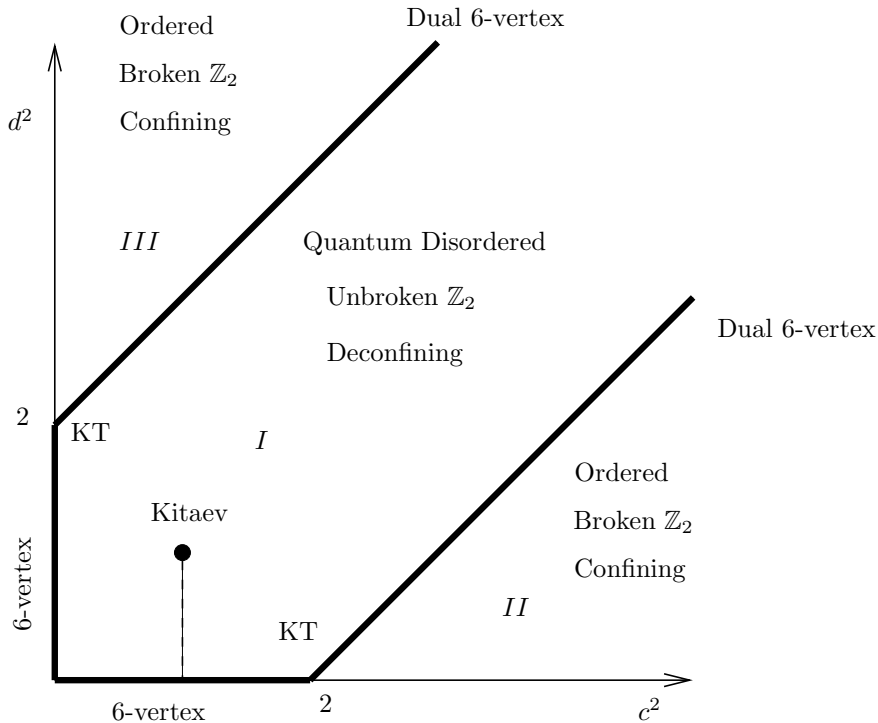} 
\caption{Phase diagram of the quantum eight-vertex model: phases $I$ and $II$ are separated by 
a dual $6$-vertex transition (same with $I$ and $III$); $6$-vertex denotes the $6$-vertex model critical lines 
and  KT are $2D$ Kosterlitz-Thouless transitions; the dotted line shows that the Kitaev point is smoothly 
connected to the critical regime of the eight-vertex model.  } 
\label{eight} 
\end{center} 
\end{figure}

Now that we have the exact ground-state wave function for the quantum 
eight-vertex Hamiltonian of \eqref{q8vham}, we can use it to determine 
the phase diagram. Given the fact that the probability of a 
configuration of arrows in the ground state wave function is equal to 
the Boltzmann weight of a classical eight-vertex model, we can deduce 
much of the physics of the quantum theory (at least its equal-time 
properties) directly from the Baxter solution of the classical 2D 
eight-vertex model [\onlinecite{baxbook}], as well as from Kadanoff's 
classic work on its critical behavior
[\onlinecite{kadanoff77,kadanoff79,kadanoff-brown79}]. The only 
change is that the weights must be squared here, since in quantum mechanics 
we weigh configurations with $|\Psi|^2$. 
The phase diagram is displayed in figure 
\ref{eight}; note that the axes are labeled by $c^2$ and $d^2$.

We will now use this knowledge, as well a simple perturbative
arguments in the quantum theory, to determine the phase diagram, the
behavior of physical observables in the different phases, and (much
of) their critical behavior.  A useful fact is that for $cd=1$, the
classical model with partition function $Z(c^2,1/c^2)$ is equivalent
to two decoupled Ising models.  This decoupling is a property only of
the wave function, not of the Hamiltonian of the full
$2+1$-dimensional quantum theory.  Since this line $cd=1$ goes through
both ordered and disordered phases, much of the physics of the quantum
eight-vertex model can be described at least qualitatively in terms of
decoupled Ising models.  In particular, for any values of $c$ and $d$
except those on the critical line, correlators decay exponentially
fast with distance with a correlation length $\xi$ which diverges as
the phase boundary is approached, in a manner given in Eq.\
\eqref{critexp}. This exponential decay occurs in general, not just on
the decoupling curve.

As with the 
dimer models discussed in section \ref{sec:2dwavefn}, the partition 
function for $cd=1$ can be expressed in terms of Grassmann 
variables with only quadratic terms, i.e.\ free fermions. 
Duality means that $Z(c^2,d^2)$ is free-fermionic for 
$c^4+d^4=2$ as well. The 
correlators in the continuum limit of the (critical) square-lattice 
quantum dimer model are therefore identical to those obtained for 
$c=\sqrt{2}$, $d=0$. The special point $c=d=1$ discussed above and in 
Ref.\ [\onlinecite{kitaev97}] (labeled in fig.\ \ref{eight} as 
``Kitaev'') is also free fermionic. At this point one does not need 
the Pfaffian techniques to compute correlators exactly, and one finds 
that the model is in a disordered phase in the Ising-spin 
language. However, the expectation value 
$\langle\mu(A)\mu(B)\rangle$ is non-vanishing, so there is 
topological order at this point.

Let us now discuss the different phases of this system. 
\begin{enumerate} 
\item 
{\em The Ordered (Confined) Phase}:\\ {}From the known phase diagram 
of the classical eight-vertex model [\onlinecite{baxbook}], we 
conclude that the ground state of the {\em quantum} model with the 
Hamiltonian of Eq.\ \eqref{q8vham} has an {\em ordered} phase for 
$c^2>d^2+2$ (and also for $d^2>c^2+2$).  That this is an ordered phase 
can be seen easily by considering the limit $c \to \infty$ (with $d$ 
fixed). In this limit the ground state is dominated by just two 
configurations, related to each other by a lattice translation of one 
lattice spacing, which have a $c$ vertex on every site, as shown in 
fig. \ref{ddw}. In this phase the staggered 
polarization operator $\langle \tau(A)\tau(B)\rangle$ has a 
non-vanishing expectation value. This result can also be obtained 
directly from the Hamiltonian of the ${\mathbb Z}_2$ gauge theory, 
Eq.\ \eqref{q8vham2}, since for $c$ large the potential energy term 
$H_V$ dominates and in it the piece associated with the $c$ projection 
operators. 
 
\begin{figure} 
\begin{center} 
\includegraphics[width=0.2 \textwidth]{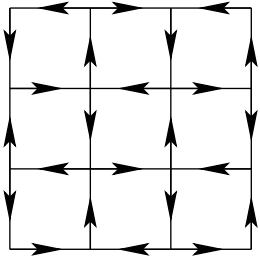} 
\caption{The ordered phase ``antiferroelectric" or ``staggered flux" phase of 
the eight-vertex model  for $c \gg 1$. In this limit there are only $c$ vertices in the ground state.} 
\label{ddw} 
\end{center} 
\end{figure} 
 
The ordered phase is also {\em confining}. Below we will 
discuss the behavior of the Wilson loop operator and show that in this 
regime it obeys an area law, the hallmark of confinement 
[\onlinecite{fradkin78,kogut79}]. We will also show that the energy of 
a state with two static sources grows linearly with their 
separation. We should also note here that the equal-time fermion 
correlation function has an exponential decay in this phase, 
suggesting that this phase may support massive fermionic excitations. 
 
\item 
{\em The Disordered (Deconfined) Phase}:\\ {}From the exact solution
of the classical model, we know that there is a {\em disordered}
phase for $c^2<d^2+2$ ($d^2<c^2+2$) (see Fig. \ref{eight}). This is
also a {\em deconfined} phase. This is most easily seen by taking a
point deep in this phase, such as the Kitaev point $a=b=c=d=1$
[\onlinecite{kitaev97}]. In fact, all along the line $a=b=1$ and
$c=d$, the correlation length of the eight-vertex model is zero
[\onlinecite{baxbook}]. All points in the disordered phase have this
line as their RG stable fixed point (all points on the line are
equivalent). This is the infinite temperature limit of the classical
2D eight-vertex model. In this regime the expectation value of the
polarization operator $\langle \tau(A)\tau(B)\rangle$ is zero (and its
correlation vanishes at a length scale of the lattice spacing.). Thus
this state does not exhibit long-range order, but since the dual
variable $\langle \mu(A)\mu(B)\rangle$ is non-vanishing, it exhibits
topological order. Quantum mechanically, the wave function is the
equal-amplitude superposition of all configurations of arrows
consistent with the eight-vertex restrictions. This state is
deconfined since in this limit the Hamiltonian reduces to the flip
term (plaquette). It is well known [\onlinecite{fradkin78}] that in
this regime the Wilson loop operator has a perimeter law behavior and
that the energy of two static sources is finite and independent of
their separation. The low energy ({\it i. e.\/} long-distance) sector
of this phase describes a {\em topological} ${\mathbb Z}_2$ deconfined
theory, equivalent to the Kitaev point.
\item 
{\em Critical Behavior}:\\ 
The 2D {\em quantum} theory has  lines of critical points $c^2=d^2+2$ 
(and $d^2=c^2+2$). There are also lines of critical points for $d=0$ and 
$0\leq c^2 \leq 2$ (and $c=0$ and $0\leq d^2 \leq 2$). 
All of these critical lines are (up to duality transformations) 
equivalent to the six-vertex model. 
Notice that, since the energy of this quantum state 
is exactly zero (by construction), the ground-state energy does not 
have singularities at the phase transitions. This is of course a 
peculiarity (or rather a pathology) of this model and it certainly 
non-generic: any perturbation leading to a non-vanishing energy will 
lead to singularities in the energy of the ground state. 
 
\end{enumerate}

In section \ref{critft}, we discussed how the ground-state correlators 
of the (multi) critical square-lattice quantum dimer model were the 
same as those of the quantum Lifshitz theory, Eq.\ \eqref{eq:Hboson}, 
at $\kappa^{-1}=2\pi$ [\onlinecite{henley97a,moessner02a}]. The same 
equivalence holds for the quantum eight-vertex model for all values of 
$\kappa$. Moving along the critical line results in changing $\kappa$ 
as given in (\ref{eq:kappa8v},\ref{eq:kappa8v2}).  Kadanoff 
[\onlinecite{kadanoff77,kadanoff79,kadanoff-brown79}] showed that the 
critical behavior of the classical eight-vertex model can be mapped 
exactly to the critical behavior of the two-dimensional Gaussian 
model, the free boson described by Eq.\ \eqref{eq:Zboson} and Eq.\ 
\eqref{eq:probab}.  The 
equal-time correlators in the quantum Lifshitz model 
summarized in Appendix \ref{app:gaussian} are therefore identical to 
those of the quantum eight-vertex model on its 
critical lines. Moreover, when $d=0$ and $c^2\le 2$, the 
eight-vertex model becomes the six-vertex model, which has a height 
description. Thus the same heuristic arguments applied in the last 
section can be applied here, again implying that the 
quantum six-vertex model and the quantum Lifshitz theory are in the 
same universality class. In 
particular, the dynamics of the quantum eight-vertex model along its 
critical lines will obey scaling with a dynamic exponent $z=2$.

To compete the identification, we need to express the coupling 
constant $\kappa$ of the quantum Lifshitz model in terms of the 
eight-vertex parameters $c$ and $d$.  This is easiest to do by 
computing the dimension $x$ of the ``energy'' operator, which when 
added to the action of (classical) critical theory, moves it away from 
criticality.  Combining universality with Baxter's exact result 
(\ref{critexp}) for the correlation-length exponent gives 
$x=2(1-\mu/\pi)$.  We have normalized the boson $\varphi$ in Appendix 
\ref{app:gaussian} so that the energy operator is given by 
$\cos(2\varphi)$, which has dimension $x=1/(2\pi\kappa)$. Note that at 
the free-fermion point $\kappa^{-1}=2\pi$, the energy operator has 
dimension 1, the dimension of a 2D fermion mass term. Combining the 
two and using a simple trigonometric identity yields 
\begin{equation} 
\kappa^{-1}=8 \cot^{-1}\left(cd\right).\qquad\qquad \hbox{for }|c^2-d^2|=2 
\label{eq:kappa8v} 
\end{equation} 
and by duality 
\begin{equation} 
\kappa^{-1}=8 \cot^{-1}\left(\sqrt{\frac{4}{c^4} -1}\right)\qquad\qquad 
\hbox{for } c^2\le 2, d=0. 
\label{eq:kappa8v2} 
\end{equation} 
At the Kosterlitz-Thouless transition point $c^2=2,d=0$ where the two 
critical lines meet, the dimension $x=0$ as expected, and both 
formulas give $\kappa^{-1}=4\pi$ here.

This equivalence to a scalar field theory can be extended to the 
scaling region near the critical lines, by repeating the arguments of 
section \ref{offcritft}. The perturbing operator is $\cos(2\varphi)$; 
symmetry forbids $\cos(\varphi)$ from being added to the action of the 
classical theory. Thus near to the critical lines the effective field 
theory will be of the form (\ref{Qsg}) and Eq.\ (\ref{eq:HQQ}); these 
field theories will have the correct critical exponents.

\subsection{Operators}
\label{sec:operators} 
 
Equal-time correlation functions of operators which are diagonal in 
the arrow representation of the eight-vertex model are identical to 
correlation functions of the classical eight-vertex model. 
Unfortunately, these correlation functions can be computed exactly 
only along the free-fermion lines, by using Pfaffian 
techniques. Likewise correlation functions of off-diagonal 
operators ({\it e.g.\/} flip operators) can only be determined at 
some special points deep in the ordered and disordered phases by using 
directly the quantum eight-vertex Hamiltonian.  However, at or near 
the critical lines, the asymptotic behavior of the correlators can be 
found by using universality.  The operators of interest in the eight-vertex 
model can all be expressed in terms of the charge and vortex operators 
of the Gaussian model [\onlinecite{kadanoff-brown79}].

The operators of interest in the quantum eight-vertex model can be 
readily identified in the quantum Lifshitz model. Thus we have a full 
identification of a $2+1$-dimensional quantum critical theory. As a 
direct consequence of being able to identify these operators we can 
also study how these operators perturb the critical theory, at least 
within a renormalization group argument along the lines of Ref.\ 
[\onlinecite{grinstein81a}]; the analog in the quantum dimer model is 
our analysis of $H_w$ for $w$ small at the end of section 2. 
 
To identify the eight-vertex operators in the bosonic language as in Ref.\ 
[\onlinecite{kadanoff-brown79}], we first study the six-vertex line, where 
there is a height description. There the height is local in terms of 
spin variables, so we expect the boson $\varphi$ to be local as 
well. Standard bosonization techniques 
[\onlinecite{friedan-martinec-shenker87}] show that only products of 
spin operators from both Ising models can be written in terms of the 
boson. One finds that $\cos(\varphi)\sim \tau(A)\tau(B)$. 
Note that this is consistent with the identification of 
$\cos(2\varphi)=2\cos^2(\varphi)-1$ with the perturbing operator; the 
usual Ising operator product gives the fusion rule $\tau(A)\tau(A)\sim 
1 + {\cal E}(A)$, where ${\cal E}(A)$ is the energy operator. 
 
To study the operators in more depth, we utilize a $2+1$-dimensional 
${\mathbb Z}_2$ gauge theory of the quantum eight-vertex model derived 
in Appendix \ref{app:z2gt}. The degrees of freedom of this gauge 
theory reside on the links of the square lattice and have a one-to-one 
correspondence to arrow configurations of the eight-vertex model. The 
Hamiltonian of the ${\mathbb Z}_2$ gauge theory is 
\begin{equation} 
H_{\rm q8v} = H_{\rm V} + H_{\rm flip} 
\label{q8vham2} 
\end{equation} 
where $H_{\rm flip}$ has the form of a sum over plaquettes 
of ${\mathbb Z}_2$ flip operators\footnote{Throughout 
this section and in Appendix \ref{app:z2gauge} we use the standard 
notation used in lattice gauge theories [\onlinecite{kogut79}]. The 
$\sigma$'s are Pauli matrices; the superscript is a Pauli matrix label 
and the subscript indicates the spatial direction, $1=$ horizontal and 
$2=$ vertical. Notice that the notation used by Kitaev 
[\onlinecite{kitaev97}] is somewhat different. Please see Appendix 
\ref{app:z2gauge}, and figures \ref{sigmaconfig} and \ref{fig:dual}, for 
details of the notation that we use here.}: 
\begin{equation} \label{sflip} 
H_{\rm flip} = - \sum_{\vec x} \sigma^3_1(\vec x) \; \sigma^3_2(\vec x+\vec e_1) \; \sigma^3_2(\vec x) \; \sigma^3_1(\vec x+\vec e_2) 
\end{equation} 
The potential energy terms are combinations of operators which project 
onto the allowed select $a$, $b$, $c$ and $d$ vertices, and assign 
the weights $a$, $b$, $c$ and $d$ to different contributions to the 
wave function. The explicit form of the potential energy terms $H_{\rm 
V} $ is given in Eq.\ \eqref{q8vpotsig}.  As in all ${\mathbb Z}_2$ 
gauge theories [\onlinecite{fradkin78,kogut79}], the physical states 
of this theory satisfy the constraint of gauge invariance (``Gauss 
Law") 
\begin{equation} \label{scon8v} 
G(\vec x)=\sigma^1_1(\vec x) \; \sigma^1_2(\vec x) \; \sigma^1_1(\vec x-\vec e_1) \; \sigma^1_2(\vec x-\vec e_2) = 
1 \qquad \forall \vec x \ 
\end{equation} 
which in this context simply expresses the restrictions on the configurations allowed in the eight-vertex model. 
The operator $G(\vec x)$ is the generator of local time-independent gauge transformations [\onlinecite{fradkin78,kogut79}]. 
 
In Appendix \ref{app:duality} we derive the dual theory of the quantum 
eight-vertex Hamiltonian of Eq.\ \eqref{q8vham}. The degrees of 
freedom of the dual theory are defined on the sites of the dual square 
lattice. The Hamiltonian is 
\begin{equation} 
H_{\rm q8v, dual} = H_{\rm V,dual} - \sum_{\vec r} \tau^1(\vec r) \ , 
\label{q8v-dual2} 
\end{equation} 
where $H_{\rm V,dual}$ is given in Eq.\ \eqref{q8vpotsig} and  Eq.\ \eqref{vms-dual}. 
We will use the both the gauge theory and its dual to investigate these phases. 
 
Let us discuss briefly the observables of this model, how they behave 
in the ordered (confined) and disordered (deconfined) phases and what 
critical behavior they exhibit at the phase transition lines.  As we 
note in Appendix \ref{app:duality}, the quantum eight-vertex model 
retains the two-sublattice structure of the classical model.  The 
${\mathbb Z}_2 \times {\mathbb Z}_2$ symmetry of the classical model also 
survives in the quantum theory. 
\begin{enumerate} 
\item 
{\em Order and disorder operators}: We denote by $\tau(A)$ and
$\tau(B)$ the $\tau^3$ (order) operator for sublattice $A$ and $B$
respectively. In the gauge theory this operator is a disorder or kink
operator [\onlinecite{fradkin78}]; in this 2D gauge theory this is
just the flux (monopole) operator of Refs.\
[\onlinecite{fradkin-raby79,kogut79}], or the vison operator of Refs.\
[\onlinecite{senthil00,senthil01}]. The order operators are just the
order parameters of the sublattice Ising models. As such they acquire
an expectation value in the ordered phase (where the staggered
polarization has an expectation value as well). In the disordered
phase their equal-time correlation functions decay exponentially with
distance (as do their connected equal-time correlators in the ordered
phase). The precise behavior of this correlation function is only
known on the decoupling curves where they reduce to the correlation
functions of the 2D classical Ising model, and on the critical lines!
The scaling dimension of this spin field (which is known as the
``twist'' field in conformal field theory
[\onlinecite{friedan-martinec-shenker87,yellow}]), is $1/8$ not only
at the decoupling point (as is well known from the 2D Ising model) but
along the entire phase boundary. We note however, that the spin fields
can not be represented directly in terms of the boson, but only in its
orbifold [\onlinecite{ginsparg88}].
 
Similarly, we will denote by $\mu(A)$ and $\mu(B)$ the ``frustration" 
(or ``fractional domain wall") or {\em disorder operator} 
[\onlinecite{kadanoff-ceva71}] for sublattices $A$ and $B$ 
respectively. In the 2D classical theory, this is the dual of the 
order operators and in conformal field theory it is also represented 
by a twist field. It has the same scaling dimension as the spin field. 
On the other hand, in the gauge theory, the disorder operators 
correspond to a local violation of the Gauss-Law constraint, Eq.\ 
\eqref{scon8v}, by imposing that $G(\vec x)=-1$ just at site $\vec x$, 
{\it i.e.\/} a static source ({\it i.e.\/} an ``electrode", or 
Polyakov loop) at site $\vec x$. This operator is analogous to the 
holon (or ``monomer") operator in the Rokhsar-Kivelson quantum dimer 
model. This expectation value of products of two operators of this 
type serves as a test for {\em confinement}. The disorder operators 
have a non-zero expectation value in the quantum disordered phase. A 
simple minded calculation shows that, in {\em in the ordered phase}, 
the ground state with two such defects (or sources) separated at a 
distance $R$, has a non-zero energy $U(R)$, which grows {\em linearly} 
with $R$, $U(R) \sim \sigma R$, consistent with the fact that the 
defects create a fractional domain wall in an ordered state. We expect 
that the string tension $\sigma$ vanishes as the critical lines are 
approached, with a behavior dictated by that of the gap scale, {\it 
i.e.\/}, $\xi^{-z}$. Thus, the excitations created by the disordered 
operator are {\em confined} in the ordered phase\footnote{The holon 
operator of the quantum dimer model has the same behavior in both the 
columnar and staggered phases.}. In contrast, in the quantum 
disordered phase, the energy of a pair of defects saturates to a 
finite value if $R \gg \xi$. Hence, the disordered phase is 
deconfined. 
 
It is also interesting to ask what is the {\em energy cost} of a set 
of vortex configurations on the quantum critical lines. This can be 
done in a number of ways. We note here that using path-integrals the 
change of the ground state energy can be found by computing the path 
integral in a background of vortices, normalized by the path-integral 
without vortices. Consider the simple (an general) case of two 
vortices of magnetic charges $\pm m$ separated a distance $R$. Let is 
denote this amplitude by $W(R,T)$, where $T$ is the (infinite) 
time-span of the system. In the dual gauge theory this is the same as 
the computation of two static Wilson (or Polyakov) loops, 
corresponding to two static sources with electric charges $\pm m$ at a 
distance $R$ from each other. The energy cost of these defects is 
\begin{equation} 
U(R)=\lim_{T \to \infty} \left(-\frac{1}{T} \ln W(R,T)\right) 
\label{energy-cost} 
\end{equation} 
We can compute $U(R)$ by using the path-integral for the quantum Lifshitz theory of Eq.\ \eqref{eq:Zboson} in a 
background of vortices, which amounts to modifying the action by the minimal coupling shift 
\begin{equation} 
(\nabla^2 \phi)^2 \to ( \vec \nabla \cdot (\vec \nabla \phi-\vec A))^2 
\end{equation} 
By explicit calculation we find that 
\begin{equation} 
U(R)=0 
\label{eq:zero} 
\end{equation} 
In other terms, the interaction energy of vortices is zero! A simple
way to understand this results is to recall that, in imaginary time,
the quantum Lifshitz model is a theory of a smectic liquid
crystal. The vortex state is just a configuration of screw
dislocations running along the $z$ axis (imaginary time). Since the
smectic is isotropic and has no resistance to shear in the $xy$ plane,
these defects do not cost any energy. From a quantum mechanical point
of view this result is consistent with the known fact that at the RK
point of the quantum dimer model, the interaction energy of a pair of
monomers is zero\footnote{We thank Shivaji Sondhi for this
remark.}. In gauge theory language this means that at this critical
point vortices are completely free.\footnote{This result,
Eq. \eqref{eq:zero} holds strictly only at the quantum Lifshitz fixed
point (or rather line) fixed point, Eq. \eqref{eq:Zboson}. However,
marginally irrelevant operators such as $\left(\nabla \phi\right)^4$,
discussed extensively in Refs.[\onlinecite{fradkin03,vishwanath03}],
give rise to corrections to scaling which change the behavior of
$U(R)$ from Eq.\eqref{eq:zero} to a $(\ln R)/R^2$
law[\onlinecite{vishwanath03}].  We thank T. Senthil for pointing this
out to us.}  In contrast, at quantum critical points of Lorentz
invariant gauge theories, which have $z=1$, it is known
[\onlinecite{peskin80}] that at their critical point the effective
interaction has the universal law $U(R)\sim1/R$, in {\em all
dimensions}.
\item 
{\em Polarization}: The polarization operator, the local arrow
configuration on a link, is the natural order parameter. In terms of
the order operators it reads $P=\tau(A) \tau(B)$. As we saw above in
the classical ``antiferroelectric" ordered phase, $P$ is a natural
order parameter, although appropriately-staggered expectation values of
$\tau(A)$ and $\tau(B)$ also do not vanish in this ordered phase.  The
behavior of the polarization operator on the critical lines of the
classical model was studied by Kadanoff and Brown
[\onlinecite{kadanoff-brown79}], who found that its scaling dimension
is $\Delta_P=1/8\pi \kappa$, with $\kappa$ given by Eq.\
\eqref{eq:kappa8v}. From the results of Appendix \ref{app:gaussian} we
see that we can identify the polarization operator with the boson
operator $P \sim \cos(\varphi)$ in the quantum Lifshitz theory, which
has the same scaling dimension.
\item 
{\em Mass term}: In the classical theory the mass term, or ``energy
density", is the product of two (dual) spin variables $\tau$ on
nearest neighboring sites on the same sublattice. We will denote them
by ${\mathcal E}(A)$ and ${\mathcal E}(B)$ respectively. Although in
the quantum theory these operators no longer correspond to an energy
density, we will define ${\mathcal E} =\left( {\mathcal
E}(A)+{\mathcal E}(B)\right)/\sqrt{2}$, as usual, and refer to this
operator as to the {\em mass term} since it drives the theory away
from criticality. In Ref.\ [\onlinecite{kadanoff-brown79}] it is found
that the scaling dimension of this operator is $\Delta_{\mathcal
E}=1/2\pi \kappa=4 \Delta_P$, which led to the identification
${\mathcal E} \sim \cos(2\varphi)$.  If we regard this operator as
acting on the critical line along the $d=0$ axis, {\it i.e.\/} the
$6$-vertex model, we see that this is a relevant operator for the
quantum Lifshitz model as well.  Grinstein's RG arguments
[\onlinecite{grinstein81a}] show that this perturbation drives the
system into a massive phase with a residual unbroken ${\mathbb Z}_2$
symmetry.  This is just the disordered phase of the quantum
eight-vertex model, and this perturbation drives the system towards
the Kitaev point. A similar analysis works on the $c^2=d^2+2$ critical
line.
\item 
{\em Wilson loops}: As it is well known gauge theories, the natural
test for confinement is the behavior of the Wilson loop operator
$W_\gamma=\prod_{\ell \in \gamma} \sigma^3(\ell)$, where $\{ \ell \}$
is a set of links that belong to a closed path $\gamma$ on the
(direct) lattice (see Ref.\ [\onlinecite{kogut79}] and references
therein).  Since these operators are off-diagonal in the arrow
representation, {\it i.e.\/} in the representation in which $\sigma^1$
is diagonal, they do not have an analog in the classical 2D
eight-vertex model.  A simple argument [\onlinecite{fradkin78}] shows
that in the ordered state, which for $c \to \infty$ is essentially an
eigenstate of $\sigma^1$, the expectation value of the Wilson loop is
zero and that first non-zero contribution comes from acting
$N(\gamma)$ times with the flip operator, where $N(\gamma)$ is the
number of plaquettes enclosed inside the path $\gamma$, and that the
behavior thus found has the form $\exp(-{\rm constant}\; N(\gamma))$,
{\it i.e.\/} an {\em area law}. Thus the ordered state is a confining
phase. Conversely, deep in the disordered phase, the ground state is
essentially an eigenstate of the flip operator and we get a perimeter
law, {\it i.e.\/} deconfinement.
\item 
{\em Spinors}: The classical 2D eight-vertex model has two Majorana 
fermion operators $\psi(A)$ and $\psi(B)$, constructed as usual as 
products of order and disorder operators 
[\onlinecite{kadanoff-ceva71}], $\psi\sim \tau \mu$. We can also 
define an analog of the spinors in the quantum theory (see the 
discussion in Appendix \ref{app:gaussian} on charge and vortex 
operators in the $2+1$-dimensional quantum Lifshitz model). The 
scaling dimension of the spinors is [\onlinecite{kadanoff-brown79}] 
$\Delta_\psi=\left(2\pi \kappa+1/2\pi\kappa\right)/4$, a result 
familiar from the Luttinger model. 
\item 
{\em Other operators}. The classical 2D eight-vertex model contains a 
marginal operator $E={\mathcal E}(A) {\mathcal E}(B)$, with scaling 
dimension $\Delta_E=2$, which is responsible for the lines of fixed 
points with varying critical exponents. The quantum eight-vertex model 
discussed here has a similar behavior.  In the quantum Lifshitz model 
the marginal operator is\footnote{As we noted above, on symmetry 
grounds a lattice model may generate additional operators which break 
rotational invariance, absent at this RK point, which may drive the 
transitions to be first order, see 
[\onlinecite{vishwanath03,fradkin03}].} $E \sim \left(\nabla^2 
\varphi\right)^2$. Another operator of interest is the crossover 
operator $\left( {\mathcal E}(A)-{\mathcal E}(B)\right)/\sqrt{2}$ 
which breaks the sublattice symmetry and has scaling dimension $2\pi 
\kappa$, The operator product expansion of two crossover operators 
generates an operator symmetric under sublattice exchange, and is 
identified with a charge-two vortex operator. The scaling dimension of 
this operator is $8\pi \kappa$ and it is becomes marginal at 
$\kappa=1/4\pi$, driving the Kosterlitz-Thouless transition along the 
$d=0$ line.\footnote{Notice that the charge-one vortex operator is not 
allowed in this case, which is the operator driving the KT transition 
in the 2D classical XY 
model [\onlinecite{kosterlitz73,kosterlitz74,berezhinskii70,jose77}].} 
Other interesting operators are the ``mixed" operators ${\tilde 
P}=\tau(A)\mu(B)$ and ${\tilde P}^*=\mu(A) \tau(B)$, both with scaling 
dimension $\Delta_{\tilde P}=\pi \kappa/2$. Hence ${\tilde P}$ and 
${\tilde P}^*$ are vortex operators as well; these are non-local with 
respect to $P$. They have been identified as the holon creation 
operators in the context of the quantum dimer model 
[\onlinecite{fendley02}].

\end{enumerate}

\subsection{Another quantum eight-vertex model} 
\label{another} 
 
Let us end this section by noting that there is a  simpler 
quantum eight-vertex model that one can write down whose behavior is quite different from the one we 
have discussed here. 
Basically, this model will give weights to all the vertices, according 
to their type. In addition, there will be a flip term. In its dual form the model is 
\begin{equation} \label{nonrk} 
H = \sum_{\vec r}  \bigl( 
\varepsilon_a \mathcal{P}_a(\vec r) + \varepsilon_b \mathcal{P}_b(\vec r) + 
\varepsilon_c \mathcal{P}_c(\vec r) + \varepsilon_d \mathcal{P}_d (\vec r) 
\bigr) - \sum_{\vec r} \tau^1(\vec r) \ , 
\end{equation} 
where the projection operators are defined in Eq.\ \eqref{vms-dual}. 
The conventional Baxter weights of the classical model ({\it i.e.\/} when the flip term is absent) are  $a=e^{-\frac{\varepsilon_a}{k_B T}}$, etc. 
 
We first note that this model is not of the Rokhsar-Kivelson type
(except when
$\varepsilon_a=\varepsilon_b=\varepsilon_c=\varepsilon_d$, which is a
dual version of the Kitaev model). The ground state wave function does
not look like the statistical weights of the classical two-dimensional
model, {\it i.e.\/} it is not a linear superposition of states
weighted according to the vertices present. Thus, strictly speaking we
can not use the machinery of this paper to study the ground state of
this model. We do expect, however, that this model will have the same
ordered phases, just like the ferro-electric and anti-ferro-electric
phases of the eight-vertex model, as well as the same
confinement-deconfinement properties discussed above. In addition, we
expect that these ordered phases have the same spectrum and general
properties as the ones we found in the Rokhsar-Kivelson type model we
studied above. Similarly, in the regime where the flip term dominates,
the ground state of this model is a uniform quantum disordered state,
with the same properties as those of the topological phase of the
model studied here. 

However, the quantum critical properties of Eq.\ (\ref{nonrk}) are
completely different from the quantum eight-vertex model discussed in
the rest of this section.  They instead have a number of features in
common with well-known three-dimensional classical models.  By setting
$\varepsilon_a+\varepsilon_b=\varepsilon_c+\varepsilon_d$, this model
reduces to two interpenetrating two-dimensional Ising models in a
transverse field, as the four-body interactions cancel. Thus along
these curves this model has second-order phase transitions in the
universality class of the classical three-dimensional Ising model, as
opposed to the quantum eight-vertex model discussed above. The latter
has $z=2$, while the model of Eq.\ \eqref{nonrk} therefore has
dynamical exponent $z=1$ and is Lorentz invariant in the critical
regime. It is easy to see that this behavior extends beyond the
decoupling point. Indeed, the universality class of the 3D classical
Ising model is controlled by the Wilson-Fisher fixed point and is
accessible by the $4-\epsilon$ expansion. Away from the decoupling
point we have then two 3D Ising models which are coupled through their
energy density. This theory is in the universality class of a
two-component real field $(\varphi_1,\varphi_2)$, whose $U(1)$
symmetry is explicitly broken down to a discrete ${\mathbb Z}_2 \times
{\mathbb Z}_2$ symmetry by quartic operators of the form $\varphi_1^2
\varphi_2^2$. This is a cubic symmetry-breaking perturbation and it is
well known to be perturbatively relevant at the Wilson-Fisher fixed
point of the two decoupled Ising models, but irrelevant at the
Wilson-Fisher fixed point for the classical $XY$ model\footnote{This
is a classic example of a dangerously-irrelevant operator, since it
removes the would-be Goldstone boson from the broken-symmetry
phase.}. It is well known from perturbative $4-\epsilon$ RG studies
that the transition near the decoupling point that the transition may
also become a fluctuation-induced first order transition depending on
the sign of the effective coupling (see for instance Refs.\
[\onlinecite{cardy96},\onlinecite{amit84}]).

\section{Non-Abelian Topological States} 
\label{sec:YMCS}

In this section we broaden our scope and discuss field theories with
continuous non-abelian symmetries.  One can of course obtain a quantum
critical point with non-abelian symmetries by taking copies of the
quantum Lifshitz Hamiltonian. By choosing the charge lattice of the
vertex operators carefully one can describe a theory with a
non-abelian symmetry, say $SU(N)$.  However the equivalent conformal
field theories thus obtained always have integer central charge and
are at level $1$. Although critical, such theories only support
excitations with abelian statistics. One may try to attempt to
generalize this result to models by writing down a theory with more
structure, such as one related to a Wess-Zumino-Witten model.
However, we will find that, contrary to naive expectation, such
theories will turn out not to be critical but, instead, in a phase
with a finite gap, a topological phase, just like the fractional
quantum Hall effect.  We will see that it is not at all clear how to
find a conformal quantum critical point with a non-trivial non-abelian
symmetry.
 
Nevertheless it is possible to use the approach of the previous
sections to discuss topological non-abelian phases, even if any
associated non-trivial quantum critical points are not yet known or do
not exist. By analogy with the theory of fractional quantum Hall
states, as well as from general results in Chern-Simons gauge theory,
one expects that a time-reversal-breaking topological phase in
$2+1$-dimensions should have an effective field theory description in
terms of a Chern-Simons gauge theory at some level $k$. On closed
manifolds this is a topological field theory in the sense that the
partition function is independent of the metric of $2+1$-dimensional
space-time, and that the expectation values of its gauge-invariant
observables, the Wilson loops, depend only on the topology of the
loops such as the knot invariants [\onlinecite{witten88}].

By analogy with the approach pursued in the previous sections, we will
discuss the properties of the wave functionals of this theory.  This
will lead us to consider first the strong-coupling limit of Yang-Mills
theory with a Chern-Simons term, and later the topological sector of
this theory.  We will find the ground-state wave functional of this
theory, and show how it is related to a two-dimensional
Wess-Zumino-Witten (WZW) model for a specific choice of polarization
({\it i.e.\/}, the choice of canonically conjugate variables) which is
natural in the strong coupling limit of the Yang-Mills Chern-Simons
theory.  Even though the WZW model is a conformal field theory with
algebraically-decaying correlators, the gauge-invariant ground-state
correlators in the two-dimensional quantum theory are 
exponentially decaying. Topological invariance will only be attained
by correlators of gauge-invariant operators, Wilson and Polyakov
loops, and only at long distances. The excitations obey non-abelian
statistics: when bringing particles around one another, not only does
one pick up phases, but the order in which they are exchanged affects
the final state.

We will also study a doubled Chern-Simons theory which, contrary to
the fractional quantum Hall effect, is time-reversal invariant. Using
by now standard methods, we will construct the ground-state wave
functionals of this doubled theory for a gauge group $G$ at level $k$,
and use its connection with a gauged WZW model to understand its
properties. We note that the doubled theory is often used
to circumvent technical problems involved in the use of the formally
anomalous wave functional of the undoubled theory: see the discussion
in the subsections \ref{sec:ymcswzw} and \ref{sec:doubled}. The
topological properties of the doubled theory have been recently
discussed in depth by Freedman {\it et al.\/}
[\onlinecite{freedman03}].
 
\subsection{Field theories with continuous non-abelian symmetries}
\label{sec:ftnon-abelian} 
 
Let us attempt first to construct a non-abelian version of the quantum 
Lifshitz theory, which one might hope would still be a quantum critical 
theory. We will follow the same procedure as in the abelian case. 
Since there are many two-dimensional Euclidean critical points with 
non-abelian symmetry, the first thing to try is the above procedure 
for making a two-dimensional quantum theory from a two-dimensional 
classical theory. This procedure is indeed fairly general.  One can 
easily construct a set of a appropriate projectors $Q_i$ in a 
lattice model without local constraints.  However, the caveat 
``without constraints'' is quite important: many interesting classical 
lattice models have local constraints like in the dimer model and in 
the six- and eight-vertex models. This means that one must find a flip 
operator which respects the constraints. There is no guarantee that 
there is a set of flips which are both local and ergodic. This is 
particularly apparent in models where the two-dimensional degrees of 
freedom are closed loops. At a critical point, arbitrarily-long loops 
are an important part of the configuration space, so one must 
construct some sort of flip which still acts non-trivially but locally 
on these loops. In the continuum limit, it is easy to imagine such 
flips, but it is not always obvious how to make them act consistently 
on the lattice. Nevertheless, the Rokhsar-Kivelson Hamiltonian and the 
quantum eight-vertex model provide examples of models where one can 
solve this problem. 
 
For field theories, the issue is similar. If a theory has an action 
$S_{2d}$ and one ignores constraints such as gauge invariance, one can 
easily find projection operators $Q(x)$ which annihilate the state 
weighted by $e^{-S_{2d}}$. For a single field $\varphi$, we have 
\begin{equation} 
Q= \frac{\delta}{\delta\varphi} + \frac{\delta S_{2d}}{\delta{\varphi}} 
\label{Qdef} 
\end{equation} 
and the Hamiltonian is defined via Eq.\ (\ref{eq:HQQ}). This $Q$ is hardly 
unique: one can multiply it on the left by any operator without 
changing the ground state. It is important to note that this 
construction is not always easy to implement. Many interesting 
two-dimensional field theories do not have a simple-to-describe 
action: they are usually defined instead in terms of representations 
of an infinite-dimensional symmetry algebra such as the Virasoro 
algebra, arising from conformal invariance [\onlinecite{yellow}]. For 
example, the action of a $G/H$ coset model ({\em e.g.\/} a conformal 
minimal model) is written as a gauged WZW model such as 
we will discuss below; one must then worry about the constraints of 
gauge invariance. 
 
A point we wish to emphasize, however, is that even if one succeeds in 
constructing such a Hamiltonian, the $2+1$-dimensional theory may be 
qualitatively very different from the classical two-dimensional 
theory. For example, the classical theory may be critical and have 
algebraically-decaying correlators, but the quantum theory may have 
exponentially-decaying correlators in the ground state. The reason is 
that quantum-ground-state correlators are weighted by $|\Psi_0|^2$, 
not $\Psi_0$. For the models discussed above, this has effects which 
are easy to take account of: it is the reason that the axes in figure 
\ref{eight} are labeled by $c^2$ and $d^2$ instead of $c$ and $d$. If 
however $\Psi_0$ has a complex part, the effects can be much more 
dramatic. 
 
Let us first discuss the simplest two-dimensional classical field
theories with a non-trivial non-abelian symmetry. The $G$ principal
chiral model is written in terms of a unitary matrix field $g$ taking
values in some simple Lie group $G$, with action
\begin{equation} 
S_{PCM} = -\frac{1}{8\pi u}\int d^2 x \, \hbox{Tr}\left[ 
g^{-1}\partial^a g \, g^{-1}\partial_a g\right] \ . 
\label{spcm} 
\end{equation} 
This action has a global $G_L \times G_R$ symmetry under $g\to L g
R^\dagger$, where $L$ and $R$ are elements of separate groups dubbed
$G_L$ and $G_R$. This model looks critical: the coupling constant $u$
is naively dimensionless. However, in two dimensions, the one-loop
beta function is proportional to the curvature of the target-space
manifold $G$.  Simple Lie groups $G$ are curved: for example, $SU(2)$
as a manifold is isomorphic to a three-sphere.  Moreover, one finds
that the trivial fixed point at $u=0$ is unstable, and the model has a
finite correlation length proportional to an exponential in
$1/u$. Correlators decay exponentially, not algebraically. The
phenomenon where a naively dimensionless coupling becomes dimensionful
due to loop corrections is known in the particle physics literature as
dimensional transmutation; it is familiar in condensed-matter physics
in the Kondo problem and the Hubbard model at half-filling. Another
reason a gap should appear is that the Mermin-Wagner-Coleman theorem
does not allow Goldstone bosons in two Euclidean dimensions. One would
obtain $S_{PCM}$ in a theory as the low-energy limit of a fermionic
theory where one attempts to spontaneously break the chiral symmetry
$G_L \times G_R$ to its diagonal $G_D$ subgroup. The would-be
Goldstone bosons would take values on the manifold $G_L \times G_R
/G_D \cong G$. This satisfies the theorem by having the low-energy
excitations get a mass and restoring the symmetry to the full $G_L
\times G_R$.

A critical theory with non-abelian symmetry $G$ does occur when an 
extra term, the Wess-Zumino term, is added to $S_{PCM}$. This term is 
easiest to write in a three-dimensional space $M$ which has the 
two-dimensional space of interest as a boundary; 
we assume the two-dimensional space has no boundary. It is 
\begin{equation} 
\Gamma(g) = \frac{1}{12\pi} \int_M d^3 x\ \epsilon^{\alpha\mu\nu} 
\hbox{Tr}\left[g^{-1}\partial_\alpha g\, g^{-1}\partial_\mu g\, 
g^{-1}\partial_\nu g\right] 
\label{swz} 
\end{equation} 
One finds that different ways of extending space to three dimensions
result only in changing $\Gamma(g)$ by $2\pi$ times an integer. Thus
we can add $ik\Gamma(g)$ to $S_{PCM}$ for any integer $k$. This model
has a critical point at $u = 1/|k|$ [\onlinecite{witten84}]. This
theory, with action
\begin{equation} 
I(g) = S_{PCM}(g,u=1/|k|) - ik\Gamma(g) \ ,
\label{swzw} 
\end{equation} 
has conformal invariance and is known as the $G_k$ Wess-Zumino-Witten 
(WZW) model.  A great deal of information is known about the WZW 
model; the most important for our current purposes is that the 
correlators decay algebraically, as they must at a critical point. 
 
Since the WZW model has an explicit action, it is easy to use the trick 
in Eq.\ (\ref{Qdef}) to find a two-dimensional Hamiltonian which has a 
ground-state wave functional with amplitudes given by the WZW action, \ie 
\begin{equation} 
\Psi_0[g] = e^{-I(g)}. 
\label{psiwzw} 
\end{equation} 
The WZW model is critical, so one might guess that the quantum version 
with this wave function would also be critical, in the fashion of the 
theories in the previous sections. However, it is not. The reason is that 
the Wess-Zumino term has an $i$ in front of it, so 
\begin{equation} 
|\Psi_0[g]|^2 = e^{-2S_{PCM}(g,u=1/|k|)} 
\end{equation} 
Thus equal-time correlators in the ground state are weighted with the 
action of a theory with a finite correlation length, the principal 
chiral model. These correlators decay exponentially. 
 
We do note that when the level $k=1$, one can use an alternate method 
to find the 2d WZW correlators in the ground-state of a quantum 
theory. All the fields of $SU(N)_1$ can be written in terms of $N-1$ 
free bosons [\onlinecite{yellow}]. The combination of 
the corresponding quantum Lifshitz theories will then have the 
$SU(N)_1$ correlators in its ground state. Likewise, $O(N)_1$ and 
$SU(2)_2$ can be written in terms of free Majorana fermions. Since the 
different bosons or fermions here do not interact, the physics is the 
same as that discussed earlier; in particular the model is essentially 
free, and the excitations are abelian. 
 
\subsection{Yang-Mills, Chern-Simons, and Wess-Zumino-Witten}
\label{sec:ymcswzw} 
 
Since it is not possible to have a wave functional with $|\Psi[g]|^2
=e^{-I[g]}$, we will need to work harder to find how the physics of
the WZW model can arise in $2+1$ dimensions. A connection between the
WZW model and the wave functions of $2+1$-dimensional Chern-Simons
gauge theory has long been known [\onlinecite{witten88,elitzur89}].
Chern-Simons theory without any matter fields is a topological field
theory: the physical states are Wilson and Polyakov loops, and their
correlation functions do not depend on distance but only on
topological properties of the loops. These correlators can be
expressed in terms of the Verlinde numbers, which also describe the
(chiral) fusion rules of the WZW model [\onlinecite{verlinde88}].
These topological field theories have been discussed in detail in many
places. One discussion of issues closely related to those of interest
here is the recent work of Freedman {\it et al.\/}
[\onlinecite{freedman03}]; we will discuss the relation of our results
to this paper in the next subsection.
 
In this subsection we will describe a wave functional which after squaring and 
averaging over fields indeed yields the WZW partition function. We 
will show that this wave functional is the ground state of a 
2+1-dimensional gauge theory. This all sounds like exactly what we 
want for a quantum critical point, but in a gauge theory, physical 
observables must be gauge invariant. Gauge-invariant states will turn 
out to have a gap, and gauge-invariant correlators in the ground state 
will be exponentially decaying.  Introducing a gauge field therefore 
will not yield a quantum critical point with non-abelian symmetry. It 
will yield, however, something just as interesting: an explicit 
Hamiltonian for a theory in a topological phase. 

Before discussing construction of these wave functionals, which was
done in great detail by Witten [\onlinecite{witten92}] and whose
construction we will follow here, it is worthwhile to discuss first
its physical meaning and to emphasize some well-known facts of
Chern-Simons gauge theories [\onlinecite{witten88}]. The ground state
wave function of any field theory can be viewed as the quantum
mechanical amplitude of some arbitrary state into the vacuum (or
ground state), {\it i.e.\/} what we normally call the vacuum state in
a given representation.  As such this amplitude is the functional
integral of the quantum field theory on an open manifold bounded by
the initial time surface.  However, the partition function of a
topological gauge theory such as Chern-Simons is only gauge invariant
on a manifold without boundary. Thus, the ground state wave function
has a gauge anomaly. Nevertheless, if the wave function is used to
compute expectation values of {\em gauge-invariant observables}, the
result is gauge invariant.  The reason is that the computation of
expectation values (as well as all inner products) involves the
conjugate wave function, {\it i.e.\/} the amplitude to evolve from the
vacuum state in the remote past into the chosen state at the fixed
time surface.  Consequently, the gauge anomaly cancels in the
computation of the expectation values of gauge-invariant operators. In
this way, at a formal level, the computation of expectation values
leads to a formally ``doubled" theory even though the number of
degrees of freedom has not changed. In subsection \ref{sec:doubled} we
will discuss a physically doubled theory which has a formal relation
with what we do in this Subsection.  There is an extensive literature
on the technical issues involved in this problem: the relevant
discussion for the analysis done in this paper can be found in
Refs. [\onlinecite{witten92,spiegelglas92,grignani96,karabali99}].
 
We should note here that the gauge anomaly of the wave function is
formally (mathematically) analogous to the gauge anomaly of
Chern-Simons theory on manifold with a spatial edge. In the latter,
the anomaly is physical: to cancel the anomaly the gauge-invariant
theory must include physical degrees of freedom residing at this
$1+1$-dimensional boundary. In the incompressible fractional quantum Hall
fluid, this results in physical edge states
[\onlinecite{wen91,wen95}].

Another important fact is that, since Chern-Simons gauge theory is a
topological field theory, the partition function on a closed
space-time manifold is independent of the metric. However, the
boundary of the manifold and the choice of polarization break the
general coordinate invariance of Chern-Simons theory. In particular,
the choice of holomorphic polarization induces a conformal structure
in the wave functional [\onlinecite{witten88,elitzur89}] which is
absent in other polarizations [\onlinecite{dunne89}]. We will see that
the conformal wave function arises naturally in a specific theory, the
strong-coupling limit of the Yang-Mills Chern-Simons theory.
 
The wave functional of interest was discussed in depth by Witten 
[\onlinecite{witten92}]. It involves the WZW field $g$ coupled to a 
gauge field $A_i$ taking values in the Lie algebra of $G$. It is most 
convenient to write the gauge field in complex coordinates 
$A_z\equiv (A_1 -i A_2)/2$ and $A_{\overline z} \equiv (A_1+ iA_2)/2$. 
When the symmetry $G_R$ of the WZW model is gauged, the closest thing to a 
gauge-invariant action is 
\begin{equation} 
I(g,A) = I(g) + \frac{k}{4\pi} \int d^2z\, \hbox{Tr}\left[2A_{\overline z} 
g^{-1} \partial_z g\ -\  A_{\overline z} A_z\right] 
\label{iga} 
\end{equation} 
This action is not gauge invariant; the subgroup $G_R$ of $G_L\times G_R$ 
is anomalous [\onlinecite{witten92}]. Indeed, the fields under gauge transformations $U(z,\overline{z})$ in $G_R$ as 
\begin{equation} 
g\to gU\ ,\qquad A_i \to U^{-1} A_i U  + U^{-1}\partial_i U 
\end{equation}
so that the action $I(g,A)$ transforms as
\begin{equation}
I(g,A)\to I(g,A) + \frac{k}{4\pi} \int d^2 z\, \hbox{Tr}\left[ 
A_{z} U\partial_{\overline z} U^{-1} 
- A_{\overline z} U\partial_z U^{-1}\right] 
- i k \Gamma(U). 
\label{igagauge} 
\end{equation} 
Note, however, that the variation of $I(g,A)$ under gauge
transformations is independent of $g$. If we were to simply square a
wave functional proportional to $e^{-I(g,A)}$, the WZW term in the
action would cancel and we would be back a gauged principal chiral
model. Instead, we define a wave functional depending only on $A$ by
doing the path integral over $g$:
\begin{equation} 
\Psi[A] \equiv \int [Dg] e^{-I(g,A)}. 
\label{psiA} 
\end{equation} 
 
This wave functional does yield the full WZW partition function after 
squaring and integrating over $A$ [\onlinecite{witten92}]. We define 
this integrated $|\Psi|^2$ as 
\begin{equation} 
 |\Psi|^2\equiv 
\frac{1}{\hbox{vol }\hat{G}} \int [DA] \left|\Psi[A]\right|^2 , 
\end{equation} 
where we have divided the measure by the volume of the gauge group 
$\hat G$ because it follows from Eq.\ (\ref{igagauge}) that even though 
$\Psi[A]$ is not gauge-invariant, $\overline{\Psi[A]}\Psi[A]$ is. 
Substituting the definition Eq.\ (\ref{psiA}) gives 
\begin{eqnarray*}  
|\Psi|^2 &=& \frac{1}{\hbox{vol }\hat{G}} \int [DA][Dg][Dh^{-1}] \; 
e^{\displaystyle{-I(g,A) - (I(h^{-1},A))^*}}\ ,\\  
I(g,A) + (I(h^{-1},A))^*&=&
 I(g)+I(h)+\frac{k}{2\pi}\int d^2 z\, \hbox{Tr} 
\left[A_{\overline z} g^{-1}\partial_z g - 
A_z \partial_{\overline z} h\cdot h^{-1} - A_{\overline z} A_z\right]  
\end{eqnarray*} 
Note that we have defined $A$ so that $(A_{z})^\dagger = -A_{\overline z}$, 
{\it i.e.\/} covariant derivatives have no $i$ in them. We have also 
used the easily-proven fact that $I(h)=I^*(h^{-1})$.
Since the integral 
over $A$ is Gaussian, it can easily be done, giving 
$$
|\Psi|^2 = \frac{1}{\hbox{vol }\hat{G}} \int 
[Dg][Dh^{-1}] \exp\left(-I(g)-I(h)+ 
\frac{k}{2\pi}\int d^2 z \, \hbox{Tr}\left[g^{-1}\partial_z g\, 
\partial_{\overline{z}} h\cdot h^{-1}\right]\right) \ .
$$
This can be simplified by using the Polyakov-Wiegmann formula 
[\onlinecite{polyakov83}]
\begin{equation} 
I(gh) = I(g) + I(h) - \frac{k}{2\pi}\int d^2 z \, 
\hbox{Tr}\left[g^{-1}\partial_z g\, 
\partial_{\overline{z}} h\cdot h^{-1}\right]. 
\label{pw} 
\end{equation} 
Thus the integrand depends only on the product $f\equiv gh$, so we can 
change variables to $f$, which cancels the volume of the gauge group. This 
yields the final expression for the normalization [\onlinecite{witten92}] 
\begin{equation} 
|\Psi|^2= \int [Df] e^{-I(f)} \ . 
\label{psisquared} 
\end{equation} 
Thus, remarkably, the integrated square of the 
wave functional $\Psi[A]$ ends up giving the full WZW path integral, 
including the imaginary piece: because of 
the path integral, the right-hand-side of Eq.\ (\ref{psisquared}) is 
positive and real as it must be. 
 
To reemphasize a point made earlier, (\ref{psisquared}) 
does not mean we have now found a $2+1$-dimensional quantum 
critical point whose equal-time correlators are those of the WZW 
model. We must first find a Hamiltonian which has $\Psi[A]$ in 
Eq.\ (\ref{psiA}) as its ground state. Once having found that, the 
physically-relevant operators are only those satisfying the proper 
gauge-invariance properties and regularization.  After having done so, 
we will end up seeing that despite sharing some key properties with 
the WZW model, the $2+1$-dimensional model is gapped and has 
exponentially-decaying correlators. 
 
To find such a Hamiltonian, let us go back to the definitions Eq.\ (\ref{iga}) 
and Eq.\ (\ref{psiA}) of the wave functional $\Psi[A]$. Because the only 
dependence on $A_z$ is through the quadratic term, we have 
\begin{equation} 
\left(\frac{\delta}{\delta A_z} - \frac{k}{4\pi} A_{\overline z}\right) 
\Psi[A]=0. 
\label{wit1} 
\end{equation} 
With a little more work [\onlinecite{witten92}], one also can show that 
\begin{equation} 
\left(D_{\overline z} \frac{\delta}{\delta A_{\overline z}} + 
\frac{k}{4\pi} D_{\overline z} A_{z} 
-\frac{k}{2\pi} F_{\overline{z} z}\right)\Psi[A] = 0, 
\label{wit2} 
\end{equation} 
where we have define the covariant derivatives via $D_ig = \partial_i g 
- gA_i$ and $D_iA_j = \partial_i A_j + [A_i,A_j]$. The field strength 
is defined as $F_{\overline{z}z} = \partial_{\overline z} A_z - 
\partial_z A_{\overline z} + [A_{\overline z}, A_z]$. 
 
We would therefore like to find a Hamiltonian with a ground-state 
wave function satisfying Eq.\ (\ref{wit1}) and Eq.\ (\ref{wit2}). As noted in Ref.\ 
[\onlinecite{witten92}], these equations arise in the canonical quantization 
of Chern-Simons theory [\onlinecite{elitzur89}], so this suggests 
we look there. The precise Hamiltonian turns out to be given by the 
strong-coupling limit of Yang-Mills theory with a Chern-Simons term 
[\onlinecite{grignani96}]. In the action, we have a gauge field $A_\mu$ 
taking values in the Lie algebra of $G$; here $\mu=0,1,2$ and 
$A_\mu$ depends on space and time. 
The action for the strong-coupling limit of Yang-Mills theory 
on a three-manifold $M$ includes only the electric-field term, namely 
\begin{equation} 
S_{SC} = \frac{1}{2e^2}\int_M\,\hbox{Tr}\left[F_{0i}F^{0i}\right]\ . 
\label{ssc} 
\end{equation} 
This term is not Lorentz-invariant, but does preserve 
two-dimensional rotational symmetry. The Chern-Simons term is 
\begin{equation} 
S_{CS} = \frac{k}{4\pi} \int_M  \epsilon^{\mu\nu\alpha} 
\hbox{Tr} \left[A_\mu\partial_{\nu} A_\alpha + \frac{2}{3} 
A_\mu A_\nu A_{\alpha} \right]\ . 
\label{scs} 
\end{equation} 
Under gauge transformations $U(x)$ belonging to $G$, the integrand in 
$S_{SC}$ is invariant, but the integrand in $S_{CS}$ is not: 
\begin{equation} 
S_{CS} \to S_{CS} + \frac{k}{4\pi} \int_M 
\,\epsilon^{\mu\nu\alpha}\, \hbox{Tr} \left[ 
\partial_{\mu}\left(A_\alpha + U 
\partial_\nu U^{-1}\right)\right] + k \Gamma(U). 
\label{csgauge} 
\end{equation} 
For the manifold $M$ a three-sphere, 
the latter term turns out to be the winding number of the 
gauge transformation $U(x)$, and is an integer times $2\pi k$ 
[\onlinecite{deser82}].  The Chern-Simons term is gauge-invariant if 
$M$ has no boundary, and $k$ is an integer. If $M$ has a spatial 
boundary, one must include massless chiral fermions on the edge to 
restore gauge invariance, giving for example the famous edge modes in 
the fractional quantum Hall effect [\onlinecite{wen90}]. 
 
The Hamiltonian with (\ref{psiA}) as a ground-state wave functional comes 
from canonically quantizing the theory with action 
\begin{equation} 
S= S_{CS} + S_{SC}, 
\label{scssc} 
\end{equation} 
following, for instance, Refs.\ 
[\onlinecite{deser82,elitzur89,grignani96}]. Since this is a 
fairly standard computation, we will be brief here. The gauge 
invariance allows us to fix temporal gauge $A_0=0$, so that the 
degrees of freedom are the gauge fields $A_1$ and $A_2$. Their 
canonical momenta are 
\begin{equation} 
\Pi_i = \frac{1}{e^2}F_{0i} + \frac{k}{8\pi} \epsilon_{ij}A_j \ .
\label{canmon} 
\end{equation} 
The Chern-Simons term contributes 
nothing to the Hamiltonian, because all the terms are first-order in 
time derivatives. The classical Hamiltonian 
in the strongly-coupled limit is therefore 
\begin{equation} 
H= \frac{1}{e^2}\int d^2 x\, \hbox{Tr}\left[(F_{0i})^2\right] 
={e^2}\int d^2 x\, \hbox{Tr}\left[\left(\Pi_i - \frac{k}{8\pi} 
\epsilon_{ij}A_j\right)^2\right]. 
\end{equation} 
Expanding $A$ and $F$ in terms of generators $T^a$ 
of the Lie algebra of $G$, we impose the canonical commutation relations 
\begin{equation} 
[A_i^a(\vec{x}),\Pi_j^b(\vec{y})] = i \delta_{ij} \delta^{ab} 
\delta^{(2)}(\vec{x}-\vec{y})\ . 
\label{ccrA} 
\end{equation} 
One of the effects of the Chern-Simons term is that $A_1$ and $A_2$ 
do not commute. In the Schr\"odinger picture, $\Pi_j$ is given by 
\begin{equation} 
\Pi_j = -i \frac{\delta}{\delta A_j} 
\label{pij} 
\end{equation} 
operating on the wave functionals. 
 
The Hamiltonian of this theory in the Schr\"odinger picture 
has a very simple form. Because $F_{0i}$ includes 
$\Pi_i$, one has the usual ordering ambiguity in the quantum Hamiltonian. We 
define 
\begin{equation} 
E = \frac{i}{e^2}\left(F_{01} + i F_{02}\right) = 
\frac{\delta}{\delta A_z} - \frac{k}{4\pi} A_{\overline{z}} 
\label{Edef} 
\end{equation} 
so that 
\begin{equation} 
[E^a(\vec{x}), (E^b(\vec{y}))^\dagger] = \frac{k}{2\pi}\delta^{ab} 
\delta^{(2)}(\vec{x}-\vec{y}). 
\label{ccrE} 
\end{equation} 
The operators $E$ and $E^\dagger$ are like annihilation and creation 
operators. We then normal-order the Hamiltonian as in section \ref{sec:scale-wf}, 
subtracting the vacuum energy to give 
\begin{equation} 
H=e^2 \int d^2 x\, \hbox{Tr}\left[ E^\dagger(\vec{x}) E(\vec{x})\right] \ .
\label{HEE} 
\end{equation} 
Thus the Hamiltonian of strongly-coupled Yang-Mills theory with a
Chern-Simons term is precisely of the form Eq.\ (\ref{eq:HQQ}), like
all the others discussed in this paper. Moreover, the annihilation
relation $E\Psi[A] =0$ here is identical to the relation Eq.\
(\ref{wit1}). We thus have found an explicit Hamiltonian whose ground
state obeys the first of the two relations satisfied by $\Psi[A]$
above.
 
Finding the Hamiltonian alone does not complete the canonical 
quantization of gauge theories with a Chern-Simons term. Fixing the 
gauge $A_0=0$ still allows time-independent gauge transformations. Moreover, 
there is no time derivative $\dot{A}_0$ in Eq.\ (\ref{scssc}), so $A_0$ should 
be viewed as a Lagrange multiplier which results in a constraint. When 
we are canonically quantizing the theory in $A_0=0$ gauge, this constraint 
is implemented on the wave functions. Specifically, one has 
\begin{equation} 
\left( D_{\overline z} \frac{\delta}{\delta A_{\overline z}} + 
D_{z} \frac{\delta}{\delta A_{z}} 
+ \frac{k}{4\pi} \partial_z A_{\overline z} 
- \frac{k}{4\pi} \partial_{\overline z} A_{z}\right) \Psi[A]=0. 
\label{gausslaw} 
\end{equation} 
This operator is the generator of time-independent gauge 
transformations on $A_z$ and $A_{\overline z}$, so this condition 
amounts to requiring that the wave function be invariant under such 
transformations. Another way of viewing this constraint is as 
requiring that the wave functionals obey the non-abelian version of 
Gauss' Law.  Requiring Gauss' Law along with Eq.\ (\ref{wit1}) yields 
Eq.\ (\ref{wit2}), the other desired relation: 
adding Eq.\ (\ref{wit1}) and Eq.\ (\ref{wit2}) together yields 
Eq.\ (\ref{gausslaw}).  Moreover, the fact (\ref{igagauge}) that 
$\Psi[A]$ is not gauge invariant is precisely the effect of the fact 
(\ref{csgauge}) that the Chern-Simons action is not gauge invariant 
when there are boundaries [\onlinecite{elitzur89}]; one can think of 
the constant-time slice required to define a wave functional as a 
boundary in space time. 
 
We have seen that the wave functional Eq.\ (\ref{psiA}) indeed 
describes a zero-energy ground state of the $2+1$-dimensional theory 
with action Eq.\ (\ref{scssc}). We now need to understand the 
correlators in the ground state and the excited states. Luckily, the 
former issue has been studied in 
Refs.\ [\onlinecite{gawedzki89,karabali90,spiegelglas92,witten93}], and 
the latter in Refs.\ [\onlinecite{grignani96,karabali99}]. We will show 
in the next subsection that the ground-state are those of a 
topological field theory, and that one can prove there is a gap in the 
spectrum. 
 
\subsection{The doubled theory} 
\label{sec:doubled}
 
It is both convenient and physically important to study the wave 
function of the ``doubled'' theory, where we have two gauge fields $A$ 
and $B$. The three-dimensional action is the sum of an action of the 
form of Eq.\ (\ref{scssc}) for both fields, where the Chern-Simons term 
for the field $A$ has coefficient $k$, while that for $B$ has 
coefficient $-k$.  As opposed to Chern-Simons theory with a single 
field, the doubled theory is also invariant under time reversal and 
parity, if the fields $A$ and $B$ are exchanged under these 
transformations\footnote{For this reason, this theory has arisen for example 
in theories of quantum computation and in superconductivity 
[\onlinecite{freedman03},\onlinecite{shivajinew}], and in effective ``coset" field theories of the non-abelian 
fractional quantum Hall states [\onlinecite{wen98,fradkin-nayak98,cabra00}].}. 
The two fields are not coupled in the action, so formally the 
wave functional of the doubled theory factorizes:
\begin{equation} 
\chi[A,B] \propto 
\Psi[A]\overline{\Psi[B]} 
\end{equation} 
There is also a compelling technical reason to study the doubled
theory.  We will show that the doubled theory amounts to gauging a
non-anomalous symmetry. In particular, after integrating out one of
the gauge fields (a simple Gaussian integration), the wave functional
is gauge invariant in the remaining field. The theory can be quantized
consistently; effectively the regularization (the measure of the path
integral) couples the two copies.  Thus in spite of this factorization
of the wave function, the operators that create the physical states
are made of operators acting on each sector, carefully glued together
to satisfy the requirements of gauge invariance.

Canonically quantizing the doubled theory is straightforward, since 
the action splits into decoupled pieces. The 
ground-state wave functional of the doubled theory 
can be written as [\onlinecite{witten92}] 
\begin{equation} 
\chi[A,B] \equiv \int [Dg] e^{-I(g,A,B)} 
\label{eq:chiAB} 
\end{equation} 
where 
\begin{equation} 
I(g,A,B) \equiv I(g) 
+ \frac{k}{4\pi} \int d^2z \hbox{Tr}\left[2A_{\overline z} 
g^{-1} \partial_z g + 2B_{z} 
g \partial_{\overline z} g^{-1} -  A_{\overline z} A_z 
- B_{\overline z} B_z + 2 B_z g A_{\overline z} g^{-1} 
\right] 
\label{igab} 
\end{equation} 
One can prove that $\chi[A,B]\propto \Psi[A]\overline{\Psi [B]}$ 
indirectly by showing that $\chi[A,B]$ satisfies both (\ref{wit1}) and 
(\ref{wit2}) for $A$, and the conjugate equations for $B$. 
Directly, we prove this by first noting that the 
Polyakov-Wiegmann identity (\ref{pw}) yields 
$$
I(g,A) + (I(h,B))^* = I(h^{-1}g,A,B)  - \frac{k}{2\pi} 
\int d^2 z\, \hbox{Tr}\left[ 
(gD_{\overline z} g^{-1}) (hD_z h^{-1})\right]\ . 
$$
where the covariant derivatives are 
$D_{\overline z} g^{-1} = \partial_{\overline z} g^{-1} + 
A_{\overline z} g^{-1}$ and 
$D_{z} h^{-1} = \partial_{z} h^{-1} + 
B_{z} h^{-1}$. 
In the path integral we can split apart the last term by 
introducing an auxiliary gauge field $C$: 
\begin{eqnarray*} 
\Psi[A]\overline{\Psi[B]} &=& \int [Dg][Dh^{-1}] e^{-I(g,A) - 
(I(h,B))^*}\\ 
&=& \int [Dg][Dh^{-1}][DC] \exp\Big(-I(h^{-1}g,A,B)\\ 
&&\qquad\qquad - 
 \frac{k}{2\pi} \int d^2 z\, \hbox{Tr}\left[ 
\alpha C_{ z} (gD_{\overline z} g^{-1}) + \alpha C_{\overline z} 
(hD_z h^{-1}) + {\alpha}^2 C_z C_{\overline z} \right]\Big)\ , 
\end{eqnarray*} 
where $\alpha$ is a (small) coupling constant.  The terms linear in 
$C$ can be removed by redefining $g$ and $h$. Namely, under $g\to g + 
\delta g$ for small $\delta g$, the action 
\begin{equation} 
I(g,A) \to I(g,A) + \frac{k}{2\pi} \int d^2 z\, \hbox{Tr}\left[ 
gD_{\overline z} g^{-1} \partial_{z}(\delta g\, g^{-1})\right]\ . 
\end{equation} 
Thus if we redefine the fields $g$ and $h$ so that $\delta g$ and 
$\delta h^{-1}$ obey $\alpha C_z = \partial_z(\delta g\, g^{-1})$ and 
$\alpha C_{\overline z} = -\partial_{\overline z}(h\, \delta h^{-1})$, 
the mixed terms cancel and the field $C$ decouples. 
The integrand then only depends on the combination $h^{-1}g$, so changing 
variables to $f=h^{-1}g$ yields $\chi[A,B]$, up to a factor of the 
volume of the gauge group: 
\begin{equation} 
\Psi[A]\overline{\Psi[B]} = \int[Dg][Dh^{-1}] e^{-I(h^{-1}g,A,B)} = 
\hbox{vol }\hat{G}\ \chi[A,B]. 
\end{equation} 
 
To understand this wave functional $\chi[A,B]$, we compute 
its integrated norm. One can easily do one of the two functional integrals 
over the gauge fields in the same manner as before, yielding 
[\onlinecite{witten92}] 
\begin{equation} 
\int [DB] \left|\chi[A,B]\right|^2 =  \int[Dg] e^{-I_{G/G}(g,A)} 
\label{chisquared} 
\end{equation} 
where $I_{G/G}$ is the action of the $G/G$ gauged WZW model, namely 
\begin{equation} 
I_{G/G}(g,A) = 
I(g) + \frac{k}{2\pi} \int d^2z\, \hbox{Tr}\left[A_{\overline z} 
g^{-1} \partial_z g  + A_{z} 
g \partial_{\overline z} g^{-1} - 
A_{\overline z} A_z + A_z g A_{\overline z} g^{-1} 
\right]\ . 
\label{igg} 
\end{equation} 
As opposed to $I(g,A)$ in Eq.\ (\ref{iga}), this action is gauge 
invariant, because one is gauging the full $G_L\times 
G_R$ symmetry of the WZW model. This means the path integral of the 
doubled theory is well defined and free of anomalies, as well as 
parity-invariant. 
 
Let us compare the doubled theory to the undoubled one.  For the
undoubled theory, we found that the effective partition function
$|\Psi|^2$, the ground-state wave functional squared and integrated
over the (one) gauge field, was that of the WZW model. However,
constructing excited states and operators directly in the undoubled
theory appears to be problematic due to the gauge anomaly discussed above. 
Although it would seem that {\it a priori} one could not require that
operators be gauge invariant if the ground-state wave functional
itself is not, as we emphasized in Subsection \ref{sec:ymcswzw}, from
general considerations of Chern-Simons theory we know that the
physical observables {\em are} gauge invariant, that {\em only} gauge
invariant observables must be considered, and that their expectation
values are free of any anomalies. However, while this is apparent in
the path-integral construction of the quantum theory, it is not so
apparent if one is to use the wave function, {\it i.e.\/} in terms of
a chiral Euclidean WZW model.  As we discussed above, at this level
one is led to introduce a formal ``doubled" theory even for the
undoubled theory.  Thus, although the wave function itself factorizes
the physical states cannot be constructed in terms of arbitrary
factors from each sector of the doubled theory.  Hence, the
requirement of gauge invariance can spoil the apparent factorization
suggested by the wave function $\chi[A,B]$ of Eq.\ (\ref{eq:chiAB}).

The formally correct way to define the correlators in the undoubled
theory is in the doubled theory: one can always introduce another
field $B$ and then integrate it out. The reason for doing this is that
the doubled theory can be properly regulated, because
(\ref{chisquared}) is gauge invariant. All the formal manipulations
done above are well founded, because the path integral can be defined
properly.  In contrast in a truly doubled theory, the gauge-invariant
physical observables couple to both fields which must then be regarded
as genuine degrees of freedom.  In contrast, in the undoubled theory,
the additional degrees of freedom are a formal device used to regulate
the theory\footnote{This approach is reminiscent of stone soup.}. This
means that correlators in both theories are those of the $G/G$ gauged
WZW model, which is anomaly-free. These correlators have been studied
in great detail [\onlinecite{gawedzki89,karabali90,spiegelglas92}]. In
particular, careful discussions of the proper regularization of this
theory can be found in these papers. As common in non-abelian gauge
theories, to properly do these path integrals, one needs to introduce
fermionic ghosts. One then finds a BRST charge $Q_{BRST}$ obeying
$(Q_{BRST})^2=0$; physical states are annihilated by it.  We can thus
consistently demand that states and physical operators be gauge
invariant.

A relation between the $G/G$ gauged WZW model and topological field theory 
was conjectured in [\onlinecite{spiegelglas92}] and proven in 
[\onlinecite{witten92}].  In particular, it was shown that once we 
integrate over $A$ as well, the resulting partition function 
$|\chi|^2$ is independent of the metric of two-dimensional space. The 
correlators in the ground state therefore are independent of distance, 
and given by a topological field theory. A topological field theory is 
obtained by studying the states which are annihilated by $Q_{BRST}$ 
but not given by $Q_{BRST}$ acting on something else.  In mathematical 
language, the physical states of the topological theory are given by 
the cohomology of $Q_{BRST}$. It was derived directly in 
[\onlinecite{gerasimov93}] that the correlators of the $G/G$ topological 
field theory are given in terms of the Verlinde numbers 
[\onlinecite{verlinde88}], which give the dimensions of conformal 
blocks in the ordinary WZW model. 
 
Going to the doubled theory has therefore allowed us to not only avoid
technical problems, but also to prove that the ground state
correlators are those of a topological field theory. We should stress
once again that the doubled theory used here has twice as many degrees
of freedom as the undoubled theory. It is a {\em physically distinct}
time-reversal invariant theory, unlike the undoubled theory which
breaks time-reversal symmetry.  This physical difference is apparent
from the construction of its observables: it has twice as many
``anyons", which come in time-reversed pairs. On a torus, the Wilson
loops of the corresponding topological field theory wind around both
cycles [\onlinecite{freedman03}]. The weights of the loops can be
defined locally, and one can see how the non-abelian statistics arise
[\onlinecite{freedman03}].  The results of this section show how this
precise topological field theory arises as the ground state of a
specific Hamiltonian. We note in addition that if one specializes
these results to an abelian gauge field, the topological field theory
obtained is known as a (two-dimensional) BF theory
[\onlinecite{spiegelglas92},\onlinecite{rossini92}].  Recently,
Freedman, Nayak and Shtengel [\onlinecite{freedman03b}] have discussed
(reasonably local) lattice models of interacting fermions and bosons,
which they argue have topological ground states with some of the
topological algebraic structure of the doubled non-abelian
Chern-Simons theories. It is reasonable to expect that the universal
long-distance structure of the wave functions of these topological
ground states, at least deep in the topologically-ordered phase, has
the same structure of the wave functions we discussed in this section.
 
Since we have an explicit Hamiltonian Eq.\ (\ref{HEE}) (plus the analogous 
term for the $B$ gauge field), we can go beyond the 
topological field theory describing the ground state. Because of the 
commutation relations (\ref{ccrE}), $E^\dagger$ acts like a creation 
operator. However, in a non-abelian gauge theory, $E$ and $E^\dagger$ 
are not gauge invariant, so the appropriate states are slightly more 
complicated than just $E^\dagger \Psi[A]$. The simplest candidate is 
[\onlinecite{grignani96}] 
\begin{equation} 
\psi^a(x,A,B) = E^{\dagger b}(x) 
\int[Dg]\,\hbox{Tr}\left[T^a g(x) T^b g^{-1}(x)\right] 
e^{-I(g,A,B)} \ . 
\end{equation} 
Once this amplitude is squared and $B$ is integrated over, one obtains 
a gauge-invariant probability.  In the $2+1$-dimensional picture, one 
can think of one of these states as a Polyakov loop, which intersects 
two-dimensional space at a single point $x$.  The commutation relation 
Eq.\ (\ref{ccrE}) shows that this is indeed an eigenstate with a gap 
proportional to $k$. A more thorough treatment, taking into account 
the measure of the path integral, shows that the gap is shifted to 
$e^2(k+2c_A)/(4\pi)$, where $c_A$ is the quadratic Casimir of the 
adjoint representation of the Lie algebra of $G$ 
[\onlinecite{karabali99}].  For our purposes, the important point is 
that there is indeed a gap. 
 
We have thus seen that correlators of Wilson loops in the ground state
of strongly-coupled Yang-Mills theory are given by a topological field
theory. We have also seen that the theory has a gap, so it is indeed
in a topological phase. The last thing we would like to discuss is the
ground-state correlators for operators not in the topological
theory. In other words, we wish to consider the full set of physical
operators (i.e.\ those annihilated by $Q_{BRST}$), including those
which are given by $Q_{BRST}$ acting on something else. Because our
Hamiltonian involves only the electric and not the magnetic field, it
is not Lorentz invariant. Thus the existence of a gap does not
immediately require that correlators decay exponentially. However, in
this theory, they do, as implied by the results of
[\onlinecite{witten93}].  We noted above that proper quantization of
gauged WZW models requires introducing fermionic ghosts, and a
fermionic operator $Q_{BRST}$. A fermionic symmetry suggests the
appearance of supersymmetry, and indeed the $G/G$ topological field
theory can also be obtained from a supersymmetric field theory where
the supersymmetry charge is ``twisted'' into $Q_{BRST}$
[\onlinecite{gepner91,vafa91,intriligator91,spiegelglas92}].  The
supersymmetric field theory also describes the correlators of the full
theory, not just the topological subsector.  For the case where
$G=SU(N)$ at level $k$, the appropriate two-dimensional field theory
is a (twisted) supersymmetric sigma model, where the bosonic fields
take values on the ``Grassmannian'' manifold $U(N+k)/(U(N)\times
U(k))$ [\onlinecite{witten93}]. It is likely that the other simple Lie
groups end up giving supersymmetric sigma models on the analogous
Grassmannians.
 
This means we have now come full circle! We started without gauge
fields, and found that the equal-time correlators are those of the
principal chiral model, a two-dimensional sigma model with a curved
target space. These correlators are exponentially decaying.  We then
introduced gauge fields, in the hope of finding a quantum critical
point. After this lengthy discussion, we have ended up showing the
ground state is described by a supersymmetric two-dimensional sigma
model with a curved target space, the Grassmannian. This means that
the correlators here are exponentially decaying as well, and the
theory is in a topological phase whose properties are encoded in the
wave function $\chi[A,B]$.

\bigskip\bigskip 
\begin{acknowledgments} 
 
We are grateful to M.P.A.~Fisher, M.~Freedman, S.~Kivelson, C.~Nayak,
K.~Shtengel, T.~Senthil, S.~Sondhi, M.~Stone, A.~Vishwanath and
X.G.~Wen for many illuminating conversations on quantum dimer models,
topological phases and beyond. We are grateful to M.~Stone for
motivating us to keep track of factors of $i$. We also thank
M.~Freedman, C.~Nayak and Z.~Wang for organizing a very stimulating
conference on topological order at the American Institute of
Mathematics, during which part of this work was done.  This work was
supported in part by the National Science Foundation through the
grants NSF-DMR-01-32990 at the University of Illinois and
NSF-DMR-0104799 at the University of Virginia.  The work of P.F.\ was
also supported by the DOE under grant DEFG02-97ER41027.
 
\end{acknowledgments} 
 
\appendix 
 
\section{Operators of the quantum Lifshitz field theory} 
\label{app:gaussian} 
 
In addition to the products of field operators $\varphi(\vec x)$, in
what follows we will be interested in two types of local operators:
charge and vortex operators. The charge operators are
\begin{equation} 
{\mathcal O}_n(\vec x)=e^{\displaystyle{-i n\; \varphi(\vec x)}} 
\label{eq:charge} 
\end{equation} 
where $n \in {\mathbb Z}$. This operator creates a boson coherent-state which we will refer to as a {\em charge} $n$ excitation. 
The vortex operators are 
\begin{equation} 
{\tilde{\mathcal O}}_m(\vec x)=e^{\displaystyle{i  \int d^2z  \; \varphi(\vec z)\; \Pi(\vec z)}} 
\label{eq:vortex} 
\end{equation} 
where 
\begin{equation} 
\varphi(\vec z)=m \arg(\vec z-\vec x) 
\end{equation} 
where $0\leq \arg(\vec z-\vec x) \leq 2\pi$ is the argument of the vector $\vec z-\vec x$ (with a branch cut defined arbitrarily 
along the negative $x$ axis). 
The action of the operator ${\tilde{\mathcal O}}_m(\vec x)$ on an eigenstate of the field operator $\ket{[\varphi]}$ is simply a shift 
\begin{equation} 
e^{\displaystyle{i  \int d^2z  \; \varphi(\vec z)\; \Pi(\vec z)}} \ket{[\varphi]}= 
\ket{[\varphi(\vec x)-\varphi(\vec x)]} 
\label{eq:gauge} 
\end{equation} 
In other words, it amounts to s singular gauge transformation. Therefore, its action is equivalent to coupling the field 
$\varphi$ to a vector potential whose space components $\vec A$ satisfy 
\begin{equation} 
\oint_\gamma d \vec z \cdot \vec A[\vec z]=2\pi m 
\label{eq:circulation} 
\end{equation} 
for all closed paths $\gamma$ which have the point $\vec x$ in their interior, and zero otherwise. 
In particular, the wave function of the state resulting from the action of the vortex operator on the ground state is: 
\begin{equation} 
\Psi_m[\vec x]=\me{[\varphi]}{{\tilde{\mathcal O}}_m(\vec x)}{{\rm vac}} 
=\frac{1}{\sqrt{\mathcal Z}} 
e^{\displaystyle{-\frac{\kappa}{2} \int d^2z \; \left(\vec \nabla \varphi-\vec A\right)^2}} 
\label{eq:vortex-wf} 
\end{equation} 
where $\vec A$ is any vector field which satisfies Eq.\ (\ref{eq:circulation}). 
The (equal-time) ground state expectation value of a product of vortex operators with 
magnetic charges $\{m_l\}$, {\it i.e.\/} the {\em overlap} of the state with k vortices at 
locations $\vec x_l$ and magnetic charge $m_l$ with the vortex-free ground state wave function, is therefore 
\begin{equation} 
\me{{\rm vac}}{{\tilde{\mathcal O}}_{m_1}(\vec x_1)\ldots {\tilde{\mathcal O}}_{m_k}(\vec x_k)}{{\rm vac}}= 
\frac{1}{{\mathcal Z}} \int {\mathcal D} \varphi \; 
e^{\displaystyle{-\kappa \int d^2z \; \left(\vec \nabla \varphi-\vec A\right)^2}} 
\label{eq:multivortex} 
\end{equation} 
where ${\mathcal Z}$ is given by Eq.\ (\ref{eq:Zboson}). The vector potential in Eq.\ (\ref{eq:multivortex}) satisfies 
\begin{equation} 
\varepsilon_{ij} \nabla_i A_j=2\pi \sum_{l=1}^k m_l \delta^2 (\vec z-\vec x_l) 
\label{eq:vortex-config} 
\end{equation} 
This result is equivalent to the expectation value of the vortex operator in the 2D classical $c=1$ compactified 
free bose field discussed extensively by Kadanoff [\onlinecite{kadanoff79}] (see also  [\onlinecite{yellow}]). 
 
The boson propagator of this theory, in imaginary time $t$, is 
\begin{equation} 
G(\vec x- {\vec x}^\prime, t-t^\prime)=\langle \varphi(\vec x, t) \; \varphi({\vec x}^\prime, t^\prime)\rangle 
= \int \frac{d\omega}{2\pi} 
\int \frac{d^2q}{(2\pi)^2} 
\; 
\frac{ 
e^{\displaystyle{i \omega (t-t^\prime)-i\vec q \cdot (\vec x -{\vec x}^\prime)}} 
}{\omega^2+\kappa^2 \left({\vec q}^{\;2}\right)^2} 
\label{eq:phiprop} 
\end{equation} 
which has a short-distance logarithmic divergence. From now on we will use instead the regularized (subtracted) propagator 
\begin{equation} 
G_{\rm reg}(\vec x,t) \equiv G(\vec x,t)-G(a,0) 
=-\frac{1}{8\pi \kappa}\left[ \ln\left(\frac{\vert \vec x \vert^2}{a^2}\right) 
+\Gamma\left(0,\frac{\vert \vec x \vert^2} 
{4\kappa \vert t \vert}\right)\right] 
\label{eq:regprop} 
\end{equation} 
where $a$ is a short-distance cutoff and $\Gamma(0,z)$ is the incomplete Gamma function 
\begin{equation} 
\Gamma(0,z)=\int_z^\infty \frac{ds}{s} \; e^{\displaystyle{-s}} 
\label{eq:incomplete} 
\end{equation} 
The regularized propagator has the asymptotic behaviors 
\begin{equation} 
G_{\rm reg}(\vec x,t)= 
\begin{cases} 
-\displaystyle{\frac{1}{4\pi \kappa}} \ln\left(\displaystyle{\frac{\vert \vec x\vert}{a}}\right), & \textrm{for} \;  \vert t \vert \to 0 \\ 
-\displaystyle{\frac{1}{8\pi \kappa}} \ln\left(\displaystyle{\frac{4 \kappa \vert t \vert}{a^2\gamma}}\right) , & \textrm{for} \; \vert \vec x \vert \to a 
\end{cases} 
\label{eq: asymptotic} 
\end{equation} 
where $\ln \gamma=\mathbf{C}=0.577\ldots$ is the Euler constant. 
 
The correlation functions of the charge operators are 
\begin{equation} 
\langle {\mathcal O}_n(\vec x,t)^\dagger {\mathcal O}_n({\vec x}^{\; \prime},t^\prime)\rangle= 
e^{\displaystyle{\;  n^2 \; G_{\rm reg}(\vec x-{\vec x}^{\; \prime},t-t^\prime)}} 
\label{eq:charge-corr} 
\end{equation} 
At equal (imaginary) times, $|t-t^\prime|\to 0$, it behaves like 
\begin{equation} 
\langle {\mathcal O}_n(\vec x,0)^\dagger {\mathcal O}_n({\vec x}^{\;\prime},0)\rangle= 
\left(\displaystyle{\frac{a}{\vert \vec x-{\vec x}^{\; \prime} \vert}}\right)^{\displaystyle{\frac{ n^2}{4 \pi\kappa}}} 
\label{eq:charge-eqtime} 
\end{equation} 
which implies that the operator ${\mathcal O}_n$ has (spacial) scaling dimension 
\begin{equation} 
\Delta_n=\frac{ n^2}{8 \pi \kappa} 
\label{eq:dimension-electric} 
\end{equation} 
For \mbox{$\vert \vec x-{\vec x}^{\; \prime} \vert \to a$}, its asymptotic behavior is instead given by 
\begin{equation} 
\langle {\mathcal O}_n(\vec 0,t)^\dagger {\mathcal O}_n({\vec 0},t^\prime)\rangle= 
\left(\displaystyle{\frac{a^2\gamma}{4\kappa \vert t-t^\prime\vert}}\right)^{\displaystyle{\frac{ n^2}{8\pi \kappa}}} 
\label{eq:charge-auto} 
\end{equation} 
This behavior is manifestly consistent with a dynamical critical exponent $z=2$. 
 
It is straightforward to show by an explicit calculation of the overlap of Eq.\ \eqref{eq:vortex-config}, 
which is completely analogous to the classical vortex correlation functions of the 2D Gaussian 
model [\onlinecite{kadanoff79}], that the (spacial) scaling dimension of the vortex of magnetic charge $m$ is, 
\begin{equation} 
\Delta_m=2\pi \kappa m^2 
\label{eq:dimension-magnetic} 
\end{equation}

\section{$U(1)$ Gauge-theory for the quantum six-vertex model} 
\label{app:gauge} 
 
In this Appendix, we describe the quantum six-vertex model 
in the language of gauge theory. The quantum eight-vertex model will be 
described in appendix \ref{app:z2gauge}. 
We will follow closely the gauge theory description of the 
quantum dimer model, which is described in detail in [\onlinecite{fradkin91}]. 
We note that this gauge theory is {\em not} simply an abelian 
version of the gauge theory discussed in section \ref{sec:YMCS}. 
 
We define 
link variables $E_i (\vec x)$, where $\vec x$ labels the vertices and 
$i=1,2$ indicates the direction; the unit vector in direction $i$ is denoted 
by $\vec e_i$. The link variables are integer valued, and 
can be viewed as the eigenvalues of angular momentum operators which we 
will also denote by $E_i (\vec x)$. 
We assign the values $E_1=1$ and $E_2=1$ to the right and up going arrows 
respectively, while for the left and down going arrows we have $E_1=-1$ and 
$E_2=-1$. 
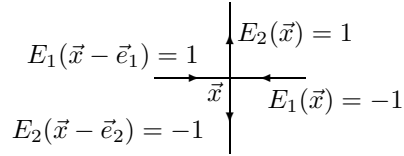
\begin{figure}[h] 
\setlength{\unitlength}{1 mm} 
\begin{center} 
\begin{picture}(20,20)(0,0) 
\put(0,10){\line(1,0){20}} 
\put(10,0){\line(0,1){20}} 
\put(10,10){\vector(0,1){6}} 
\put(10,10){\vector(0,-1){6}} 
\put(0,10){\vector(1,0){6}} 
\put(20,10){\vector(-1,0){6}} 
\put(7,7){$\vec x$} 
\put(-19,2){$E_2(\vec x -\vec e_2)=-1$} 
\put(15,6){$E_1(\vec x) = -1$} 
\put(11,15){$E_2(\vec x) = 1$} 
\put(-17,12){$E_1(\vec x - \vec e_1)=1$} 
\end{picture} 
\caption{The `electric' fields associated to a particular vertex} 
\label{econfig} 
\end{center} 
\end{figure} 
An example is given in figure \ref{econfig}. 
In the Hamiltonian, we need a term which will lead to a restriction to those 
states in which all the $E_i(\vec x)$'s have values $\pm 1$. Such a term is 
\begin{equation} 
H_{\rm E} = \frac{1}{k_1} \sum_{\vec x, i} 
\left( 
E_i^2 (\vec x)-1 
\right)^2 \ , 
\end{equation} 
in the limit $k_1 \rightarrow 0$. 
 
The six-vertex model also requires that the same number of arrows point 
in and out at a vertex. This is precisely a lattice version of Gauss' law. 
Defining the lattice divergence as 
$\dmin_i E_i^{\vphantom{-}} (\vec x) \equiv 
E_1(\vec x) - E_1(\vec x -\vec e_1) + E_2(\vec x) - E_2(\vec x -\vec e_2)$, 
one easily finds that only for the six vertices of type $a,b$ and $c$, 
we have 
\begin{equation} \label{con6v} 
\dmin_i E_i^{\vphantom{-}} (\vec x) = 0 \ .
\end{equation} 
The lattice differentiation $\dmin_i$ is defined by 
$\dmin_i f (\vec x) \equiv f(\vec x) - f(\vec x -\vec e_i)$. 
In the following, when dealing with sums over the plaquettes of the 
lattice, we will frequently use 
$\Delta_i f (\vec x) \equiv \dplus_i f (\vec x) \equiv 
f(\vec x+ \vec e_i) - f(\vec x)$. 
Of course, the constraint has to commute with all the terms in the 
Hamiltonian. In the following, we will find that this is indeed the case. 
 
The two main ingredients in the gauge theory description are the flip term 
which flips the flippable plaquettes (see below) and a potential term, which 
give a finite weight to (only) the flippable plaquettes. 
Flippable plaquettes are those which have both $n_d=\widetilde{n}_d =0$. 
Pictorially, a flippable plaquette here is one 
where the arrows around the plaquette 
point either all clockwise or all counterclockwise: 
\begin{equation} 
\setlength{\unitlength}{.5 mm} 
\begin{picture}(20,20)(0,0) 
\put(5,0){\line(0,1){20}} 
\put(15,0){\line(0,1){20}} 
\put(0,5){\line(1,0){20}} 
\put(0,15){\line(1,0){20}} 
\put(5,5){\vector(1,0){6}} 
\put(15,5){\vector(0,1){6}} 
\put(15,15){\vector(-1,0){6}} 
\put(5,15){\vector(0,-1){6}} 
\end{picture} 
\raisebox{4 mm}{$\Longleftrightarrow$} 
\begin{picture}(20,20)(0,0) 
\put(5,0){\line(0,1){20}} 
\put(15,0){\line(0,1){20}} 
\put(0,5){\line(1,0){20}} 
\put(0,15){\line(1,0){20}} 
\put(15,5){\vector(-1,0){6}} 
\put(15,15){\vector(0,-1){6}} 
\put(5,15){\vector(1,0){6}} 
\put(5,5){\vector(0,1){6}} 
\end{picture} \ . 
\end{equation} 
In terms of the electric field, this can be written as 
\begin{equation} \label{hpot6v} 
H_{\rm V} = \frac{V}{64} \sum_{\mysquare[1.5]} 
(\Delta_2 E_1)^2 (\Delta_1 E_2)^2(E_1-E_2)^2 \ , 
\end{equation} 
where the sum is over all the plaquettes of the lattice. 
The factors $(\Delta_2 E_1)^2$ and $(\Delta_1 E_2)^2$ make sure that the 
arrows on opposite links of the plaquette are anti-parallel. The factor 
$(E_1-E_2)^2$ checks if two arrows on one vertex are both pointing clockwise 
or both counterclockwise. 
 
As the $E_i(\vec x)$'s have the integers as their eigenvalues, the 
canonically conjugate operators, $a_i(\vec x)$ are phases, 
i.e.\ $0\leq a_i(\vec x)<2\pi$.  Using the commutation relations 
\begin{equation} 
\left[a_j(\vec x),E_{j'}(\vec x ')\right] = 
i \delta_{jj'}\delta_{\vec x,\vec x'} \ , 
\end{equation} 
it is easy to show that the operators $e^{\pm i a_j (\vec x)}$ act as 
raising and lowering operators on the $E$'s, and thus e.g.\ the 
operator $e^{-2 i a_1 (\vec x)}$ will flip the arrow pointing outward 
from $\vec x$ to the arrow pointing inward. 
We can use these raising and lowering operators to write the term in the 
Hamiltonian which flips the flippable plaquettes 
\begin{equation} \label{hflip6v} 
H_{\rm t} = 
-2 t \sum_{\mysquare} 
\cos (2\textstyle\sum\nolimits_{\vsquare[2]} a_j(\vec x)) \ , 
\end{equation} 
where 
$\sum_{\vsquare[2]} a_j(\vec x)=\Delta_1 a_2 (\vec x) -\Delta_2 a_1 (\vec x)$ 
is the oriented sum of the $a$'s around 
a plaquette. The total Hamiltonian of the gauge theory version of the 
6-vertex model is therefore 
\begin{equation} 
H_{\rm 6v} = H_{\rm E} + H_{\rm t} + H_{\rm V} \ . 
\end{equation} 
As usual, the Rokhsar-Kivelson point is located at $t=V$. 
 
We will proceed by going to the dual formulation of the theory, and 
show that the theory is equivalent to a height model (which is well 
know). Doing the duality basically amounts to solving the 
(electrostatic) constraint. In the process, it gets replaced by a 
magnetic constraint.  To solve the constraint, we introduce the new 
variables $S(\vec r)$, which live on the sites of the dual lattice 
(or plaquettes of the direct lattice); these operators have the 
integers as their spectrum.  In addition, we need the fields 
$B_i (\vec r)$, which live on the links of the dual lattice. We can now 
write the ``electric" fields $E_i$ as follows: 
\begin{equation} \label{dual} 
E_i (\vec x) = 
\epsilon_{ij} \left(\dmin_j S (\vec r) + B_j (\vec r)\right) \ . 
\end{equation} 
Substituting this in the constraint $\dmin_i E_i^{\vphantom{-}}=0$ gives the 
`magnetic' constraint 
$\epsilon_{ij}\dmin_j B_k^{\vphantom{-}} (\vec r) = 0$, so 
$B_k$ is curl free, and can be written as a gradient. But, as the 
there are no sources, we can do even better, as becomes clear when we 
interpret $S (\vec r)$ as a height variable which lives on the 
plaquettes of the direct lattice. 
 
We will first recall the known fact that the configurations of the 6-vertex 
model can be mapped onto height configurations. The rules are as follows. 
First, pick a reference site, and give it a reference height, say 
$S (\vec 0)=0$. 
Then, if one crosses an outgoing arrow clockwise (both seen from the vertex), 
the height decreases by one, while crossing an incoming arrow (again in a 
clockwise manner), the height increases by one. As all the vertices have two 
incoming and two outgoing arrows, this indeed gives a consistent height 
configuration. So, as an example, around an $a$ vertex we have 
\begin{center} 
\setlength{\unitlength}{1 mm} 
\begin{picture}(10,10)(0,0) 
\put(5,0){\line(0,1){10}} 
\put(0,5){\line(1,0){10}} 
\multiput(5,0)(0,5){2}{\vector(0,1){3.5}} 
\multiput(0,5)(5,0){2}{\vector(1,0){3.5}} 
\put(7,7){0} 
\put(6,1){-1} 
\put(1,1){0} 
\put(1,7){1} 
\end{picture} \ . 
\end{center} 
We now assume that the $S(\vec r)$'s appearing in Eq.\ (\ref{dual}) 
can in fact be interpreted as the heights of the plaquettes. Because of the 
duality Eq.\ (\ref{dual}), the $B_i$ are now 
completely determined by the $E_i$, because they determine $S$ via 
the height rule. So we found that interpreting the $S (\vec r)$ as heights 
amounts to picking a gauge for the $B_i (\vec r)$. Combining the height rule 
of the previous paragraph with Eq.\ (\ref{dual}), we easily find that 
$B_j (\vec r) \equiv 0$. 
 
To complete the duality transformation, we need to transform the flip 
term.  This will involve the canonically conjugate variable to $S(\vec 
r)$. Let us call this the momentum $P (\vec r)$, which satisfies 
$\left[ P(\vec r),S(\vec r')\right] = i \delta_{\vec r,\vec r'}$. 
Again, acting with $e^{i P(\vec r)}$ on the plaquette at $\vec r$ 
will increase the eigenvalue of $S(\vec r)$ by one. Flipping a 
plaquette changes $S$ by $\pm 2$, so we find that we can write the 
flip term of the Hamiltonian as 
$H_{\rm t} = -2 t \sum_{\vec r} \cos (2 P(\vec r))$. 
In other words, we find that the circulation around 
the plaquette at $\vec r$ is given by 
$\sum_{\vsquare[2]} a_j(\vec x) = P (\vec r)$. 
 
In principle, we could go on to describe the eight-vertex model in a similar 
fashion. The main difference with the six-vertex model is that the 
constraint \eqref{con6v} is no longer satisfied. One can introduce a 
matter field which has no-zero values on the $d$ vertices. In addition, a 
flip term which flips every plaquette has to be constructed. This flip 
term has to commute with the new constraint. It turns out that this is 
indeed possible, but the flip term will involve the conjugate of the matter 
field. Also, one would need a potential term which gives weights plaquettes 
according to the vertices present on the plaquette. 
However, as there is a more natural gauge description, which is based on 
the $\mathbb{Z}_2$ symmetry of the eight-vertex model, we will discuss and 
use that description of the quantum eight vertex model in the next appendix.

\section{A $\mathbb{Z}_2$ gauge theory for the quantum eight vertex model} 
\label{app:z2gauge} 
 
In this appendix, we will discuss a $\mathbb{Z}_2$ gauge theory which
can be viewed as an extension of the Kitaev model which incorporates
vertex weights differing from unity. The model we will discuss is of
the Rokhsar-Kivelson type, but it is not the simplest quantum
generalization of the classical eight-vertex model, as we pointed out
in section \ref{sec:q8v}.
 
\subsection{The $\mathbb{Z}_2$ gauge theory} 
\label{app:z2gt} 
 
In this model, spins living on the bonds of the square lattice are the
degrees of freedom. Thus, on every link of the square lattice, we
define a Pauli algebra of $2 \times 2$ Hermitian matrices
$\sigma^a_j(\vec x)$, where $a=1,2,3$ labels the three Pauli matrices,
and for a lattice site $\vec x$, we denoted the orientation of the
link by $j=1,2$ ($1=$horizontal and $2=$vertical).  (Thus, the degrees
of freedom live half-way between the lattice sites $\vec x$ and $\vec
x+\vec e_j$, where $\vec e_j$ is a unit vector along the direction
$j$.) In what follows we will take the states $|\uparrow\; \rangle$
(an ``up" spin) and $|\downarrow\;\rangle$ ( a ``down" spin) as the
states which diagonalize $\sigma^1$ (instead of $\sigma^3$, as it is
customary).  The relation with the eight-vertex model is
simple. Up-spins correspond to arrows pointing up or to the right,
while down spins correspond to arrows pointing down or to the left.
Around each vertex, the number of up-spins has to be even. In this
section, we will denote the vertex by $\vec x$ and the associated
plaquette by its south-west corner $\vec x$.
 
For details, see Ref.\ [\onlinecite{fradkin78,kogut79}].  The
constraint can written in a simple form\footnote{Throughout we use the
upper label to indicate the Pauli matrix and the lower label to
indicate direction.}
\begin{equation} \label{scon8v2} 
 \sigma^1_1(\vec x) \; \sigma^1_1(\vec x-\vec e_1) \; \sigma^1_2(\vec x) \; \sigma^1_2(\vec x-\vec e_2)=1, \qquad \forall \vec x 
\end{equation} 
The term which flips all the arrows around a plaquette can be written in 
terms of $\sigma^{3}$'s 
\begin{equation} \label{sflip2} 
H_{\rm flip} =-\sum_{\vec x} \sigma^3_1(\vec x) \; \sigma^3_1(\vec x+\vec e_2)\; \sigma^3_2(\vec x)\; \sigma^3_2(\vec x+\vec e_1) 
\end{equation} 
This flip operator commutes with the constraint, because a vertex and
a plaquette have either $0$ or $2$ common bonds. Hence, both operators
can be diagonalized simultaneously. In the model considered by Kitaev
[\onlinecite{kitaev97}], equal weight is given to all
types of vertices, so no term in the Hamiltonian is required. To go
beyond the point $a=b=c=d=1$, we need a term which will weight the
plaquettes according to the types of vertices present. Thus, we will
need to introduce operators which can discern among the various
vertices.  In addition, the ``weight" of the vertex depends on its
position in the plaquette under consideration. 
\begin{figure} 
\begin{center} 
\psfrag{1}{$\sigma^a_1(\vec x)$} 
\psfrag{2}{$\sigma^a_2(\vec x)$} 
\psfrag{3}{$\sigma^a_1(\vec x-\vec e_1)$} 
\psfrag{4}{$\sigma^a_2(\vec x-\vec e_2)$} 
\psfrag{x}{$\vec x$} 
\includegraphics[width=0.3 \textwidth]{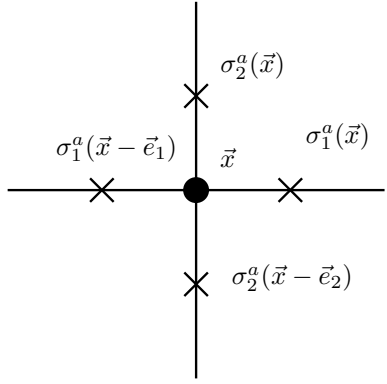} 
\caption{Notation for the $\sigma$'s; $a=1,2,3$ is the Pauli matrix label.} 
\label{sigmaconfig} 
\end{center} 
\end{figure} 
The notation that we use is summarized in figure \ref{sigmaconfig}.

Let us start by giving
the terms which give a non-zero contribution for $a,b,c$ and $d$
vertices respectively (but zero otherwise). 
Let us define the ``vertex magnetizations" 
\begin{eqnarray} 
S_a(\vec x)&=&\frac{1}{4}\left(\sigma^1_1(\vec x)+\sigma^1_2(\vec x)+\sigma^1_1(\vec x-\vec e_1)+\sigma^1_2(\vec x-\vec e_2)\right) 
\nonumber \\ 
S_b(\vec x)&=&\frac{1}{4}\left(\sigma^1_1(\vec x)-\sigma^1_2(\vec x)+\sigma^1_1(\vec x-\vec e_1)-\sigma^1_2(\vec x-\vec e_2)\right) 
\nonumber \\ 
S_c(\vec x)&=&\frac{1}{4}\left(\sigma^1_1(\vec x)-\sigma^1_2(\vec x)-\sigma^1_1(\vec x-\vec e_1)+\sigma^1_2(\vec x-\vec e_2)\right) 
\nonumber \\ 
S_d(\vec x)&=&\frac{1}{4}\left(\sigma^1_1(\vec x)+\sigma^1_2(\vec x)-\sigma^1_1(\vec x-\vec e_1)-\sigma^1_2(\vec x-\vec e_2)\right) 
\label{vms} 
\end{eqnarray} 
With this notation, the projectors onto the vertices $a$, $b$, $c$ and $d$ are just the squares of the vertex
 magnetizations of Eq.\ (\ref{vms}): 
\begin{equation} 
{\mathcal P}_a=S_a^{\;2}, \qquad {\mathcal P}_b=S_b^{\;2}, \qquad {\mathcal P}_c=S_c^{\;2}, \qquad {\mathcal P}_d=S_d^{\;2} 
\label{P-S} 
\end{equation} 
It is straightforward to show that these operators act as projection operators on vertices of type $a$, $b$, $c$ and $d$
 respectively, {\it i.e.\/} they yield $1$ when acting on the corresponding vertex and zero otherwise. 
 It is elementary to check that 
\begin{equation} 
{\mathcal P}_a+{\mathcal P}_b+{\mathcal P}_c+{\mathcal P}_d={\mathcal I} 
\label{identity} 
\end{equation} 
where ${\mathcal I}$ is the identity operator. 
 
We can now write down the potential term which assigns weights to the 
plaquettes in the same way as done by the projectors of section 
\ref{sec:q8v}. Flipping a plaquette will change a $d$ vertex to an 
$a$ vertex if the $d$ vertex is in the south-west 
or north-east ``even'' corner of the plaquette. A $d$ at 
the other ``odd'' corners will go to a $b$ under the flip. In short, 
$d_e \leftrightarrow a_e$, $d_o \leftrightarrow b_o$. For $c$'s this is 
opposite, namely 
$c_e \leftrightarrow b_e$, $c_o \leftrightarrow a_o$. 
The potential term will involve all four vertices around a plaquette, and thus 
we need to distinguish between the different position in the plaquette. 
This is achieved by the following, albeit rather cumbersome, term 
\begin{equation} \label{q8vpotsig} 
\begin{split} 
H_{\rm V} &= \sum_{\vec x} 
\left( 
\frac{d}{a} {\mathcal P}_a (\vec x) + 
\frac{c}{b} {\mathcal P}_b (\vec x) + 
\frac{b}{c} {\mathcal P}_c (\vec x) + 
\frac{a}{d} {\mathcal P}_d (\vec x) 
\right) \times \\ 
&\left( 
\frac{c}{a} {\mathcal P}_a (\vec x+e_1) + 
\frac{d}{b} {\mathcal P}_b (\vec x+e_1) + 
\frac{a}{c} {\mathcal P}_c (\vec x+e_1) + 
\frac{b}{d} {\mathcal P}_d( \vec x+e_1) 
\right) \times \\ 
&\left( 
\frac{c}{a} {\mathcal P}_a (\vec x+e_2) + 
\frac{d}{b} {\mathcal P}_b (\vec x+e_2) + 
\frac{a}{c} {\mathcal P}_c (\vec x+e_2) + 
\frac{b}{d} {\mathcal P}_d (\vec x+e_2) 
\right) \times \\ 
&\left( 
\frac{d}{a} {\mathcal P}_a (\vec x+e_1+e_2) + 
\frac{c}{b} {\mathcal P}_b (\vec x+e_1+e_2) + 
\frac{b}{c} {\mathcal P}_c (\vec x+e_1+e_2) + 
\frac{a}{d} {\mathcal P}_d (\vec x+e_1+e_2) 
\right) \ . 
\end{split} 
\end{equation} 
The potential term \eqref{q8vpotsig} assigns potential energies to the
plaquettes in the same way as is done by the projectors of section
\ref{sec:q8v}. That is, the potential is the product of vertex weights
obtained by flipping the plaquette, divided by the product of the
vertex weights of the plaquette itself. As mentioned before, the
two-body terms only couple spins on the same sublattice.  The most
non-local terms in the potential energy term consist of eight-body
interactions. Of course, a plaquette potential energy term of this
sort is needed, if one assigns weights to plaquettes\footnote{In spite
of the appearances, $H_V$ as given in Eq.\ \eqref{q8vpotsig} respects
rotational invariance; the apparent asymmetry is due to the use of the
vertex weights as labels.}.  Note that for $a=b=1$ and $c=d$ (or
$c=\tfrac{1}{d}$), the eight-body terms cancel each other. However,
there will remain four and six-body interactions, which will mix
different sublattices.
 
The total Hamiltonian for the $\mathbb{Z}_2$ 
gauge theory of the quantum eight vertex model is 
\begin{equation} 
H_{\rm q8v} = H_{\rm V} + H_{\rm flip} 
\label{hq8v} 
\end{equation} 
where the two terms are given by \eqref{sflip2} and \eqref{q8vpotsig}. The 
states in this model have to satisfy the constraint \eqref{scon8v2}. 
We can check these formulas in a few limits. For $a=b=c=d=1$,
each factor in \eqref{q8vpotsig} will be $1$ when acting on states
satisfying the constraint. 
Hence, the potential term became the identity operator, which merely results 
in an energy shift, and thus we find back the Kitaev model, as we must.

Another interesting limit is $d \rightarrow 0$. As was discussed in
section \ref{sec:8vqh}, this limit gives a slight generalization of the
six-vertex model, in the sense that plaquettes which have
$n_d=\widetilde{n}_d=1$ or $n_d=\widetilde{n}_d=2$ will have finite energy,
and they can be interpreted as static defects.
As can be seen from the potential term
\ref{q8vpotsig}, plaquettes which have $n_d>\widetilde{n}_d$ will
be suppressed as they receive infinite energy. 
The flip term has to be modified, because we need to have a flip term which
commutes with the constraint, which has become a stronger statement
in the six-vertex case, namely 
(c.f.\ Eq.\ \eqref{con6v}) 
\begin{equation} \label{scon6v} 
\sigma_1^1 (\vec x) - \sigma_{1}^1 (\vec x-\vec e_1) + 
\sigma_2^1 (\vec x) - \sigma_{2}^1 (\vec x-\vec e_2) = 0 \qquad \forall \vec x \ . 
\end{equation} 
The flip term which preserves the six-vertex constraints is
\begin{equation} \label{sflip6v} 
H_{\rm flip,6v} = - \sum_{\vec x} 
 \bigl(\sigma^-_1 (\vec x) \; \sigma^-_2 (\vec x + e_1) \;
\sigma^+_1 (\vec x + e_2) \; \sigma^+_2 (\vec x) + {\rm h.c.} \bigr) \ , 
\end{equation} 
where the raising and lowering operators (in the representation we use) 
are given by 
\begin{equation} 
\sigma^\pm = \tfrac{1}{2} (\sigma^3 \mp i\sigma^2) 
\end{equation} 
To make contact with the flip term for the eight-vertex model, we
rewrite Eq.\ \eqref{sflip} as
\begin{equation} 
\begin{split} 
H_{\rm flip,q8v} = & - \sum_{\vec x} 
\bigl(\sigma^+_1 (\vec x) + \sigma^-_1 (\vec x)\bigr) 
\bigl(\sigma^+_2 (\vec x + e_1) + \sigma^-_2 (\vec x + e_1)\bigr) \times \\ 
&\bigl(\sigma^+_1 (\vec x + e_2) + \sigma^-_1 (\vec x + e_2)\bigr) 
\bigl(\sigma^+_2 (\vec x) + \sigma^-_2 (\vec x)\bigr) \ . 
\end{split} 
\end{equation} 
The flip term for the six-vertex model is therefore precisely the flip
term for the eight-vertex model minus the terms which cause the
six-vertex constraint to be violated.  It is easily checked that the
six-vertex flip term Eq.\ \eqref{sflip6v} commutes with the constraint
\eqref{scon6v}.  We thus find that on the level of the wave function,
the limit $d \rightarrow 0$ is smooth, as the amplitude of the
configurations which contain $d$ vertices goes to zero. In addition,
for $d \neq 0$, the flip term commutes with constraint
\eqref{scon8v2}, while for $d=0$, the flip term commutes with the
$U(1)$ constraint \eqref{scon6v}.  Thus, the symmetry is enhanced from
$\mathbb{Z}_2$ for $d\neq 0$ to $U(1)$ for $d=0$, as was to be
expected.

\subsection{The dual of the gauge theory} 
\label{app:duality} 
 
\begin{figure}[h] 
\begin{center} 
\psfrag{x}{$\vec x$} 
\psfrag{R1}{$\vec r$} 
\psfrag{R2}{$\vec r-\vec e_1$} 
\psfrag{R3}{$\!\!\!\!\!\!\!\!\!\vec r-\vec e_1-\vec e_2$} 
\psfrag{R4}{$\vec r-\vec e_2$} 
\includegraphics[width=0.21 \textwidth]{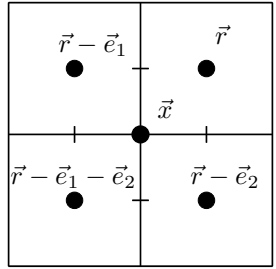} 
\caption{The dual lattice sites are labeled by $\vec r$.} 
\label{fig:dual} 
\end{center} 
\end{figure} 
 
We now have a Rokhsar-Kivelson generalization of the eight-vertex model, in 
a gauge-theory language. We can use this representation of our model to 
study the various phases. However, this is more easily done in a dualized 
version, as the dual takes the form of an Ising model. 
In the dual picture, the spin degrees of freedom will live on the
sites of dual square lattice, {\it i.e.\/} the centers of the
plaquettes of the direct lattice.  Thus, we will label by $\vec r$ the
site of the dual lattice on the center of the plaquette labeled by
$\vec x$ (its SW corner). Of course, the potential term in the dual
language will still be quite formidable. We will denote the dual Pauli
operators by $\tau^1$ and $\tau^3$. To start with the flip term, the
product of $\sigma^3$'s around a plaquette becomes $\tau^1$ on the
plaquette [\onlinecite{fradkin78,kogut79}]
\begin{equation} \label{tx} 
\tau^1 (\vec r) = 
\sigma^3_1(\vec x) \; \sigma^3_2(\vec x+\vec e_1) \;
\sigma^3_1(\vec x+ \vec e_2) \; \sigma^3_2(\vec x) 
\end{equation} 
To see what happens with the constraint and the projector operators defined by Eq.\ (\ref{vms}) and Eq.\ (\ref{P-S}) 
we need the dual form of the $\sigma^1$\;'s living on the 
links. In term of the dual variables $\tau^3$, and using the notation of Fig. \ref{fig:dual}, the $\sigma^1$\;'s are given 
by 
\begin{equation} \label{tz} 
\sigma^1_1(\vec x)= \tau^3(\vec r) \; \tau^3(\vec r-\vec e_2), \qquad
\sigma^1_2(\vec x)= \tau^3(\vec r) \; \tau^3(\vec r-\vec e_1) 
\end{equation} 
 
\begin{figure}[h] 
\begin{center} 
\psfrag{R}{$\vec r$} 
\psfrag{G}{$\Gamma$} 
\includegraphics[width=0.25 \textwidth]{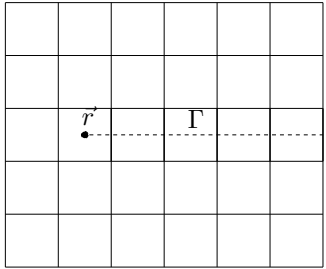} 
\caption{The dual path $\Gamma$.} 
\label{dual-path} 
\end{center} 
\end{figure}

We thus easily find that the constraint is automatically satisfied
(again, going to the dual picture amounts to solving the
constraint). Also, it is trivial to show
[\onlinecite{fradkin78,kogut79}] that the inverse relation, {\it
i.e.\/} to express the dual lattice $\tau^3$ operators in terms of the
$\sigma^1$ operators of the original lattice is
\begin{equation} 
\tau^3(\vec r)=\prod_{\ell \in \Gamma(\vec r)} \sigma^1(\ell) 
\label{tau3} 
\end{equation} 
where $\{\ell\}$ is a set of links of the direct lattice pierced by a path $\Gamma(\vec r)$ on the {\em dual} lattice ending at
 the dual site $\vec r$ (but which is otherwise arbitrary); see Fig. \ref{dual-path}. 
 
Note that we need to choose the spin on one of the plaquettes; all the 
others are subsequently determined by the $\sigma^x$'s on the links. 
In terms of the dual variables, the projection operators for site $\vec x$, defined by Eq.\ \eqref{P-S}, take the form 
\begin{equation} \label{vms-dual} 
\begin{split} 
\mathcal{P}_a(\vec x) &= \frac{1}{4} \left( 
1 + \mathcal{A}(\vec r) + \mathcal{B}(\vec r) + \mathcal{C}(\vec r) 
\right) \\ 
\mathcal{P}_b(\vec x) &= \frac{1}{4} \left( 
1 - \mathcal{A}(\vec r) - \mathcal{B}(\vec r) + \mathcal{C}(\vec r) 
\right) \\ 
\mathcal{P}_c(\vec x) &= \frac{1}{4} \left( 
1 - \mathcal{A}(\vec r) + \mathcal{B}(\vec r) - \mathcal{C}(\vec r) 
\right) \\ 
\mathcal{P}_d(\vec x) &= \frac{1}{4} \left( 
1 + \mathcal{A}(\vec r) - \mathcal{B}(\vec r) - \mathcal{C}(\vec r) 
\right) \ , \\ 
\end{split} 
\end{equation} 
where $\mathcal{A,B}$ and $\mathcal{C}$ are given by 
\begin{equation} 
\begin{split} 
\mathcal{A}(\vec r) &= \tau^3(\vec r-\vec e_1) \; \tau^3(\vec r-\vec e_2) \\ 
\mathcal{B}(\vec r) &= \tau^3(\vec r) \; \tau^3(\vec r-\vec e_1-\vec e_2) \\ 
\mathcal{C}(\vec r) &= \tau^3(\vec r) \; \tau^3(\vec r-\vec e_1) \; 
\tau^3(\vec r-\vec e_1) \; \tau^3(\vec r - \vec e_1-\vec e_2) \ . 
\end{split} 
\end{equation}

The interaction in terms of these projectors has the same
structure as in the spin representation of the classical eight-vertex
model: it consists of two-body terms on interpenetrating sublattices,
and a four-body term coupling the two sublattices. All the two-body interaction terms will only couple
spins on the same sublattice. The four- and six-body interaction terms
(of which there are many!), couple the sublattices. The same holds for
the eight-body term, naturally. The dual form of the
theory is
\begin{equation} 
H_{\rm q8v, dual} = H_{\rm V,dual} - \sum_{\vec r} \tau^1(\vec r) \ , 
\label{q8v-dual} 
\end{equation} 
where $H_{\rm V,dual}$ is given by \eqref{q8vpotsig}, but now with the
projectors given in Eq.\ \eqref{vms-dual}.  Thus, formally this theory
takes the form of a (multi-spin) Ising model in a transverse field.
However, the two-body interactions only couple spins on the same
sublattices, together with the multi-spin terms conspire to change the
quantum critical behavior from the conventional $z=1$
Lorentz-invariant criticality of the standard Ising model in a
transverse field to the $z=2$ quantum critical behavior discussed in
the rest of this paper.
 
Now that we found the dual version of our gauge theory, 
we would like to discuss the limits $a=b=c=d=1$ and 
$d=0$. Again, the first limit brings us back to the Kitaev point, because 
the potential term becomes the identity operator again, and we are left 
with the very simple spin flip term of Eq.\ \eqref{q8v-dual}, $H_{\rm f}=-\sum_{\vec r} \tau^1(\vec r)$. 
The limit $d \rightarrow 0$ is however more complicated in this dual 
gauge theory. First of all, we now do need a constraint, which was not 
present for $d \neq 0$. Moreover, the flip term now only can act, depending 
on the surrounding spins. 
Let us start by dualizing the constraint Eq.\ \eqref{scon6v}, which results 
in 
\begin{equation} \label{tcon6v} 
\bigl( \tau^3(\vec r) - \tau^3(\vec r-\vec e_1-\vec e_2) \bigr) 
\bigl( \tau^3(\vec r-\vec e_1) + \tau^3(\vec r-\vec e_2) \bigr) = 0 \qquad 
\forall \vec x \ . 
\end{equation} 
Obviously, the eight-vertex flip term $\tau^1 (\vec r)$ does not commute with 
this constraint. To find a flip term which does commute with the constraint, 
we dualize the six-vertex flip term \eqref{sflip6v}, which results in 
\begin{equation} \label{tflip6v} 
H_{\rm flip,q6v} = -\frac{1}{8}\sum_{\vec r} \tau^1 (\vec r) 
\bigl( 1-\tau^3(\vec r+\vec e_1) \tau^3(\vec r+\vec e_2) \bigr) 
\bigl( 1+\tau^3(\vec r-\vec e_1) \tau^3(\vec r+\vec e_2) \bigr) 
\bigl( 1+\tau^3(\vec r+\vec e_1) \tau^3(\vec r-\vec e_2) \bigr) \ . 
\end{equation} 
The factors in bracket can be seen to give a non zero result only on 
plaquettes which are flippable. Hence, this flip term commutes with 
the constraint \eqref{tcon6v}. Of course, this can also be checked explicitly. 
Apart from a factor $\tau^2$, there are factors depending on the $\tau^3$'s 
coming form both the flip term and the constraint. The signs in this product 
conspire in such a way to render the commutator zero.


\end{document}